%% file: main.tex
\documentclass{article}
\usepackage[english]{babel}
\usepackage[letterpaper,top=2cm,bottom=2cm,left=2.8cm,right=2.8cm,marginparwidth=1.75cm]{geometry}
\usepackage{amsmath,amsfonts}
\usepackage{graphicx}
\usepackage{array}
\usepackage{longtable}
\usepackage{caption}
\usepackage{float}
\usepackage[colorlinks=true, allcolors=blue]{hyperref}
\usepackage{natbib}
\usepackage{algorithm2e}
\usepackage{subfigure}
\usepackage{multirow}
\usepackage{makecell}
\usepackage{titlesec}
\usepackage{lineno}
\usepackage{xcolor}
\usepackage{colortbl}
\DeclareCaptionLabelFormat{cont}{#1~#2\alph{ContinuedFloat}}
\setcitestyle{authoryear,round}
\providecommand{\keywords}[1]
{\small	\textbf{Keywords:} traffic simulation, queue transmission, queue length,  time-varying free-flow speed, traffic flow propagation, urban road networks}
\date{}
\title{{A Link-Based Flow Model with Turn-Level Queue Transmission and Time-Varying Free-Flow Speed for Urban Road Networks}}

\begin{document}
\maketitle
\vspace{-4em}

\begin{center}
Lei Wei\footnotemark[1],
S. Travis Waller\footnotemark[2],
Yu Mei \footnotemark[3], Peng Chen\footnotemark[4],
Yunpeng Wang\footnotemark[4],
Meng Wang$^{*}$\footnotemark[1]
\end{center}
\footnotetext[1]{Chair of Traffic Process Automation, Technische Universität Dresden, Hettnerstraße 3, Dresden, 01069, Germany}
\footnotetext[2]{Chair of Transport Modelling and Simulation, Technische Universität Dresden, Hettnerstraße 1-3, Dresden, 01069, Germany}
\footnotetext[3]{Department of Intelligent Transportation System, Baidu, No. 10 Shangdi 10th Street, Haidian District, Beijing 100085, China}
\footnotetext[4]{Beijing Key Laboratory for Cooperative Vehicle Infrastructure Systems and Safety Control, Beihang University, Beijing 100191, China}
\renewcommand{\thefootnote}{\fnsymbol{footnote}}
\footnotetext[1]{Corresponding author: meng.wang@tu-dresden.de}
\renewcommand{\thefootnote}{\arabic{footnote}}

\begin{abstract}

Macroscopic link-based flow models are efficient for simulating flow propagation in urban road networks. Existing link-based flow models described traffic states of a link with  two state variables of link inflow and outflow and assumed homogeneous traffic states within a whole link. Consequently, the turn-level queue length change within the link can not be captured, resulting in underrepresented queue spillback. Moreover, a constant link free-flow speed was assumed to formulate models, restricting their applicability in modeling phenomena involving time-varying free-flow speed. This study proposed a new link-based flow model by introducing an additional state variable of link queue inflow and adapting the link outflow to be free-flow speed-dependent. In our model, the vehicle propagation within each link is described by the link inflow, queue inflow, and outflow, which depends on the link free-flow speed changes. A node model is further defined to capture the presence of signal control and potential queue spillback, which estimates the constrained flow propagation between adjacent road segments. Simulation experiments were conducted on a single intersection and a network with consecutive intersections to verify the proposed model performance. Results demonstrate the predictive power of the proposed model in predicting traffic operations of intersections with multiple turning movements and time-varying free-flow speed. Our model outperforms the baseline link-based flow model and preserves the computational tractability property of link-based flow models.

\end{abstract}
% \linenumbers  % Enable line numbering
% \pagewiselinenumbers  % Restart line numbering on each page
% \renewcommand\linenumberfont{\footnotesize}
\section{Introduction}

\input{Introduction}

\section{Mathematical Model Formulation}
\input{Mathematical_Model_Formulation}
\subsection{Link model}
\input{Link_model}

\subsubsection{Queue density}
\input{Queue_density}
\subsubsection{Queue inflow}
\begin{figure}[H]
\captionsetup{font={small}}
\centering
\includegraphics[width=5.3in]{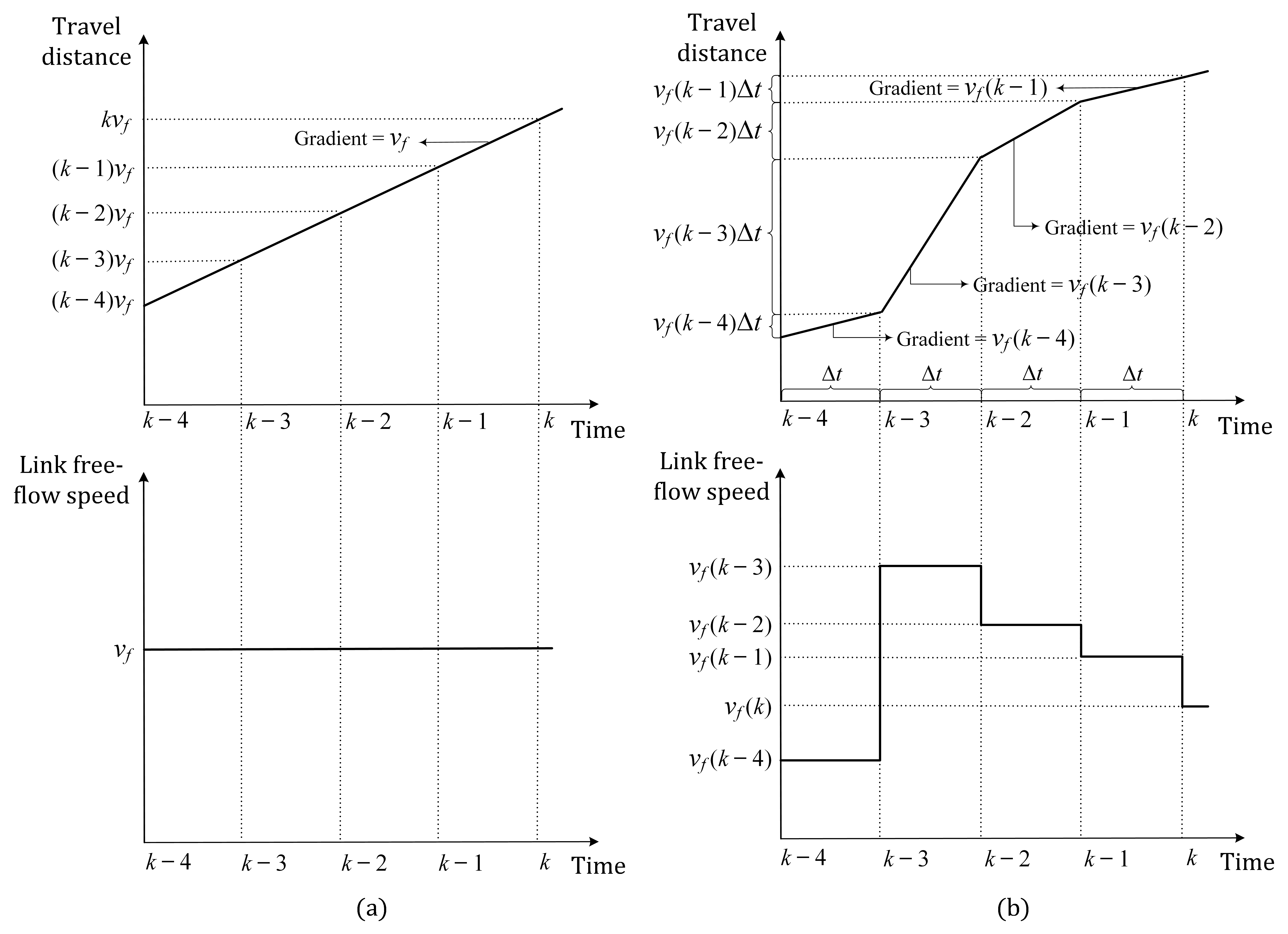}
\caption{{Illustration of distance
travelled by the entered flows on the link free-flowing part at different sampling time steps between $[k-4,k]$. (a) Fixed free-flow speed; (b) Time-varying free-flow speed.}}
\label{distance}
\end{figure}

% Although a few models attempted to capture the queue inflow dynamics by considering the shockwave speeds
% and congested traffic states \citep{ma2014continuous,raovic2017dynamic}, the variable
% free-flow speeds within an RS are not distinguished in their models, which makes it difficult to be applied in dynamic traffic management strategies involving TFS.  
Different from existing link-based flow models, the queue transmission within the link formulated by the turn-level queue inflow in our model is determined in two ways depending on whether the free-flow speed changes. As shown in Figure \ref{distance}, in the case of a fixed free-flow speed, it is possible to calculate the time steps required to pass the free-flow length. If the free-flow speed is variable, we need to determine the actual queue inflow based on the distance
travelled by the entered flows over a specific period.
% The queue inflow can be determined based on the free-flow length $ L_f(k)$ and speed. Existing models usually assumed a fixed free-flow speed to calculate the time steps required by vehicles to pass $ L_f(k)$ \citep{bliemer2007dynamic,van2018hierarchical,raovic2017dynamic,ma2014continuous}.  If the free-flow speed is time-varying, we can determine the queue inflow based on the distance traveled by the entered vehicles over a specific period, as shown in Figure \ref{distance}. We first present the queue inflow model based on constant free-flow speed, followed by its time-varying speed counterpart. 

%\noindent
\textbf{(a) Fixed free-flow speed}

{Assuming a fixed speed $v_{f}$ to travel the free-flow part of the link as shown in Figure \ref{distance}(a), the free-flow travel time $t_{f}$ at  $k$ can be calculated:
\begin{equation}
    t_{f}= \frac{ L_{f}(k)}{v_{f}}~,~n_{f}= \left\lceil {{t_{f}}/\Delta t} \right\rceil~,~\gamma_{f} = n_{f} - \frac{t_{f}}{\Delta t}
    \label{sampling time steps}
\end{equation}
where $n_{f}$ is the number of sampling time steps; $\gamma_{f}$ represents the fraction of sampling time steps and $0 \leq \gamma_{f} < 1$.}

{Based on the cumulative link inflow stored in the state variable shown in Eq. (\ref{state}), it is possible to calculate the cumulative queue inflow at $k$:
\begin{equation}
\label{N_qu}
    N_{i}^{qu}\left( k \right) = \gamma_{f} N_{i}^{in}\left( k + 1 - n_{f}\right) + \left(1 - \gamma_{f} \right) N_{i}^{in}\left( k - n_{f}\right)
\end{equation}}
 
{ Note that when $ L_f(k)$ shrinks towards zero as the queue is almost filling up the whole link, vehicles entering the link will immediately become part of the queue, i.e. $n_f = 1$. Although this situation is addressed in the link model through a workaround, the queue inflow calculated by Eq. (\ref{N_qu}) may potentially violate the Courant Friedrich and Lewy (CFL) conditions \citep{himpe2016efficient}. To satisfy the CFL conditions, the sampling time step in Eq. (\ref{N_qu}) should satisfy $k>k+1-n_f$, hence, $n_f>1$, implying $n_f\ge 2$. If $n_f = 1$, 
 the cumulative queue inflow $N_i^{in}(k)$ at $k$ estimated by Eq. (\ref{N_qu}) cannot be really updated since the current cumulative link inflow $N_i^{in}(k)$ is not available. This limitation, however, will be addressed in the subsequent time-varying free-flow speed scenario.}
 
\textbf{(b) Time varying free-flow speed}

If the free-flow speed is time-varying as shown in Figure \ref{distance}(b), it is impossible to directly determine the number of sampling time steps ${n}_{f}$ in Eq. (\ref{sampling time steps}). Instead, we use  matrix operations to select vehicles that have passed the queue tail to estimate the queue inflow. It is assumed that a minimum desired speed $v_{min}$ exists and can be used to determine the number of sampling time steps $\bar {n}_{f}$ required to pass the whole link in free-flow conditions:

\begin{equation} 
    \bar t_{f} = \frac{L_{i}}{v_{min}}~,~\bar {n}_{f} = \left\lceil {t_{f}/\Delta t} \right\rceil
    \label{t_f,min}
\end{equation}

{From Eq. (\ref{t_f,min}), we can find that the number of sampling time steps only depends on $v_{min}$ that does not relate to the change of free-flowing length, thus the potential violation of CFL conditions caused by $ L_f (k)$ shrinks towards zero can be avoided. The idea now is to calculate the potential travel distance for the incoming flows between $k - \bar {n}_{f}$ and $k$ and sum the number of vehicles that have passed the free-flow length to obtain the queue inflow. Therefore, some auxiliary matrixes and variables are defined. The potential travel distance $\boldsymbol{D}_{f}$ of flows entering the road segment at $k$ is determined based on the cumulative sum of the time-varying free-flow speeds in $
\left\lbrack {k - \bar {n}_{f},k} \right\rbrack$ as follows:
\begin{equation}
    \boldsymbol{D}_{f} = \boldsymbol{T}_{f}\begin{bmatrix}
    {{v}_{f}\left( \mathit{\max}\left( 1,k - \bar {n}_{f} \right) \right)} \\
    {{v}_{f}\left( \mathit{\max}\left( 1,k - \bar {n}_{f} \right) + 1 \right)} \\
     \vdots \\
    {{v}_{f}(k)} \\
    \end{bmatrix}\Delta t \in \left\{ \begin{matrix}
    {\mathbb{R}^{(\bar {n}_{f} + 1) \times 1},~k > \bar {n}_{f}} \\
{\mathbb{R}^{k \times 1},~k \leq \bar {n}_{f}~~~~~~} \\
    \end{matrix} \right.
\end{equation}
where $\boldsymbol{T_f}$ is an upper triangular matrix as follows:
\begin{equation}
    \boldsymbol{T}_{f} = \begin{bmatrix} 1 & 1 &\cdots & 1\\
    0 & 1 &\cdots & 1\\
    \vdots & \ddots &\ddots& \vdots\\
    0 & \cdots &0 & 1\\
    \end{bmatrix} \in \left\{ \begin{matrix}{\mathbb{R}^{{(\bar n}_{f} + 1){\times (\bar n}_{f}  + 1)},~k > \bar n_{f} } \\{\mathbb{R}^{k \times k},~k \leq \bar n_{f}~~~~~~~~~~~~~} \\\end{matrix} \right.
\end{equation}}
%     \begin{equation}
%     \boldsymbol{v}_{f} \in \left\{ \begin{matrix}
%     {\mathbb{R}^{{({\bar {n}_{f}  + 1})} \times 1},~k > \bar {n}_{f} } \\
%     {\mathbb{R}^{k \times 1},~k \leq \bar {n}_{f} ~~~~~~~~~~~} \\
%     \end{matrix} \right.
% \end{equation}

A binary variable $ \delta_{qu}(j)$ is used to select the incoming vehicles at different time steps which potentially passed the free-flow part of the link:
\begin{equation}
     \delta_{qu} (j) =  D_{f}(j) >  L_{f}(k)
\end{equation}
where $j=1,2,...,p$; $p= \left\{ \begin{matrix}
    {{\bar {n}_{f} + 1 },~k > \bar {n}_{f}} \\
{k},~k \leq \bar {n}_{f}~~~~~~~~ \\
    \end{matrix} \right.$, which represents the number of rows in $\boldsymbol{D_{f}}$. 

If $ D_{f}(j) >  L_{f}(k)$,  $\delta_{qu}(j) = 1$, otherwise $\delta_{qu}(j)  = 0$, and $ \boldsymbol \delta_{qu} \in \left\{ \begin{matrix}
    {\mathbb{R}^{(\bar {n}_{f}  + 1) \times 1},~k > \bar {n}_{f} } \\
    {\mathbb{R}^{k \times 1},~k \leq \bar {n}_{f}~~~~~~} \\
    \end{matrix} \right.$.

The link inflow change $\boldsymbol{\mathrm{\Delta}N}_{in}$ up to $k$ can be again derived from the former cumulative link inflows by:
\begin{equation}
    \boldsymbol{\mathrm{\Delta}N}_{in} = \boldsymbol{T}_{in}\begin{bmatrix}
   \left.  N_{i}^{in}\left( {\mathit{\max}\left( 1,k - \bar {n}_{f} \right.} \right) \right) \\
    \left. N_{i}^{in}\left( {\mathit{\max}\left( 1,k - \bar {n}_{f} \right.} \right) + 1 \right) \\
    \begin{matrix}
     \vdots \\
    {{N_{i}^{in}}(k)} \\
    \end{matrix} \\
    \end{bmatrix}\in \left\{ \begin{matrix}
    {\mathbb{R}^{(\bar {n}_{f}  + 1) \times 1},~k > \bar {n}_{f}} \\
    {\mathbb{R}^{k \times 1},~k \leq \bar {n}_{f} ~~~~~~~} \\
    \end{matrix} \right.
\end{equation}
where $\boldsymbol{T}_{in} = \begin{bmatrix}
 0 &0&0 &\cdots & 0\\
{- 1} & 1 &0  &\cdots  &0  \\
 0& {- 1} & 1 &\ddots  &\vdots \\
 \vdots & \ddots & \ddots & \ddots &0  \\
 0&\cdots &  0& -1 & 1 \\
\end{bmatrix} \in \left\{ \begin{matrix}
{\mathbb{R}^{(\bar n_f +1){\times (\bar n_f  + 1)},~k > \bar n_f}} \\
{\mathbb{R}^{k \times k},~k \leq \bar n_f ~~~~~~~~~~} \\
\end{matrix} \right.$.

% where $\boldsymbol{T}_{in} = \begin{bmatrix}
%  0 & 0 & 0 & \cdots & 0& 0 \\
% {- 1} & 1 & 0 & \cdots & 0& 0 \\
% 0 & {- 1} & 1 & \cdots & 0& 0 \\
%  \vdots & \vdots & \vdots & \ddots & \vdots& \vdots \\
% 0 & 0 & 0 & \cdots & -1& 1 \\
% \end{bmatrix} \in \left\{ \begin{matrix}
% {\mathbb{R}^{(\bar n_f +1){\times (\bar n_f  + 1)},~k > \bar n_f}} \\
% {\mathbb{R}^{k \times k},~k \leq \bar n_f ~~~~~~~~~~} \\
% \end{matrix} \right.$.

{Thus, the cumulative queue inflow of $i$ at $k$ under time-varying free-flow speed is now determined by:
\begin{equation}
\label{N_qu_LT3}
     N_{i}^{qu}\left( {k} \right) =  N_{i}^{in}\left( {k - \bar {n}_{f}} \right) + \boldsymbol{\mathrm{\Delta}N}{_{in}}^\mathsf{T}\boldsymbol\delta_{qu}
\end{equation}}
\vspace{-\baselineskip}
\vspace{-2em}
\subsubsection{Link inflow}
The link inflow of $i$ is constrained by the upstream link outflow and the inflow limit of $i$, and hence depends on the queue lengths and backward wave speeds. The time $t_{sh}$ determined for the backward wave in the queue to reach the tail of the queue is determined by $w_i$:
\begin{equation}
    t_{sh} = \frac{ L_{q}(k)}{w_i},~{n}_{sh} = \left\lceil {{t_{sh}}/\Delta t} \right\rceil~,~{~\gamma}_{sh} = {n}_{sh} - \frac{t_{sh}}{\Delta t}
\end{equation}

{Based on the cumulative link outflow stored in the state variable shown in Eq. (\ref{state}), 
we can determine the backward wave travels upstream in the queue so that the vehicle can enter the link after ${n}_{sh}$-th steps:
\begin{equation}    
{N}_{i}^{in,sh}\left( {k + 1} \right) = {\gamma}_{sh}{N}_{i}^{out}\left( {k + 2 - {n}_{sh}} \right) + \left( 1 - {\gamma}_{sh}{){N}}_{i}^{out}\left( {k + 1 - {n}_{sh}} \right) \right.
\label{sh}
\end{equation}}

{Thus, the cumulative inflow limit $\bar{N}_{i}^{in}$ from the origin or the upstream node is given by:
\begin{equation}
    \bar{N}_{i}^{in}\left( {k + 1} \right) =  N_{i}^{in,sh}\left( {k + 1} \right) + {  \rho_{i}^{q}(k)  L_{q}(k) + \rho_{i}^{jam}  L_{f}(k) }
\end{equation}}

{The allowed maximum inflow rate of $i$ at $k$ can be obtained:
\begin{equation}
     \bar {q}_{i}^{in} (k)= \left( { \bar{N}_{i}^{in}\left( {k + 1} \right) -  N_{i}^{in}(k)} \right)/\Delta t
  \label{in_max}
\end{equation}}

{The actual link inflow rate at $k$ is the minimum of the maximum value $ \bar {q}_{i}^{in} (k)$ and the desired value $ q_{i}^{in,des} (k)$ in the link (from origin or upstream node):
\begin{equation}
\label{q_in}
     q_{i}^{in} (k)= \min \left( \bar {q}_{i}^{in} (k), q_{i}^{in,des} (k)\right)
\end{equation}
where the detailed estimation about desired link inflow rate $ q_{i}^{in,des} (k)$ will be introduced later. Thus, the cumulative link inflow at $k+1$ can be obtained:
\begin{equation}
\label{cum_in}
     N_{i}^{in} (k+1)=    N_{i}^{in} (k)+   q_{i}^{in} (k) \Delta t
\end{equation}}
\vspace{-3em}
\subsubsection{Link outflow}
The link outflow can be limited either by an outflow limit of the link representing a bottleneck or limited inflow capacity of a downstream link or by the vehicles on the link that have not been able to traverse the complete link since their entry. Similar to the queue inflow, for the time-varying free-flow speed, the maximum potential outflow, i.e., vehicles which would have traveled the link in free-flow conditions, is first estimated. A binary variable ${\delta}_{out}(j)$ is used again to select the vehicles which potentially passed the link $i$:
\begin{equation}
 \delta_{out}(j)= {D}_{f}(j) > L_{i}
\end{equation}
If $ D_{f}(j) > L_{i}$, $ \delta_{out}(j) = 1$, otherwise $ \delta_{out}(j) = 0$. {The desired cumulative outflow of $i$ is now determined by:
\begin{equation}
{N}_{i}^{out,des}\left( {k + 1} \right) = {N}_{i}^{in}\left( {k - \bar n_{f}} \right) + { \mathrm{\boldsymbol \Delta}N}{_{in}}^\mathsf{T}\boldsymbol \delta_{out}
\label{inflow limit}
\end{equation}}

{The maximum cumulative outflow of $i$ at $k$ can be calculated:
\begin{equation}
\label{max_out}
\bar {N}_{i}^{out}\left( {k} \right) = \mathit{\min}\left( {N}_{i}^{out}(k) + {q}_{sat}(k){b}_{i}(k),{N}_{i}^{out,des}\left( {k} \right) \right)
\end{equation}
where ${q}_{sat}(k)$ represents the saturation flow rate; ${b}_{i}(k)$ represents the effective fraction of green time and ${b}_{i}(k) \in (0,1]$, which indicates the impact of signal control on the node's incoming link outflow.}

{The maximum outflow rate $\bar {q}_{i}^{out}$ of $i$ at $k+1$ can be calculated:
\begin{equation}
\label{q_out_des}
\bar {q}_{i}^{out}\left( {k} \right) = \left( {\bar {N}_{i}^{out}\left( {k + 1} \right) - \bar {N}_{i}^{out}(k)} \right)/\Delta t
\end{equation}}
% Using ${Q}_{k}^{out,des}(k)$ to represent the desired total outflow rates of all RSs in the network at the sampling instant $k$:
% \begin{equation}
% {Q}_{k}^{out,des}(k) = \left\lbrack {q_{1,\Sigma}^{out,des}(k),q_{2,\Sigma}^{out,des}(k),\ldots,q_{m,\Sigma}^{out,des}(k)} \right\rbrack
% \end{equation}

\subsection{Node model}
\input{Node_model}

\section{Model verification}
This section presents some numerical simulation results to verify our proposed LQM with queue transmission and TFS for traffic flow simulation. {To this end, we perform two case studies: (1) In case study 1, simulation experiments are performed using a simulated intersection with spillback risks to assess the flow propagation using the proposed LQM. The superiority of incorporating the turn-level queue transmission and TFS in link-based flow model for capturing flow propagation can thus be demonstrated in detail. The proposed LQM is validated by comparing the outputs with those obtained from the microscopic traffic simulation by SUMO, and also compared with the traditional link-based flow model; (2) In case study 2, we expand the test scenario to encompass a field network featuring three consecutive intersections. The unbalanced traffic demands on the west/east and north/south incoming links are further employed to verify the performance of our proposed model in real-life network structures. The performance of the proposed LQM is compared with simulation results by SUMO due to the limitations of field traffic data collection.}
\input{Simulation_experiments}

\section{Conclusion}
This study proposed a macroscopic link-based flow model with queue transmission and TFS to capture the flow propagation for road networks. Different from existing link-based flow models that only focused on the inflow and outflow at the road segment boundaries, the flow propagation in our model is determined by using the link inflow, queue inflow and outflow across the entire road segment, which enables a more accurate representation of flow propagation providing queue length information. By determining the potential travel distance for the incoming flows under time-varying free-flow speeds,  the proposed model can be applied without the assumption of constant free-flow speed. We further developed a node model to formulate vehicle propagation among adjacent links. This node model allocates link supply, e.g., maximum allowed link inflow, among feeding links to regulate upstream desired link outflow, enabling derivation of link outflow without assuming downstream infinite capacity.

Simulation experiments were conducted on both a single intersection and a network with consecutive intersections to verify the performance of the proposed model using SUMO. The results show that the average cumulative flow difference can be reduced from 6.7veh with a baseline link-based flow model to 2.45veh with the proposed model in a standard four-arm intersection with three turning directions on each approach. Furthermore, the proposed model yields reliable flow propagation results in a real-sized network.

% we can make the LTM applied to intersections with multiple turning movements and TFS.

We acknowledge that the urban traffic flow contains multiple travel modes (e.g., private vehicle, bus, tram, cyclist, pedestrian), but the proposed model only focuses on motor traffic. To this end, the interactions among different travel modes should be further investigated \citep{he2014multi,cheng2019applying} to formulate a multimodal link-based flow model. Moreover, the proposed model assumed fixed vehicle turning rates at the intersections, hence the uncertainty of vehicle routing at the network level is overlooked. The model in large-scale networks with multiple travel modes and uncertain vehicle route choices warrant more studies in the future.

\begin{figure}[H]
\captionsetup{font={small}}
\centering
\includegraphics[width=3.1in]{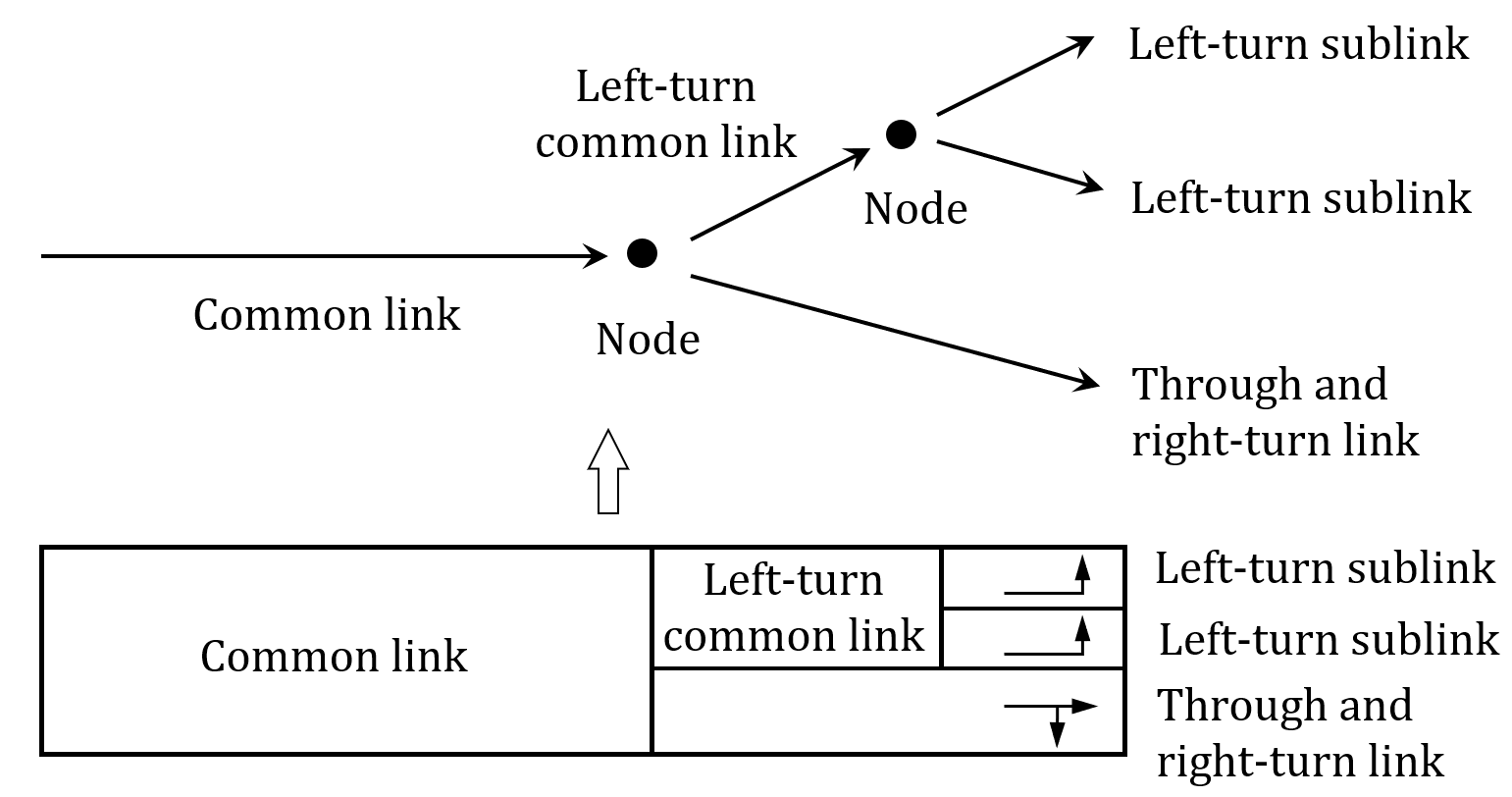}
\caption{\parshape=1 2em \dimexpr\linewidth-2em {Example of a road segment with lane configuration (L)$|$L$|$TR}}
\label{lane2}
\end{figure}

{\section*{Appendix A: Lane Configuration Discussion}}
{This appendix discusses the formulation of the proposed link model when dealing with other lane configurations in the networks. Figure \ref{LTM_structure} describes the situation where the separate lanes are designed, namely the following lane configuration: L$|$T$|$R. If there is lane sharing on the road, such as L$|$T$|$TR, where through traffic can choose the middle lane and the right lane. The queue length of TR consists of through and right-turn vehicles. We assume that the vehicle turning rates on the shared lane TR are known. The outflow for a certain turning direction of TR is equal to the realized total outflow multiplied by the corresponding turning rate that is similar to the existing link-based flow models \citep{yperman2005link,van2018hierarchical}. For other lane configurations, e.g., (L)$|$L$|$TR, where an additional left lane only becomes available near the intersection but is not available at the beginning of the link. This configuration also can be simulated by the proposed LQM, in which the left-turn link can be further divided into a common link upstream, followed by two parallel turn lanes as shown in Figure \ref{lane2}. Then, the common link and turn links are connected by the node model that is similar to the illustration in Figure \ref{LTM_structure}.}

\parshape=1 2em \dimexpr\linewidth-2em

{\section*{Appendix B: Proof of Invariant Holding-Free Solution}}

Algorithm \ref{algotithm1} is an invariant holding-free algorithm that satisfies the IP and can provide a holding-free solution \citep{jabari2016node}.

{{\bf Proof:} The holding-free solution by Algorithm \ref{algotithm1} is first proved. It is assumed that for some links $j$ that have:
\begin{equation}
    s_j=\min_{e_{j,i}>0} \frac{s_i}{e_{j,i}}<\delta_{j},~i\in \boldsymbol I^{out}
    \label{p1}
\end{equation}
where $\delta _{j}=\bar{N}_{j}^{out}(k+1)$. Let $o$ represent any outgoing link that the minimum is \textit{not} attained. That is,
\begin{equation}
    \frac{s_{o}}{e_{j,o}}>s_j=\min_{e_{j,i}>0} \frac{s_i}{e_{j,i}}
    \label{p2}
\end{equation}
where $s_{o}$ can be regarded as having a residual supply after the demands on incoming links $1,2,...,j$ have been assigned. We can have the residual supply of $o$:
\begin{equation}
    s_{o}'=s_{o}-{e_{j,o}}\min_{e_{j,i}>0} \frac{s_i}{e_{j,i}}>s_{o}-{e_{j,o}}\frac{s_{o}}{e_{j,o}}=0
\end{equation}}

{Now let $o$ denote an outgoing link where the minimum \textit{is} attained, indicating that ${e_{h,o}} > 0$ for link $h$, and $N_h^{out}(k+1)=s_{o}/{e_{h,o}}$. If $\Delta s_o=0$, then $\Delta N_h^{out}(k+1)=0$; in such instances, the entire supply of link $o$ has been exhausted by incoming links $h$ ($h < j$), and there is no further demand that can utilize link  $o$. If $\Delta s_{o}>0$, then
\begin{equation}
    s_{o}'=s_{o}-{e_{j,o}}q_j=s_{o}-{e_{j,o}}\frac{s_{o}}{e_{j,o}}=0
\end{equation}}

Hence, as shown in Eq. (\ref{p2}), due to $e_{j,o} > 0$, there is no available supply along $o$ that can be utilized by any $j$ ( $j > h$). On the other hand, if
\begin{equation}
    s_j=\min_{e_{j,i}>0} \frac{s_i}{e_{j,i}}\ge \delta_{j}
\end{equation}
we can have $q_j-\delta_j=0$. Therefore, Algorithm \ref{algotithm1} provides holding-free solutions. The proof that the Algorithm \ref{algotithm1} generates solutions that satisfy the IP is straightforward: for any $j$, we use $N_j$ to represent the increased demand, then
\begin{equation}
   \min \left( N_j,\min_{e_{j,i}>0} \frac{s_i}{e_{j,i}} \right)=\min\left(\delta_j,\min_{e_{j,i}>0} \frac{s_i}{e_{j,i}}\right)=\min\left(\bar{N}_{j}^{out}(k+1), s_j\right)=N_j^{out}(k+1)
   \label{p3}
\end{equation}
since $\delta_j < N_j$. Therefore, the solution remains invariant under increases in demand within the constraints imposed by the supplies. On the contrary, if the solution is constrained by demand, increasing the supply for outgoing links does not impact the results since Eqs. (\ref{p2}) and (\ref{p3}) still holds. Lastly, as $j$ is arbitrary, these properties hold for other solutions of Algorithm \ref{algotithm1}. $\Box$ 

\begin{figure}[H]
\captionsetup{font={small}}
\centering  
\subfigure[Cumulative flow of L10 by LM]{
\includegraphics[width=5cm,height = 4.1cm]{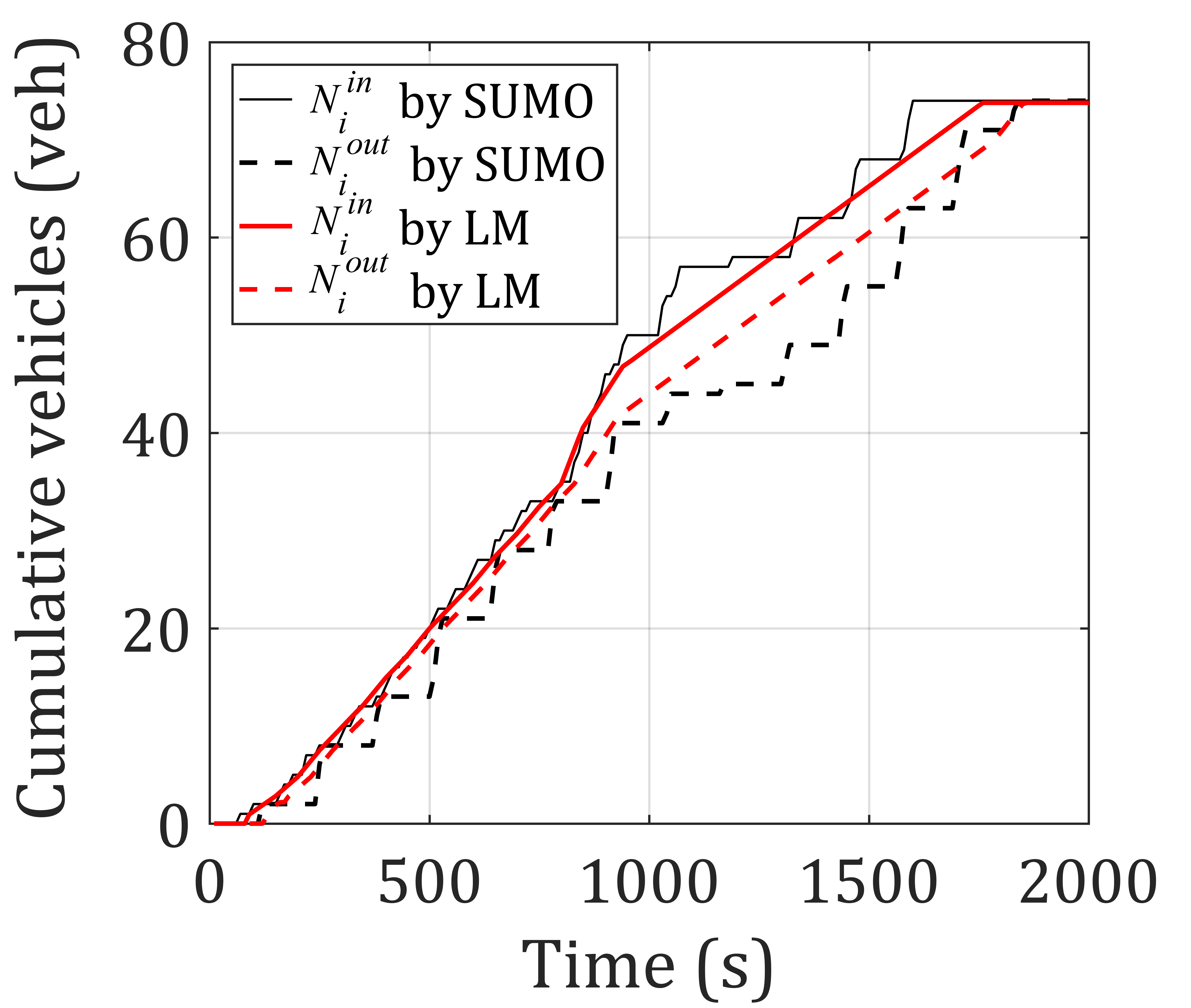}}\subfigure[Cumulative flow of L10 by LQM]{
\includegraphics[width=5cm,height = 4.1cm]{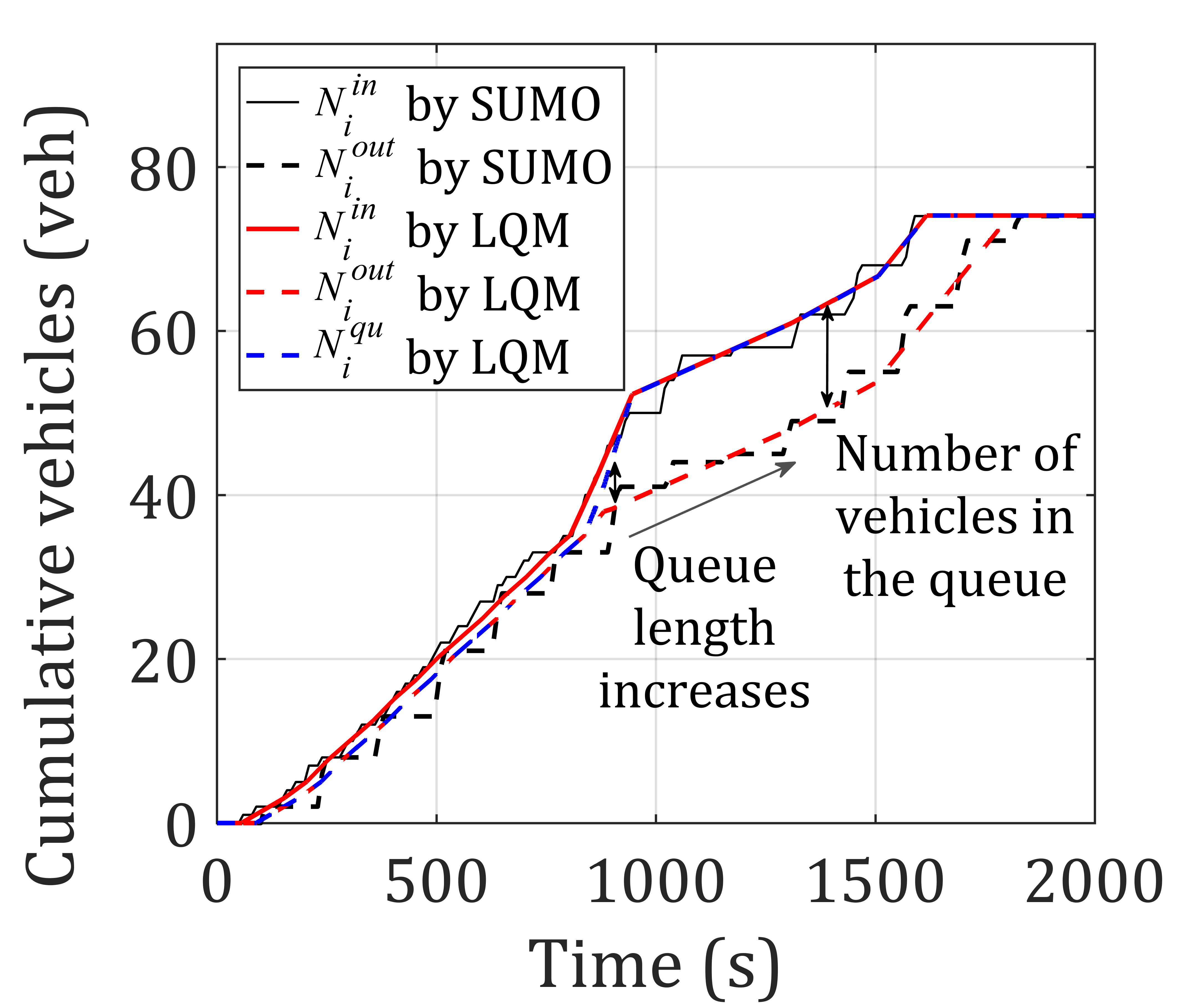}}\subfigure[Queue length of L10 by LQM]{
\includegraphics[width=4.9cm,height = 4.1cm]{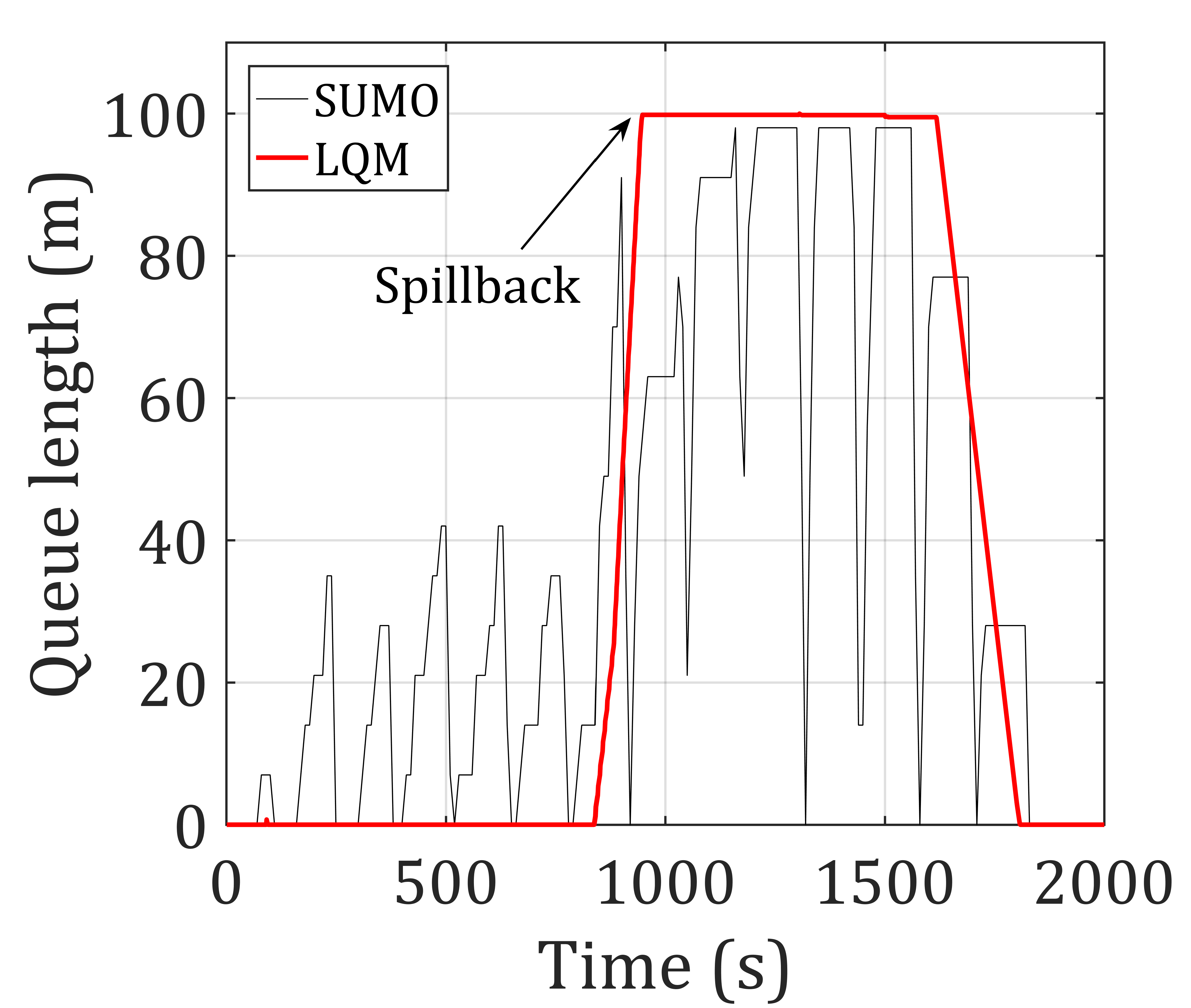}}\\
\subfigure[Cumulative flow of L28 by LM]{
\includegraphics[width=5cm,height = 4.1cm]{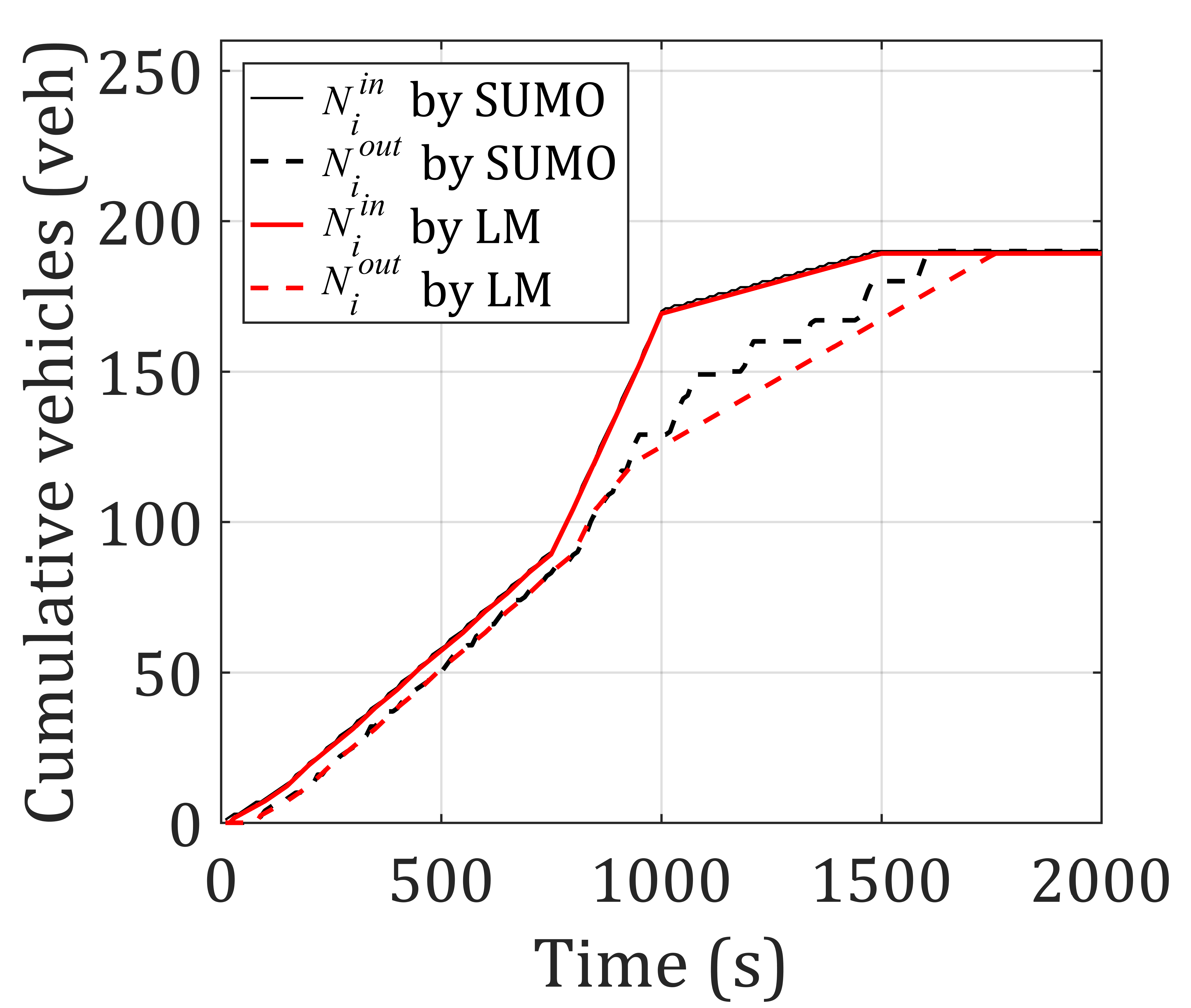}}
\subfigure[Cumulative flow of L28 by LQM]{
\includegraphics[width=5cm,height = 4.1cm]{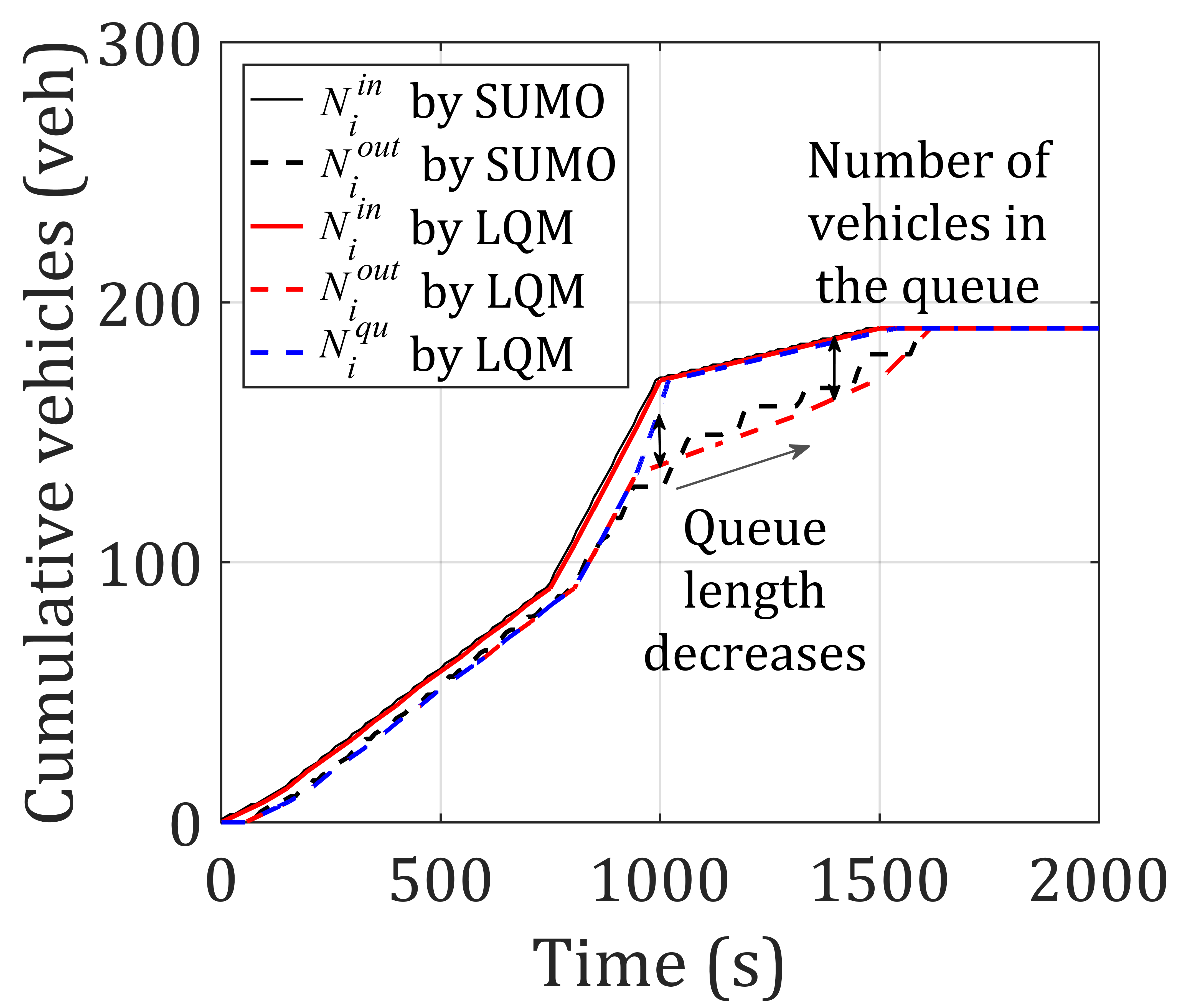}}\subfigure[Queue length of L28 by LQM]{
\includegraphics[width=4.9cm,height = 4.1cm]{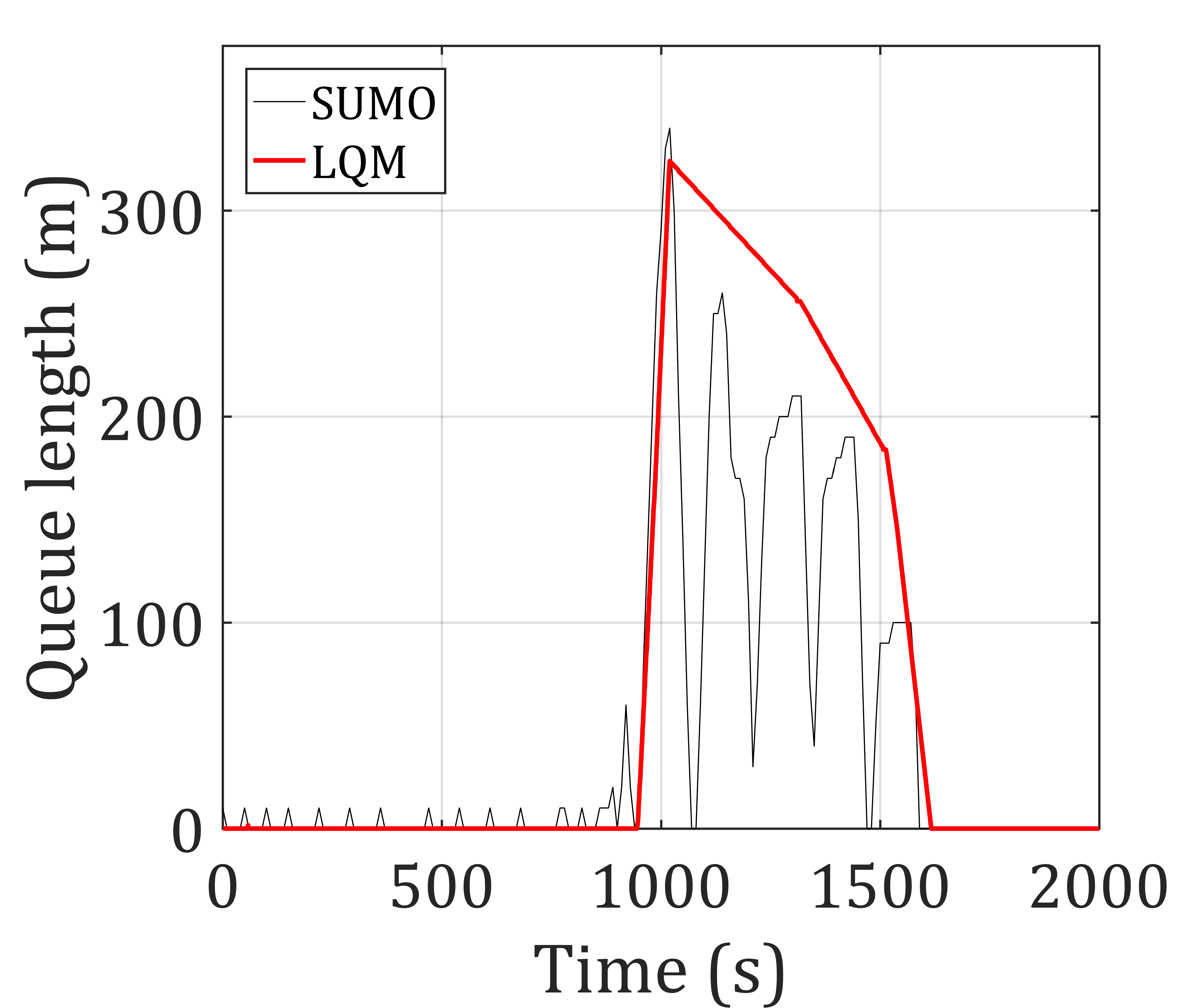}}
\subfigure[Flow rate (m/s) of L10 by LQM]{
\includegraphics[width=6.4cm,height = 4.3cm]{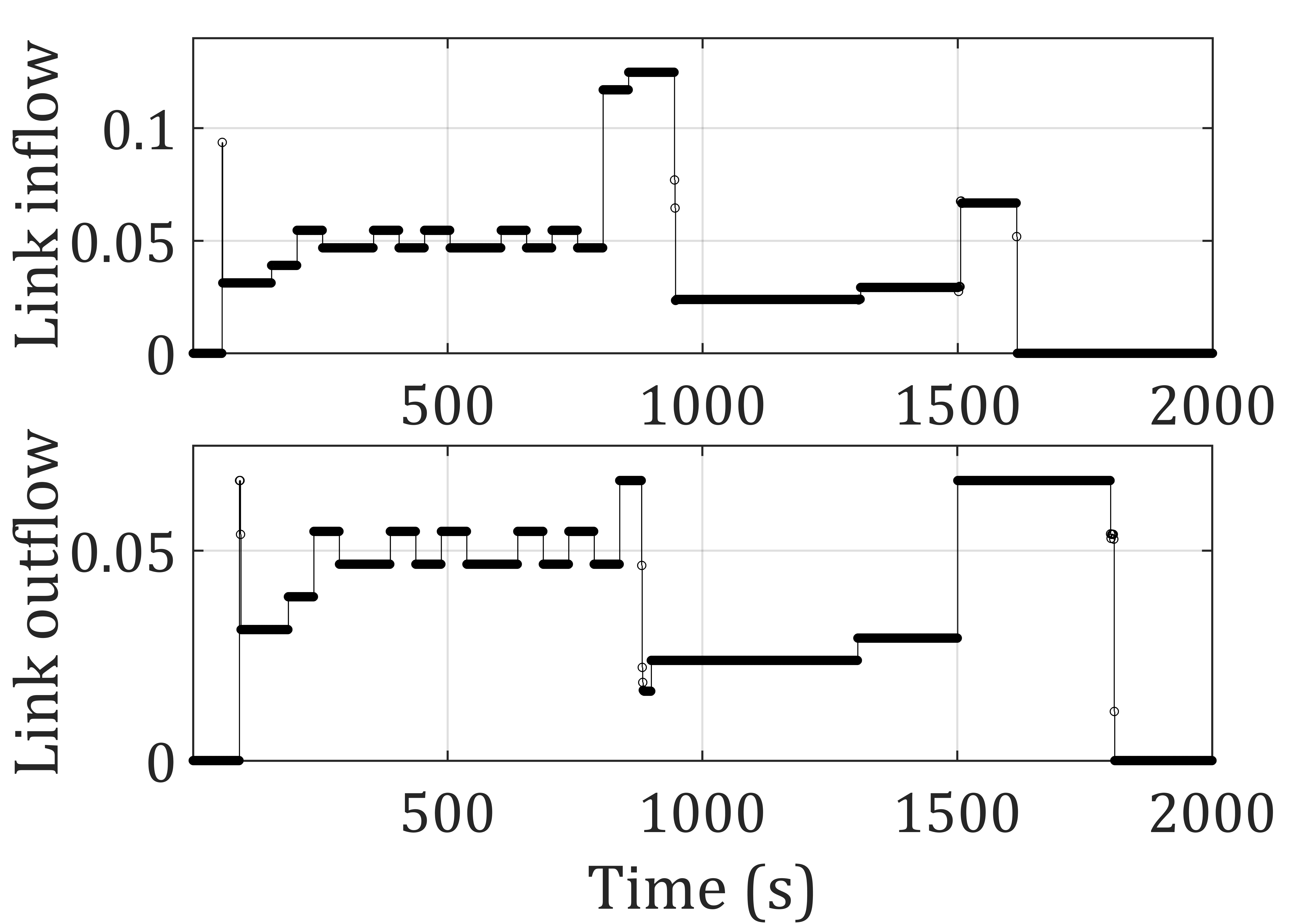}}~~~\subfigure[Flow rate (m/s) of L28 by LQM]{
\includegraphics[width=6.4cm,height = 4.3cm]{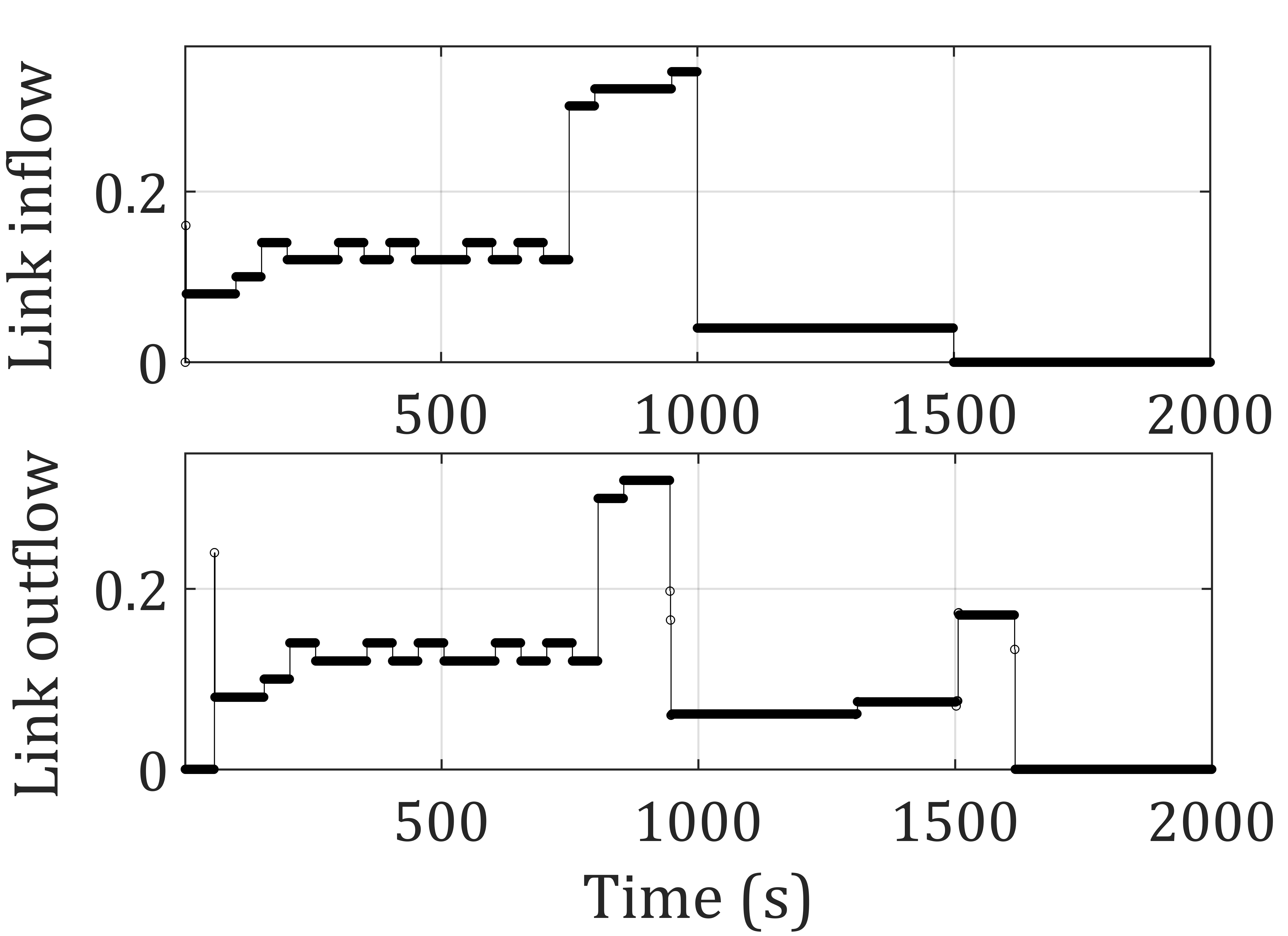}} \\
\caption{{Comparison of the upstream left-turn direction related to bottleneck spillback from LM, LQM and SUMO simulations, where L10 is the turn link 10; L28 is the corresponding upstream common link 28.}}
\label{b l flow}
\end{figure}

\begin{figure}[H]
\captionsetup{font={small}}
\centering  
\setcounter {subfigure} {0}
\subfigure[Cumulative flow of L6 by LM]{
\includegraphics[width=5cm,height = 4.1cm]{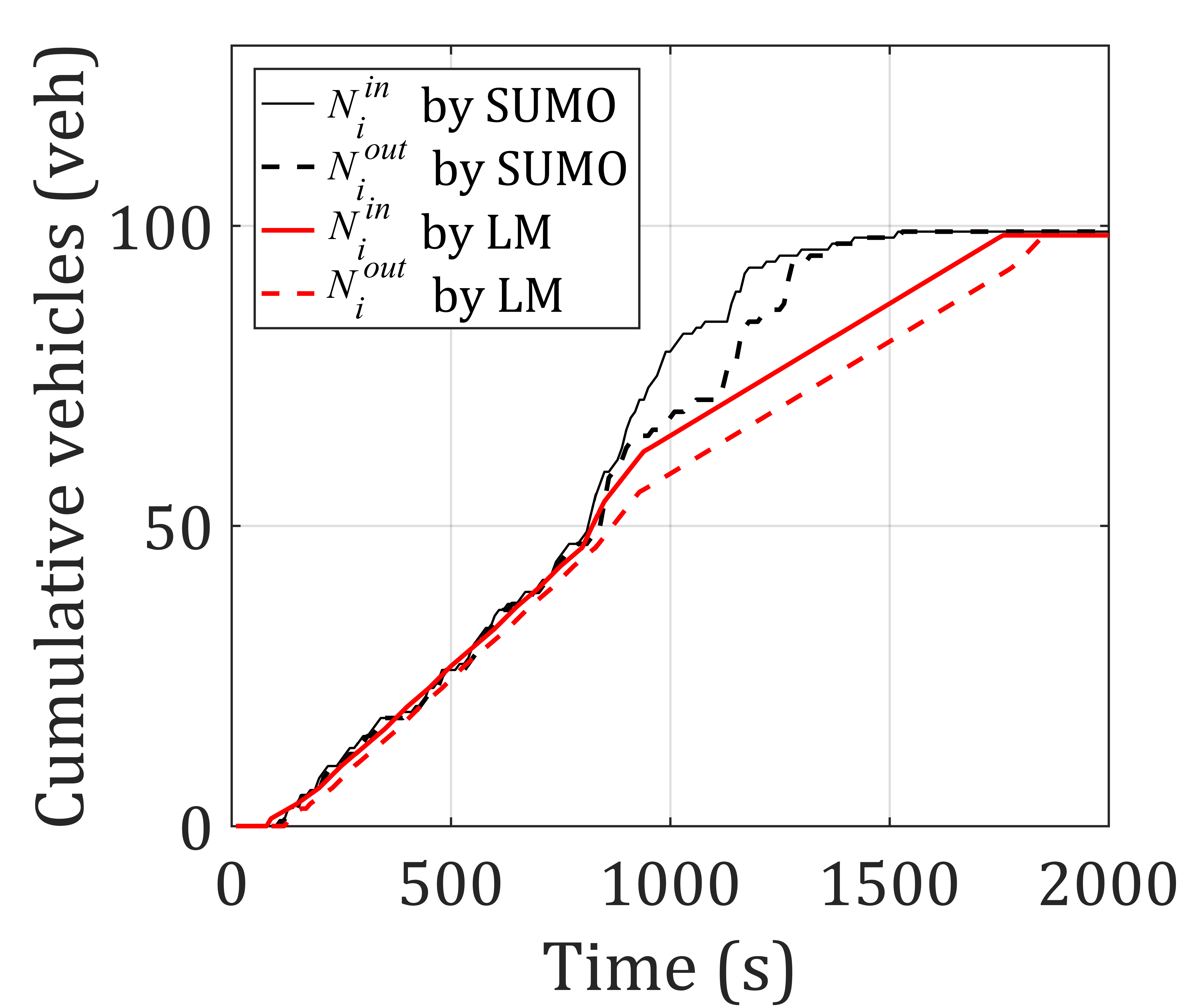}}\subfigure[Cumulative flow of L6 by LQM]{
\includegraphics[width=5cm,height = 4.1cm]{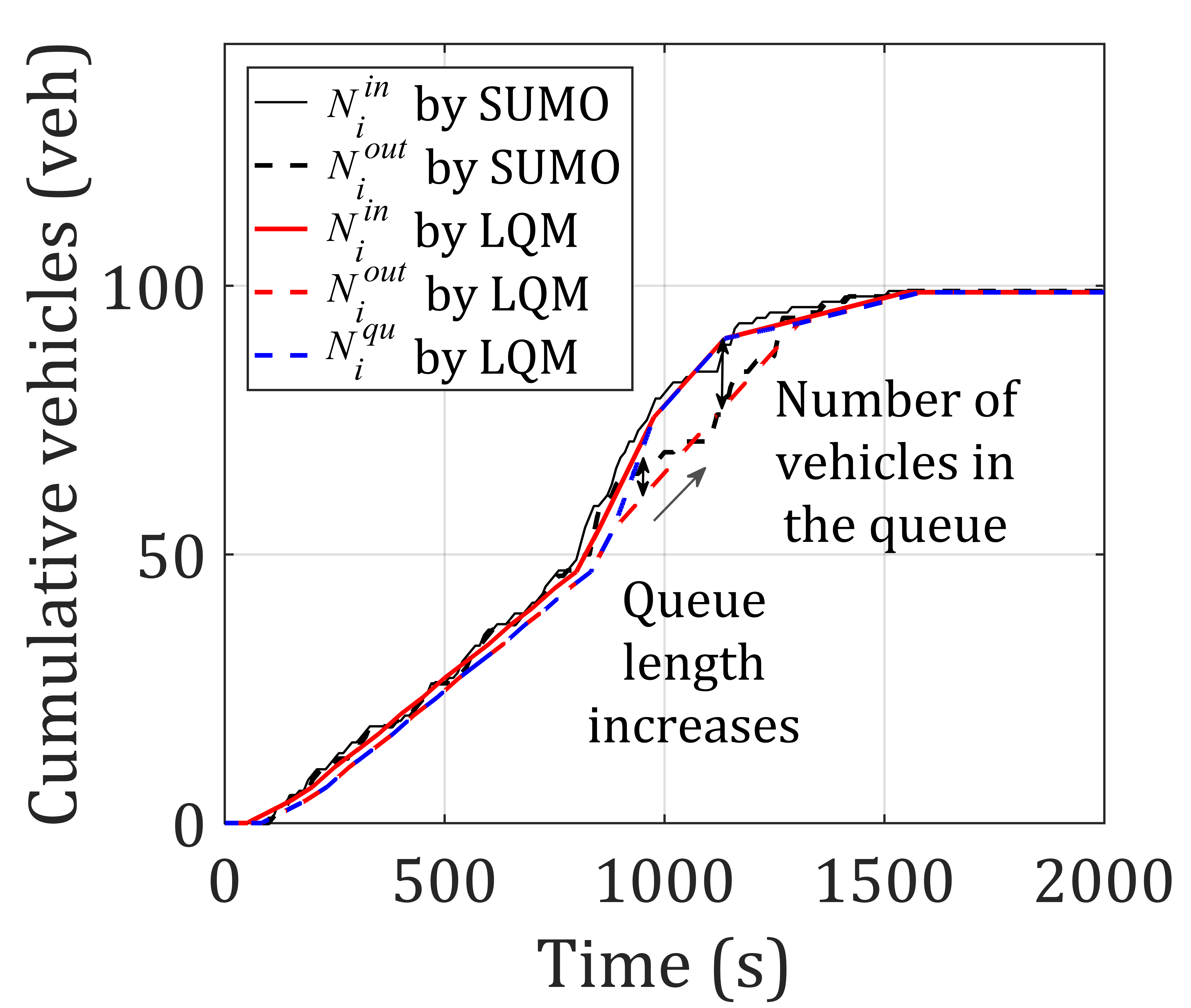}}\subfigure[Queue length of L6 by LQM]{
\includegraphics[width=4.9cm,height = 4.1cm]{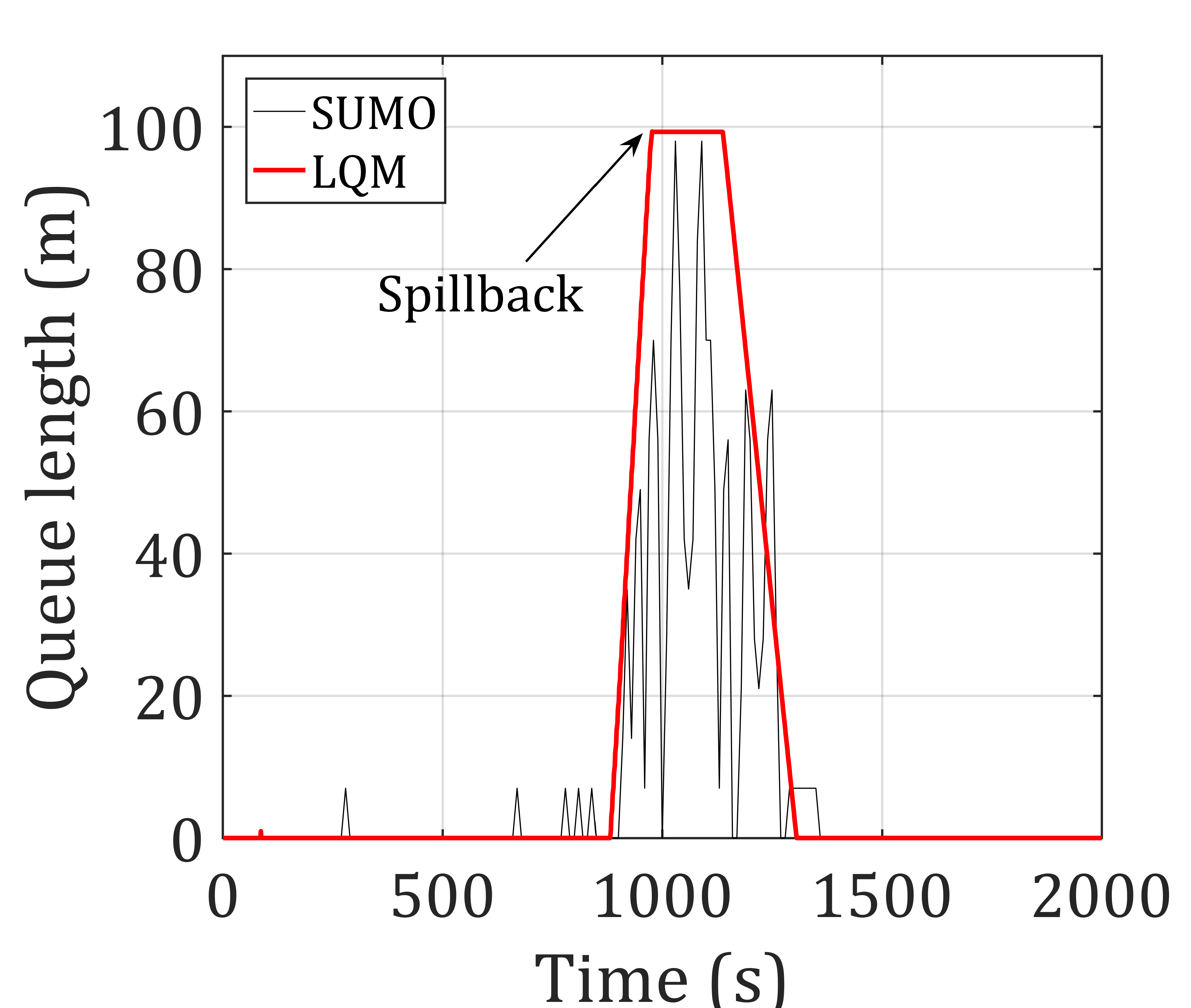}}\\
\subfigure[Cumulative flow of L26 by LM]{
\includegraphics[width=5cm,height = 4.1cm]{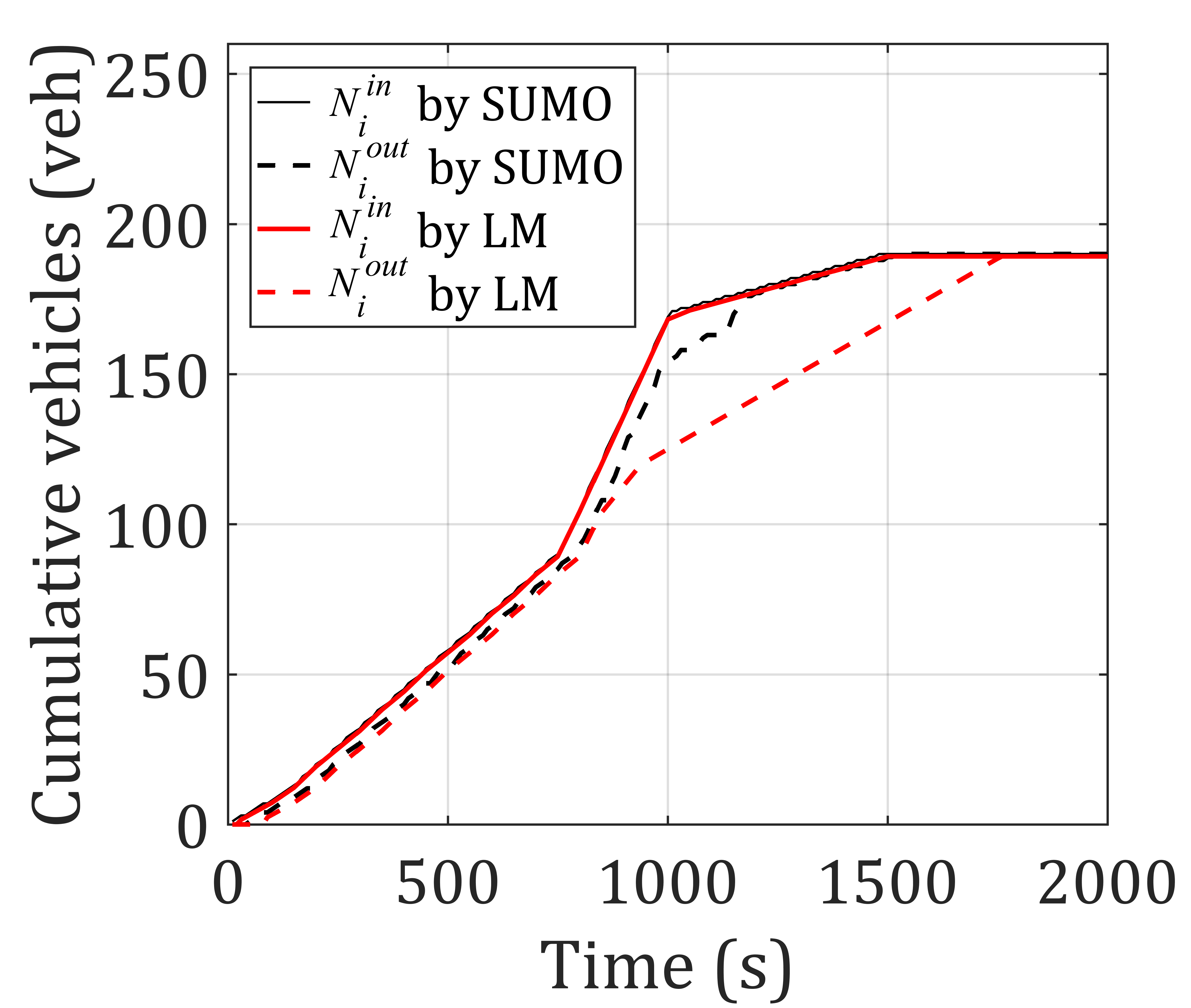}}\subfigure[Cumulative flow of L26 by LQM]{
\includegraphics[width=5cm,height = 4.1cm]{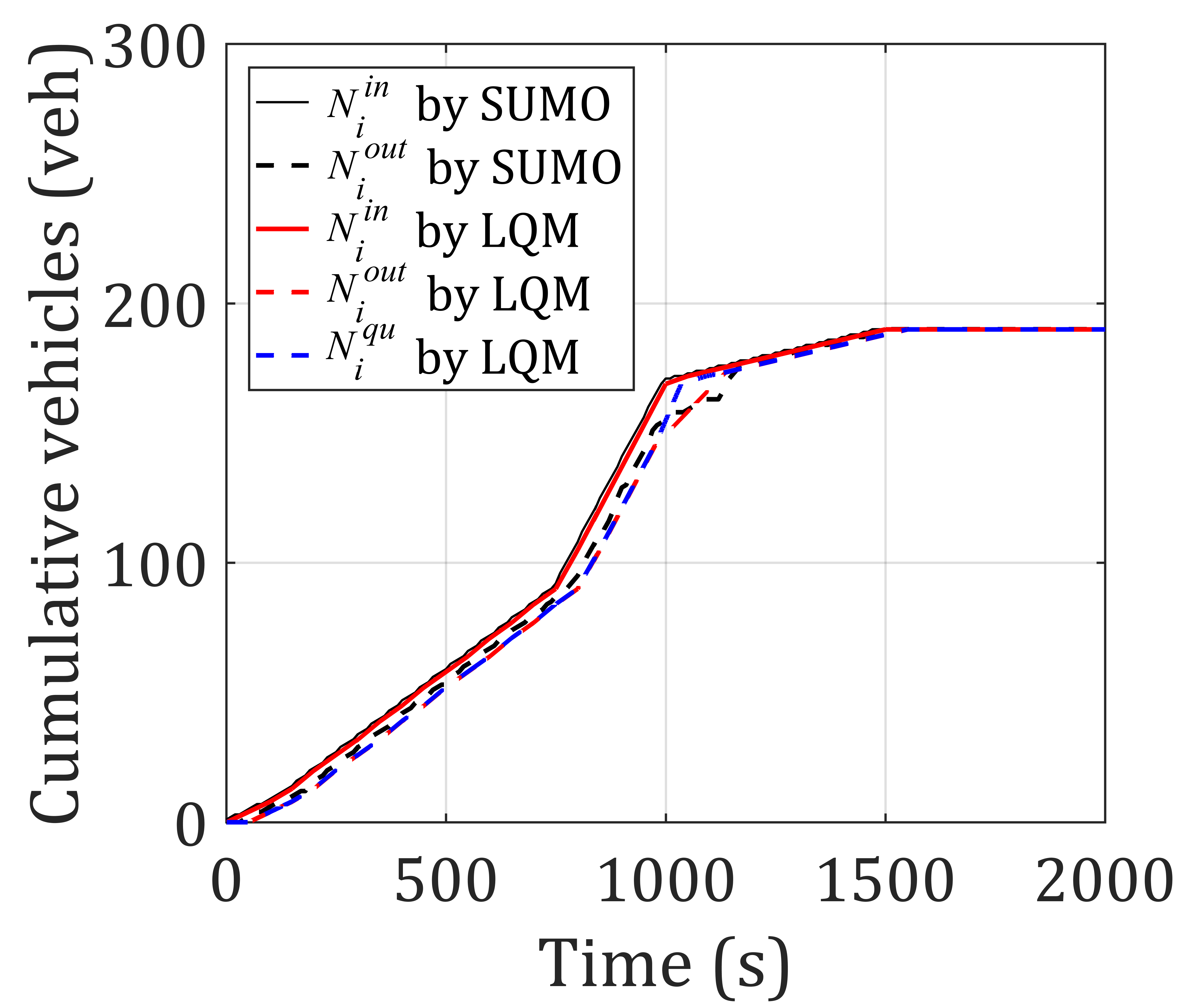}}\subfigure[Queue length of L26 by LQM]{
\includegraphics[width=4.9cm,height = 4.1cm]{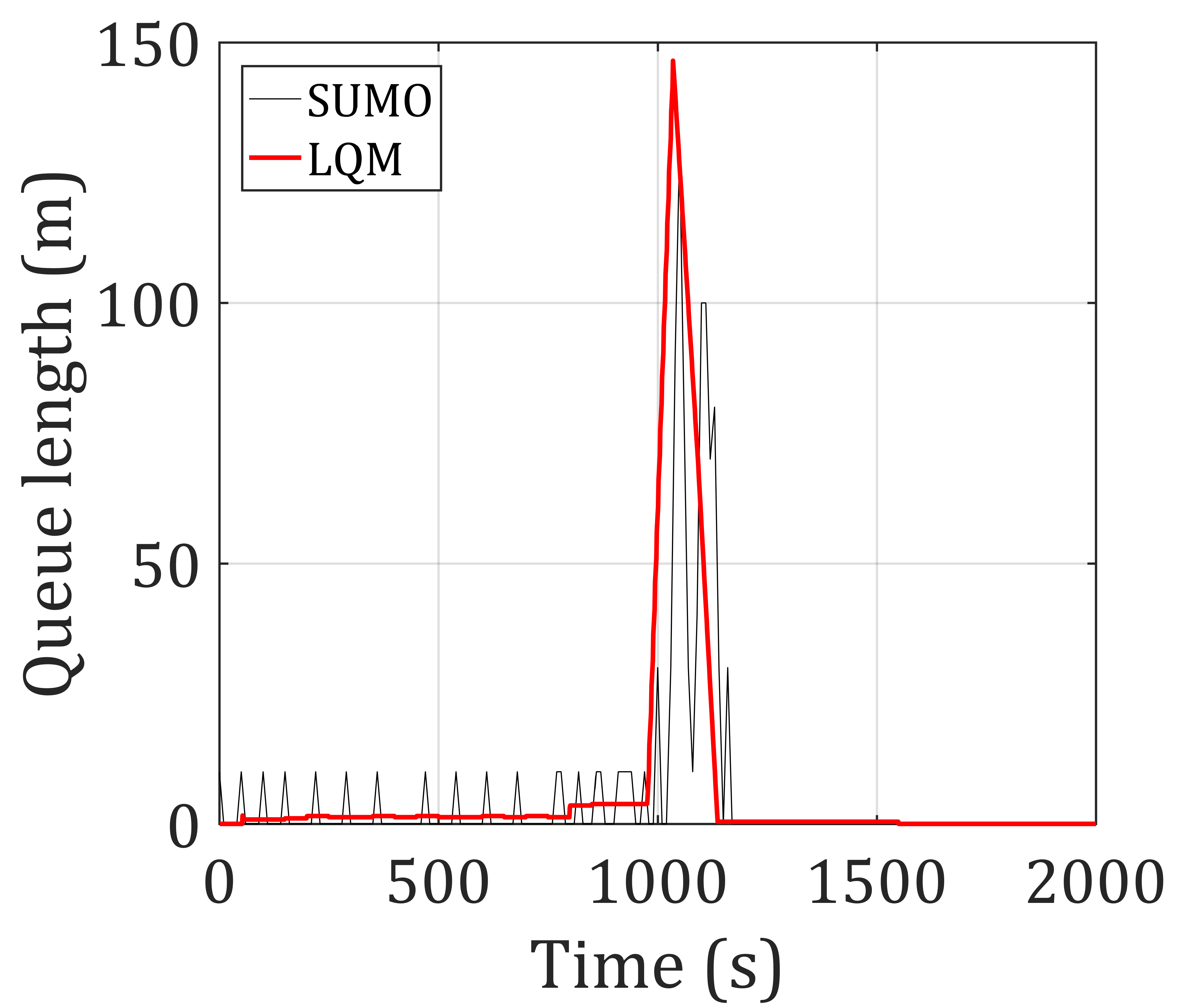}}
\subfigure[Flow rate (m/s) of L6 by LQM]{
\includegraphics[width=6.4cm,height = 4.3cm]{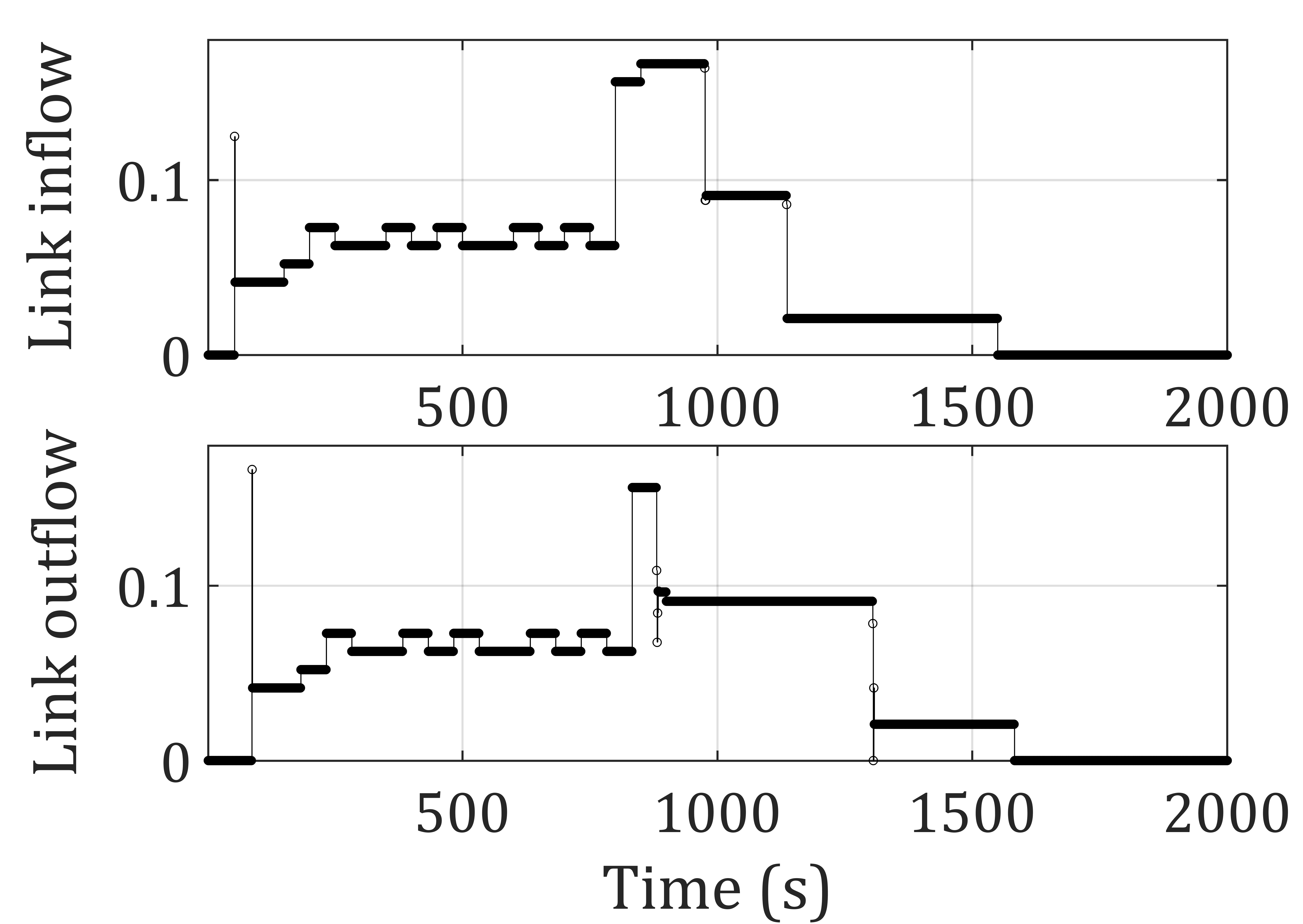}}~~~\subfigure[Flow rate (m/s) of L26 by LQM]{
\includegraphics[width=6.4cm,height = 4.3cm]{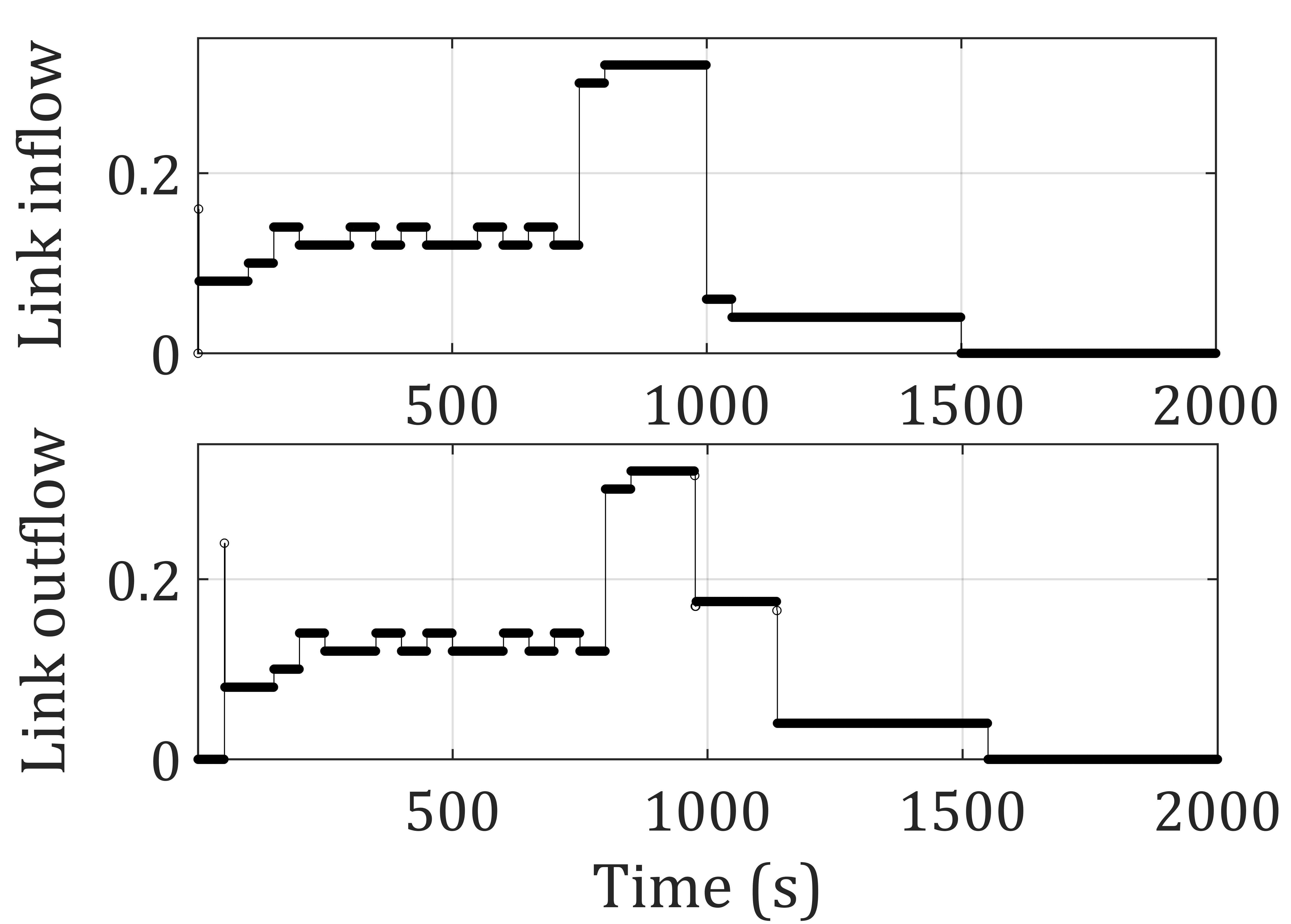}} \\
\caption{{Comparison of the upstream right-turn direction related to bottleneck spillback from LM, LQM and SUMO simulations, where L6 is the turn link 6; L26 is the corresponding upstream common link 26.}}
\label{b r flow}
\end{figure}

{\section*{Appendix C: Supplementary Simulation Results}}
{This appendix supplements the simulation results in Section 3.1.1. The flow operation results for Link 31's upstream left-turn and right-turn incoming directions, are illustrated in Figures \ref{b l flow} and \ref{b r flow}, where the common links 26 and 28 are also the origin links.}

{Since LM overestimates the cumulative link inflow and outflow in the upstream through direction during certain periods, as demonstrated in Link 2 of Figure \ref{b t flow}(a), the cumulative link inflow and outflow values for the upstream left-turn and right-turn directions are occasionally smaller than those in SUMO. This is evident in Links 6 and 10 of Figures \ref{b l flow}(a) and \ref{b r flow}(a) respectively, especially when queues appear. In contrast, the simulated cumulative flow curves of LQM closely resemble those of SUMO across all upstream incoming directions. The queue length changes and maximum values estimated by LQM also consistently align with SUMO, as demonstrated in Figures \ref{b l flow}(c) and (f), and \ref{b r flow}(c) and (f).}

\vspace{\baselineskip}
\bibliographystyle{elsarticle-harv}
\bibliography{Reference}
\end{document}

%% file: Introduction.tex
Modelling traffic flow dynamics is indispensable to developing effective traffic management and control strategies \citep{sharma2021assessing,berhanu2004models,zegeye2013integrated,rios2018impact}. Traffic flow models can be categorized into microscopic and macroscopic levels. Since macroscopic models describe the aggregate traffic as fluids with constrained demand and supply, they provide an efficient and continuum solution to capture flow propagation over networks compared to microscopic models, e.g., car following \citep{newell2002simplified,hidas2005modelling}. Therefore, macroscopic models are increasingly used in dynamic traffic assignment \citep{gentile2007spillback,gentile2015using,peeta2001foundations} and traffic control \citep{van2016urban,van2018hierarchical,mohebifard2018distributed,mohebifard2019optimal}. The work sets its scope to modelling macroscopic flows for traffic management and control applications. We first review existing work, which motivates our study.

% \noindent
% 1.1 Related work
\subsection{Related work}
Among the existing macroscopic models, cell-based models, e.g., cell transmission models (CTMs) and link-based models, e.g., link transmission models (LTMs), are widely used to describe link dynamics \citep{daganzo1994cell,daganzo1995cell,yperman2005link,yperman2006multi,bliemer2019continuous,jin2015continuous,himpe2016efficient}. Cell-based models describe the traffic flow dynamics over spatial cells across one road segment (RS) based on the available capacity of each cell \citep{daganzo1994cell,daganzo1995cell}. {This formulation transforms the kinematic wave model into a discretized partial differential equation, and the solution is obtained using the Godunov scheme \citep{lebacque1996godunov,csikos2017variable}. While cell-based models are easy to implement thanks to spatiotemporal discretization, they become problematic at  long links due to the assumption of a homogenous flow of vehicles within each cell \citep{daganzo1994cell,carey2014implementing}. To capture the inhomogeneous flow and queue propagation within a link, long links have to be discretized to multiple spatial cells at the expense of computational efficiency. It is also difficult to capture node behavior with cell-based models. \citep{jabari2016node,yahyamozdarani2023continuous}.}

As an alternative macroscopic modelling regime, link-based models take each RS/link as a unit to describe the traffic flow propagation based on the kinematic wave theory \citep{newell1993simplified,newell1993bsimplified}, which makes a tradeoff between computational efficiency and accuracy \citep{yperman2005link,yperman2006multi}. Because link-based models need less variables to describe the traffic state of a link and can be simulated in larger time steps than the cell-based models, the solutions are more efficient \citep{gentile2010general,raadsen2016efficient}. Many link-based models for simulating traffic flow have been developed over recent years. \citet{yperman2005link,yperman2006multi} formulated an LTM, in which the traffic demand and supply functions are defined by the cumulative link inflow and outflow according to the LWR model \citep{newell1993simplified,newell1993bsimplified,lighthill1955kinematic,daganzo2005variational}. \citet{jin2015continuous} further proposed a continuous LTM for a simplified traffic network without signalized intersections, in which the Hopf-Lax formula was used to derive a kinematic wave model with a given cumulative inflow and outflow at the link boundaries. With the link-based model, the stationary states of the network can be estimated, but the model efficiency warrants more studies due to the computational complexity. To this end, \citet{bliemer2019continuous} used the multi-step linearization method to simplify the expansion fans to ensure that the cumulative link inflow and outflow curves are piecewise linear. {\citet{durlin2008dynamic} proposed the link and intersection models with traffic wave tracking that estimates the travel times on networks for dynamic traffic assignment purposes. \citet{van2003macroscopic} developed a macroscopic model that is an extended version of the METANET flow model for mixed urban
and freeway road networks for control purposes. \citet{yan2014improved} further considers the delay time of vehicles to improve the macroscopic model for model-based travel time optimization. \citet{himpe2016efficient} introduced an iterative algorithm for link-based models to achieve fast dynamic network loading (DNL). The solutions of the algorithm are formulated by a predefined sparse space-time discretized grid, which leads to numerical tractability without imposing an upper limit on the time step. \citet{jabari2016node} proposed a signalized node model for the congested networks that used a simple greedy algorithm to achieve invariant holding-free solutions. \citet{yahyamozdarani2023continuous} also formulated a signalized node model that considers the boundary conditions caused by stage timing and vehicle arrival patterns for DNL. However, the turn-level lane configurations and queue dynamics are overlooked in their models.}

To capture more exact simulation results, \citet{raadsen2016efficient} proposed an event-based link-based model and designed an algorithm to solve the simplified DNL problems. The traffic flow dynamics at the upstream and downstream boundaries of the links are used to predict the aggregated traffic states of the adjacent links. Although this method increases the possibility of yielding exact results, the free-flow speed on the link is assumed as a fixed value. To support simulating variable free-flow speed, \citet{raadsen2021variable} developed an LTM to describe flow propagation without imposing any constant speed limits across the entire road section. However, the model is only suitable for freeways and cannot be used for signal-controlled road segments due to the flow interruption caused by the green-red phases. \citet{van2016urban,van2018hierarchical} proposed a discrete-time link-based model to describe  the accumulation of vehicles based on the link inflow and outflow considering the impacts of signal control by using the discrete sampling time steps, where the fractions of green time are regarded as constraints to limit the dynamic link outflow. However, the RS with different turning directions was assumed as a homogeneous link with the same demand, density, and green time. The model neglects the dynamic queue density and queue length within the link, causing underrepresented inflow limits.

For the potential queue spillback representation, \citet{jin2021link} 
 and \citet{ma2014continuous} proposed queue models that are similar to link-based models, which consider the shockwave speeds to describe the congested traffic states and potential spillbacks. \citet{raovic2017dynamic} developed a dynamic queuing transmission model for DNL considering multiple vehicle classes. The dynamic queue model and link-based model are combined to present the spillback and shock wave propagation in a congested network. This model attempts to capture the queue length but in an aggregated manner. The varied queue lengths in different turning directions are not distinguished in their models, which makes it difficult to reflect the actual spillback at bottlenecks with multiple turning directions. In addition, they assumed a constant free-flow speed over the simulation span and hence cannot model the dynamic flow operations caused by time-varying control strategies. 
To sum up, the existing link transmission models usually focus on the formulation of inflow and outflow at the link’s upstream and downstream boundaries \citep{raovic2017dynamic,himpe2016efficient,bliemer2019continuous,van2018hierarchical,yperman2005link}, and thus the detailed queue dynamics within the road cannot be captured. Although some models estimate vehicle accumulation on the link over time, this is not equivalent to queue formation since in congested conditions, accumulation includes both queued vehicles and those still moving upstream of the queue. Consequently, current LTMs cannot capture turn-level queue density and length changes within the link, leading to underrepresented flow propagation results. Moreover, a constant link free-flow speed is usually assumed to formulate link-based models, which restricts the model application in modelling phenomena involving time-varying free-flow speed (TFS) caused by potential bottlenecks on roads. These can be a moving bottleneck (e.g., bus, truck, first-generation autonomous vehicles) or a temporary bottleneck at a fixed location (e.g., traffic incident, or pick-up/drop-off of ride-hailing vehicles), which temporarily reduces the free-flow speed on the link during particular periods. A link-based model that accounts for such free-flow speed variations in dynamic traffic environments with a computational advantage is still lacking.

% Thus, vehicle accumulation provides a macroscopic view of density changes, it does not directly equal queue length. 

% Although the accumulation of vehicles on the link over time can be estimated by some models, this accumulation is not equal to queue formation. For instance, at free-flow conditions, vehicles spread out along the link move freely not necessarily forming a queue. The accumulation of vehicles in this case represents the distributed traffic density along the entire length of the link. In congested conditions, vehicles slow down and queues may form near intersections or merges. The accumulation of vehicles on a link can then include both queued vehicles and those still moving at reduced speeds upstream of the queue. Therefore, while the accumulation of vehicles gives a macroscopic view of density change on a link, it does not directly equal queue length. Existing link-based models do not quantify the queue length as a separate state variable. Such model formulation fails to be aware of accurate queue density and queue length changes within the link, resulting in underrepresented flow propagation results. 

% \noindent1.2 Contribution
\subsection{Contribution}
% This study proposed an extended LTM with turn-level queue transmission and TFS to capture the queuing and spillback in different turning directions for signalized intersections. This is realized by a novel queue inflow model that depends on whether the free-flow speed changes. Unlike most existing LTMs focused on the inflow/outflow at RS upstream/downstream boundaries, the queue transmission in multiple turning directions in our model enables a more accurate representation of flow propagation. Our model relaxes the assumption of constant free-flow speed and incorporates   varying outflow speeds for different turns near the intersections.
% REVISED:  
This study proposed a link-based model with queue transmission and TFS to capture the flow propagation and turn-level queue length changes in road networks. This is realized by a new link-based flow model that introduces an additional state variable of link queue inflow and adapt the link outflow to be free-flow speed-dependent.  Unlike most existing link-based models focused on the aggregated flow dynamics at link boundaries, the queue inflow formulation within the link in our model enables a detailed representation of flow propagation, providing queue length changes in multiple turning directions. Such a refined representation is critical for model-based traffic control applications in urban networks \citep{zhou2016two,van2018hierarchical}. Moreover, our model relaxes the assumption of constant link free-flow speed and incorporates varying outflow speeds for different turns near the intersections.

Specifically, in our model, each link is segmented into free-flowing and queuing parts, determined by the queue density characterized through triangular fundamental diagrams (FDs). The vehicle propagation in different parts is described by the dynamic link inflow, queue inflow and link outflow, all dependent on free-flow speed changes. A node model that respects the invariant holding-free property is further specified to capture the potential spillbacks at the turn level. It computes the actual flow propagation from the node’s incoming links to outgoing links. The performance of the proposed model is analyzed by comparing the model outputs with a baseline link-based model and microscopic simulation in both a single intersection and a real-world network in Baoding City, China.

The rest of the paper is organized as follows. The model framework and specifications are presented in the next section. After that, the proposed model was implemented and validated using simulation experiments. The last section summarizes the conclusions and demonstrates some future research topics.

%% file: Mathematical_Model_Formulation.tex
Using LQM to represent the proposed link-based model with queue transmission and TFS, the model contains both a link model and a node model. We first present the model framework and assumptions, followed by the formulation of link and node models. Table \ref{notation} summarizes the description of the notations used in this study.

\begin{table}[H]
\ContinuedFloat*
\captionsetup{font={small}}
\footnotesize
\renewcommand\arraystretch{1.1}
\caption{Descriptions of notations\label{Table1}}
\label{notation}
%\begin{tabular}{cccp{10cm}}
\centering
\begin{tabular}{m{1.2cm}<{\centering} m{1.2cm}<{\centering} m{0.8cm}<{\centering} m{10.5cm}}
\hline
Type&Notation&Unit&Definition\\
\hline
Variable&$k$&s&Sampling time step\\
~&$T$&s& Simulation period\\
~&$i$&-&Link\\
~&$L_q(k)$&m&Queue length at $k$\\
~&$L_f(k)$&m&Free-flowing length at $k$\\
~&$L_{i}$&m&Link length\\
~&$N_{i}^{in}(k)$&veh&Cumulative link inflow of $i$ at $k$\\
~&$N_{i}^{qu}(k)$&veh&Cumulative queue inflow of $i$ at $k$\\
~&$N_{i}^{out}(k)$&veh&Cumulative link outflow of $i$ at $k$\\
~&$q_{i}^{in}(t)$&veh/s&Link inflow rate of $i$ at $t$\\
~&$v_{f}(k)$&m/s&Link free flow speed at $k$\\
~&$b_{i}(k)$&-&Effective fraction of green time of $i$ at $k$\\
~&$q_{i}^{cr}(k)$&veh/s&Critical flow rate of $i$ at $k$\\
~&$\rho_{i}^{cr}(k)$&veh/m&Critical density of $i$ at $k$\\
~&$\rho_{i}^{jam}$&veh/m&Jam density of $i$\\
~&$\rho_{i}^{q}(k)$&veh/m&Queue density of $i$ at $k$\\
~&$w_{i}$&m/s&Backward wave speed\\
~&$t_f$&s&Free-flow travel time\\
~&$\Delta t$&s&Sampling time interval between $k$ and $k+1$\\
~&$n_f$&-&Number of sampling time steps for $t_f$\\
\hline
\end{tabular}
\end{table}
\begin{table}[H]
\captionsetup{font={small}}
\footnotesize
\renewcommand\arraystretch{1.1}
\ContinuedFloat% continue splited float
\caption{Continued table}
\centering
\begin{tabular}{m{1.2cm}<{\centering} m{1.2cm}<{\centering} m{0.8cm}<{\centering} m{10.5cm}}
\hline
~&$\gamma_{f}$&-&Fraction of sampling time steps for $t_f$\\
~&$v_{min}$&m/s&Minimal desired speed\\
~&$\delta_{qu}(j)$&-&Binary variable, if $ D_{f}(j) >  L_{f}(k)$,  $\delta_{qu}(j) = 1$, otherwise $\delta_{qu}(j) = 0$\\
~&$t_{sh}$&s&Shockwave time in the queue\\
~&$n_{sh}$&-&Number of sampling time steps for $t_{sh}$\\
~&$\gamma_{sh}$&-&Fraction of sampling time steps for $t_{sh}$\\
~&$N_{i}^{in,sh}(k)$&veh&Link inflow with the shockwave propagation at $k$\\
~&$\bar{N}_{i}^{in}(k)$&veh&Cumulative link inflow limit at $k$\\
~&$\bar {q}_{i}^{in}(k)$&veh/s&Allowed maximum link inflow rate at $k$\\
~&$q_{i}^{in,des}(k)$&veh/s&Desired link inflow rate at $k$\\
~&$q_{i}^{in}(k)$&veh/s&Actual link inflow rate at $k$\\
~&$\delta_{out}(j)$&-&Binary variable, if $ D_{f}(j) >  L_{i}$,  $\delta_{out}(j) = 1$, otherwise,  $\delta_{out}(j) = 0$\\
~&$N_{i}^{out,des}(k)$&veh& Desired cumulative outflow at $k$\\
~&$\bar N_{i}^{out}(k)$&veh&Maximum cumulative outflow at $k$\\
~&$q_{sat}(k)$&veh/s& Saturation flow rate at $k$\\
~&$\bar {q}_{i}^{out}(k)$&veh/s&Maximum outflow rate at $k$\\
% ~&$e_{i,j}$&-&Turning rate from link $i$ to link $j$\\
~&$\beta_{i}(k)$&-& Extent of supply constraint at $k$\\
~&$s_{i}$&veh& Supply of $i$\\
~&$\xi$&-& Link with the smallest $\beta_{i}(k)$\\
~&$\tau_j(t')$&s&Travel time of vehicles entering $j$ at time $t'$\\
Vector&$\boldsymbol{x}_{i}(k)$&-&State vector of $i$ at $k$\\
or&$\boldsymbol{D}_f$&-&Potential travel distance of flows\\
matrix &$\boldsymbol{T}_f$&-&Upper triangular matrix\\
~&${\mathrm{\Delta}\boldsymbol{N}}_{in}$&-&Link inflow change up to $k$\\
~&$\boldsymbol\delta_{qu}$&-&Binary vector\\
~&$\boldsymbol I_i^{in}$&-&Feeding links of $i$\\
~&$\boldsymbol{p}^{c}$&-&Conflicting signal phases\\
~&$\boldsymbol{I}^{in}$&-&Node’s incoming links\\
~&$\boldsymbol{I}^{out}$&-&Node’s outgoing links\\
~&$\boldsymbol{I}_{\xi}^{f}$&-&Feeding links of $\xi$\\
\hline
\end{tabular}
\end{table}

\subsection{Model framework and assumptions}

The LQM encompasses both link and node models. The link model describes the propagation of traffic flow from upstream to downstream along a road segment, influencing both sending and receiving flows. Meanwhile, the node model integrates the various potential sending and receiving flows derived from the link model to determine the actual flow propagation between adjacent RSs, accounting for the constrained downstream supply. 
% Although some queue models similar to the concept of LTMs have also been proposed \citep{ma2014continuous,raovic2017dynamic}, the impacts of TFS are ignored.

% The existing LTMs usually focus on segment-level modelling, in which an RS with different turning directions is regarded as a homogeneous link with aggregated traffic characteristics (e.g., green time, density, flow, demand)\citep{van2016urban,van2018hierarchical,raovic2017dynamic,himpe2016efficient,bliemer2019continuous,yperman2005link}. Hence, the traffic operation differences among multiple turning directions are neglected. 

\begin{figure}[H]
\captionsetup{font={small}}
\centering
\includegraphics[width=3.2in]{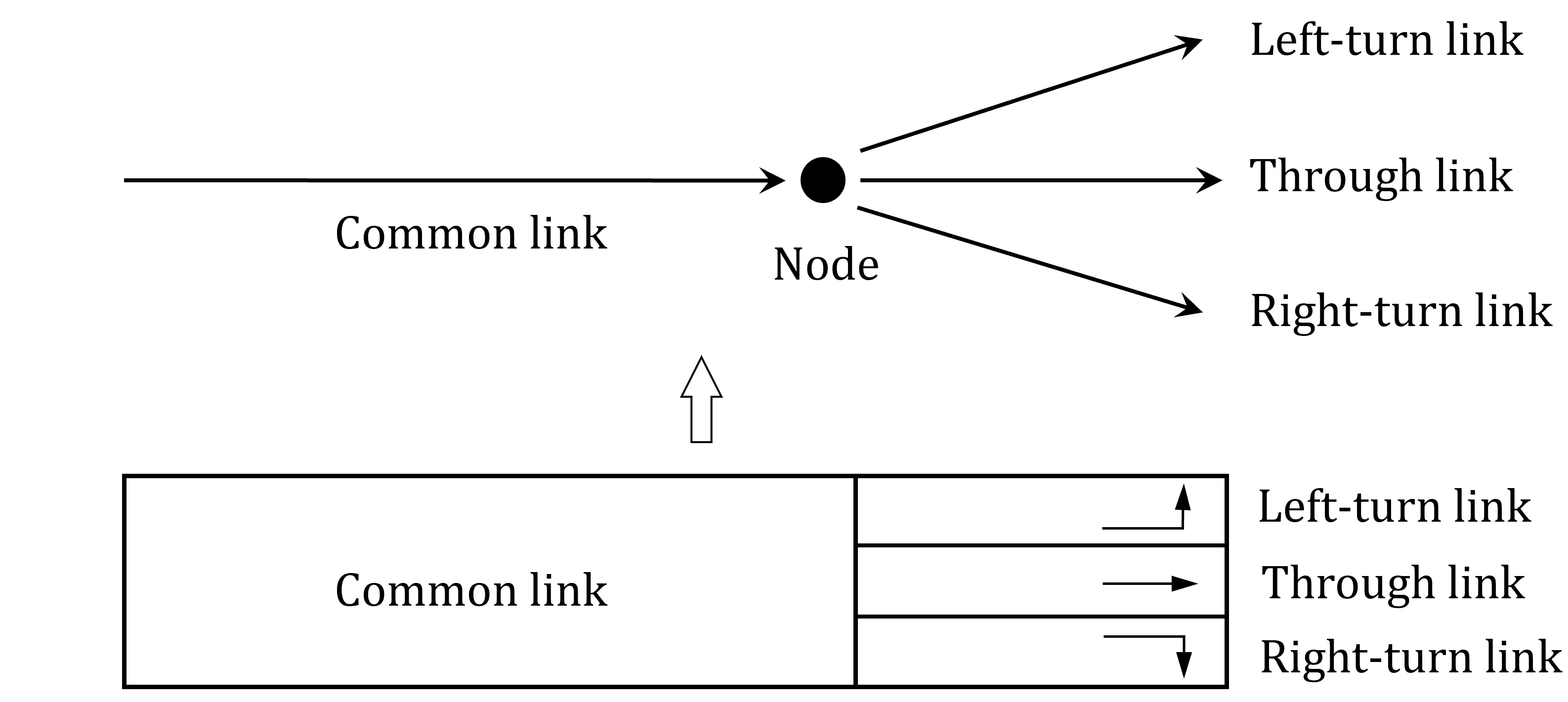}
\caption{{Example of a road with turn lanes in the LQM, where the turn links can be created at a distance equal to the real directional lanes’ length.}}
\label{LTM_structure}
\end{figure}

In this study, a road segment (RS) with multiple turning directions is modeled as exemplified in Figure \ref{LTM_structure}. The upstream section has a common link with all turns combined, followed by a node leading towards parallel turn lanes that operate independently according to the link model. Hence, an RS is divided into a common link and several turn links, where the different turn-based speeds can thus be considered.  We assumed that the vehicle turning rates on the common link are known. Note that Figure \ref{LTM_structure} describes the situation where the separate lanes are designed for the left turn (L), through (T) and right turn (R), namely the following lane configuration: L$|$T$|$R. Our model can be easily to networks with other lane configurations, as discussed in Appendix A.

%\scriptscriptstyle \rm
\begin{figure}[H]
\captionsetup{font={small}}
\centering
\includegraphics[width=3in]{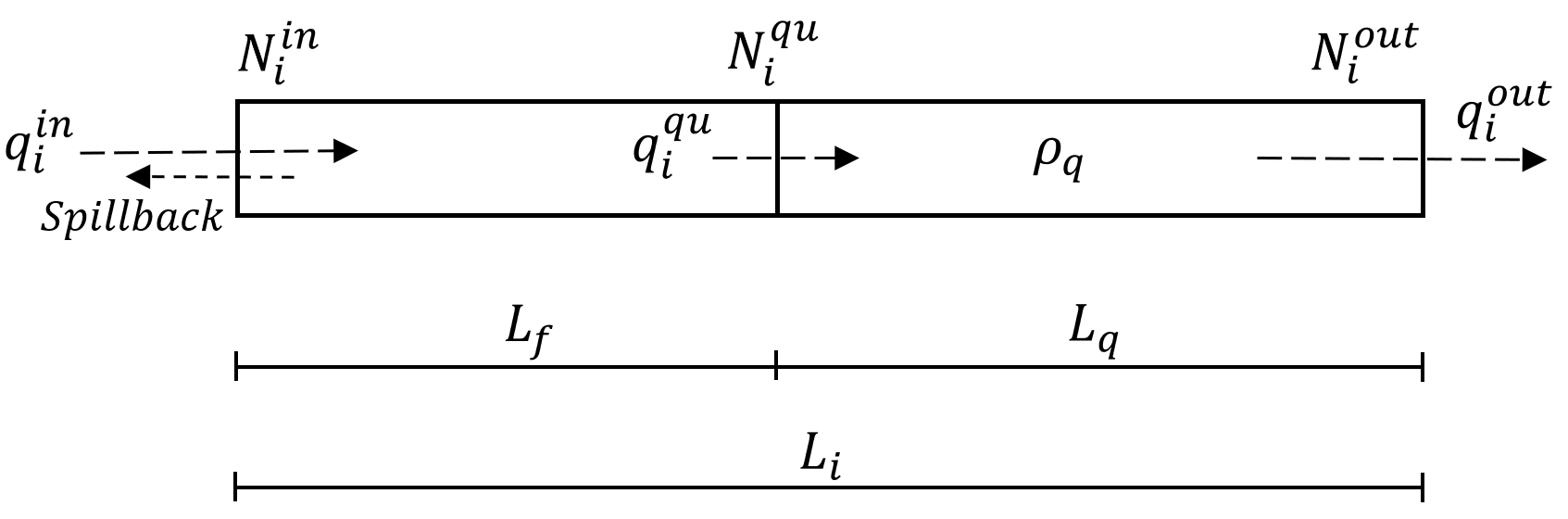}
\caption{{Illustration of the critical notations in the LQM, where $N_{i}^{in}$ is cumulative link inflow, $N_{i}^{qu}$ is cumulative queue inflow and $N_{i}^{out}$ is cumulative link outflow; $q_{i}^{in}$, $q_{i}^{qu}$ and $q_{i}^{out}$ are the link inflow rate, queue inflow rate and link outflow rate, respectively; $\rho_{q}$ is the queue density; $L_{i}$ is the link length; $L_f$ and $L_q$ are the lengths of queueing part and free-flowing part, respectively. Note that the time index $k$ is omitted from notations, but all the variables are time-dependent except $L_{i}$.}}
\label{LTM}
\end{figure}

To formulate the model in a neat way, we use an exemplary link as shown in Figure \ref{LTM}. 
We denote the length of the free-flowing and queueing parts as ${L}_{f}(k)$ and ${L}_{q}(k)$ at sampling time step $k$, respectively. {The cumulative flows including link inflow $ N_{i}^{in}(k)$, queue inflow $ N_{i}^{qu}(k)$ and link outflow $ N_{i}^{out}(k)$ are used as state variables for each link (either a common link or a turn link).  For a link $i$, the cumulative link inflow $k$ is $ N_{i}^{in}(k)$ and can be given by:
\begin{equation}
\label{entered vehicles}
 N_{i}^{in}(k) = \sum_{t = 0}^{k}  q_{i}^{in}(t)\Delta t
\end{equation}
where, $ q_{i}^{in}(t)$ represents the link inflow rate at $t$; $\Delta t$ is the sampling time interval. The discrete timing is considered in this study since the model is designed for signalized intersections. The discrete timing is beneficial for facilitating model applications, e.g., MPC-based signal control at urban intersections \citep{zhou2016two, van2018hierarchical}. The same definition as shown in Eq. (\ref{entered vehicles}) also holds for the cumulative queue inflow $ N_{i}^{qu}(k)$ and cumulative link outflow $ N_{i}^{out}(k)$. Note that the queue inflow for the common link is usually similar to its link outflow since there are few queues in the common link in most cases. However, if the queues in a turn link spill back to its upstream common link, the queue length in the common link also increases. Consequently, we assume that the outflow of the common link (which serves as the inflow for its downstream turn lanes) under turn link spillback conditions decreases even if the downstream other turn lanes have an available supply. This phenomenon is identical to real-world traffic where an excessively long queue in a specific turning direction blocks the roadway as mentioned earlier. }

{The state variable for $i$ at $k$ can be described as follows:}

{\begin{equation}
    {\boldsymbol x}_{i}(k) = \begin{bmatrix}
    { {N}_{i}^{in}(k)} \\
    { {N}_{i}^{qu}(k)} \\
    { {N}_{i}^{out}(k)} \\
    \end{bmatrix}
    \label{state}
\end{equation}}

% \begin{figure}[H]
% \captionsetup{font={footnotesize}}
% \centering
% \includegraphics[width=3.6in]{Figure/framework_F.png}
% \caption{Illustration of turn-level traffic flow dynamics determined by the link model. The red and black curves represent the cumulative number of vehicles in the TRT and LT directions, respectively. }
% \label{framework}
% \end{figure}

%% file: Link_model.tex
{An overview of road topology, vehicle trajectories and cumulative curves of a turn link is presented in Figure \ref{framework}. Due to the incorporation of queue transmission, the turn link is divided into free-flowing and queueing parts based on the queue length at $k$. The time-varying nature of the queue length allows for the trace of queue development within the link over time.  Note that the flow dynamics on the common link closely resemble those of the turn link, with the primary distinction being the absence of signal control of the downstream node. Consequently, the green time fraction $b_{i}(k)\in \lbrack 0,1\rbrack$ of the common link can be set to 1. }

%At time step $k$, six vehicles traverse this turn link, and their trajectories are depicted in the upper section of Figure \ref{framework}. 
%We assumed that the free-flow speed $ v_f(k)$ is known in advance. For clear representation, in Figure \ref{framework}, $v_f(k)$ remains constant, consistent with conventional LTMs. However, our model accommodates time-varying free-flow speed. It has the capability to simulate instances of reduced free-flow speed during particular periods resulting from disturbances. 
From the cumulative flow curves shown at the bottom of Figure \ref{framework}, the cumulative queue inflow and outflow can be inferred from the past cumulative link inflow based on the queue length and free-flow speed. The maximum cumulative queue inflow and outflow at $k$ depends on the past cumulative link inflow. Conversely, the maximum cumulative link inflow at $k$ depends on the past cumulative queue inflow and outflow. Therefore, to model the flow dynamics, the link inflow, queue inflow and link outflow in the past are required. 

The following subsections will formally specify the link model. We start by discussing queue density and queue inflow modeling, as the estimation of link inflow depends on queue length, which in turn is estimated based on queue density and queue inflow.

% As shown in Figure \ref{framework}, a link starts and ends at the boundaries of RS near upstream and downstream intersections, respectively. The RS with LT and TRT directions is divided into several free-flow and queueing parts, where the lane-changing behaviors are allowed in the free-flow parts. The overlapping region is defined as the area where the queueing part of the LT direction overlaps with the free-flow part of the TRT direction. In the overlapping region, the TRT lane may be occupied by left-turning vehicles with the probability $P$ since the longer queue length in the LT direction and permitted lane-changing behaviors. Using $L_o (k)$ to represent the length of the overlapping region: 
% \begin{equation}
%     L_{o}(k) = P\left( {L_{q,\rm LT}(k) - L_{q,\rm TRT}(k)} \right)
% \end{equation}
% where $P \in \lbrack 0,1\rbrack$, hence, $0 \leq L_{o}(k) \leq L_{q,\rm LT}(k) - L_{q,\rm TRT}(k)$.

%% file: Queue_density.tex
\begin{figure}[H]
\captionsetup{font={small}}
\centering
\includegraphics[width=3in]{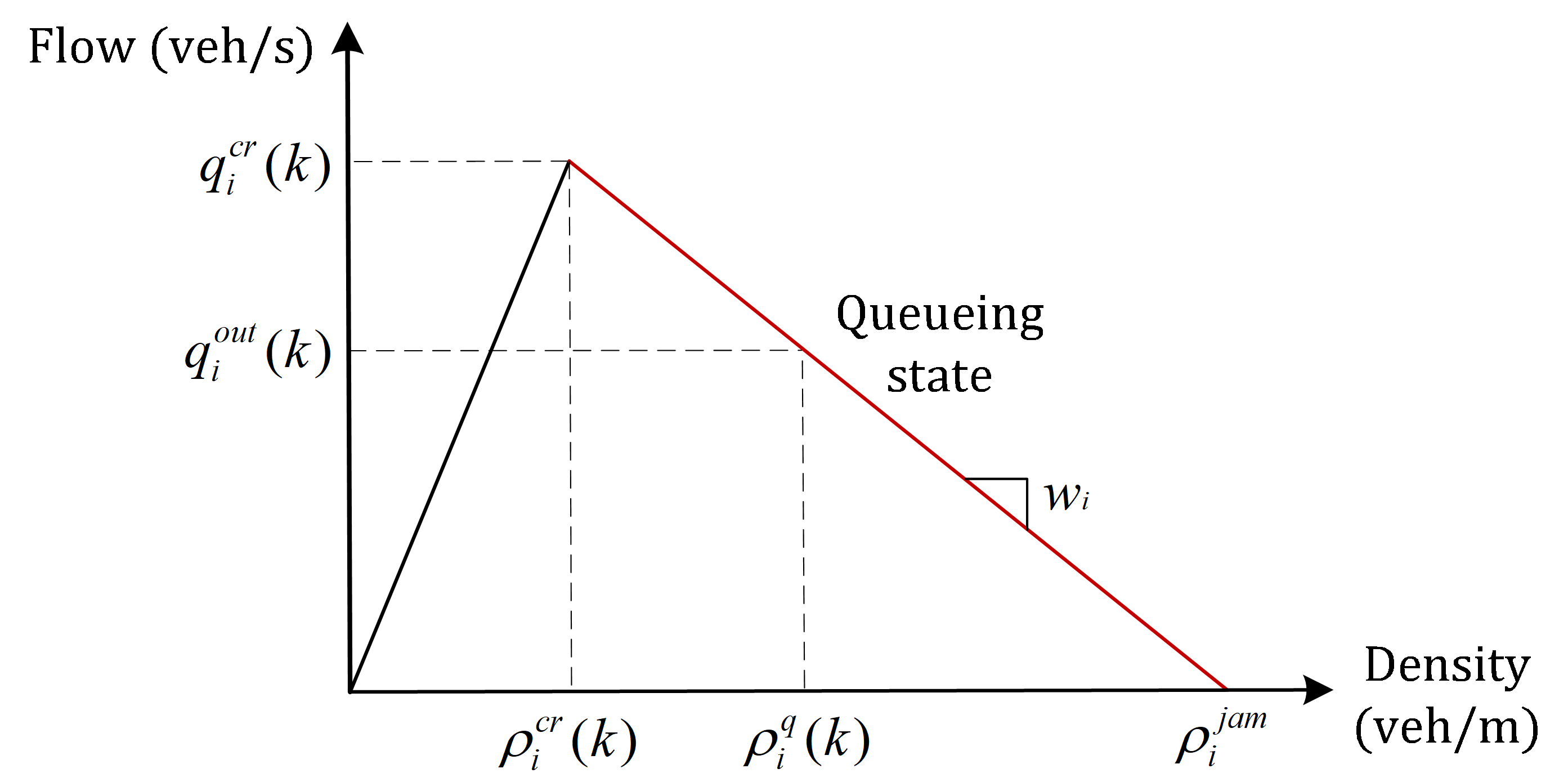}
\caption{{Fundamental diagram for link $i$ at $k$, where the red curve represents the link in queueing states.}}
\label{FD}
\end{figure}

% Existing LTMs usually assumed a predefined travel time function or a fixed travel time
% across the link to infer the outflow from the inflow, but actually the travel time is not exact and includes
% uncertainty due to the potential dynamic queues \citep{bliemer2007dynamic}. 
%To determine the dynamic queue length on link $i$, it is necessary to calculate the queue density in the queuing part of $i$. 
{The queue density can be determined based on a triangular Fundamental Diagram (FD) as shown in Figure \ref{FD}, which depicts the interrelation among the critical flow rate $ q_i^{cr}(k)$, critical density $ \rho_i^{cr}(k)$, and jam density $\rho_i^{jam}$. Note that the link free-flow speed in our LQM can be time-varying, so we do not assume the constant values for critical flow and density.} The $ q_i^{cr}(k)$ and $ \rho_i^{cr}(k)$ are timing-varying with $k$ and can be calculated based on \cite{daganzo1995cell}:

\begin{equation}
\label{q_cr}
     q_i^{cr}(k) = \rho_i^{jam}\frac{ v_{f}(k)w_i}{ v_{f}(k) + w_i}
\end{equation}
\begin{equation}
\label{rho_cr}
     \rho_i^{cr}(k) = \rho_i^{jam}\frac{w_i}{ v_{f}(k) + w_i}
\end{equation}
where $w_i$ represents the backward wave speed.

{As shown in Figure \ref{FD}, when the link density is greater than $ \rho_i^{cr}(k)$, the link can be regarded as the queueing state, and the queue density at $k$ can be represented:
\begin{equation}
     \rho_{i}^{q}(k) =  \rho_i^{cr}(k)+\Delta\rho
    \label{queue density}
\end{equation}}

Using $ q_i^{out}(k)$ represents the link outflow rate at $k$, Eq. (\ref{queue density}) can be written as \citep{raovic2017dynamic}:
\begin{equation}
\begin{split}
     \rho_{i}^{q}(k) &=  \rho_i^{cr}(k) + \left(\rho_i^{jam} -  \rho_i^{cr}(k) \right)\frac{  q_i^{cr}(k) - q_i^{out}(k)}{ q_i^{cr}(k)} 
\end{split}
    \label{queue density2}
\end{equation}

% Therefore, the queue density at $k$, as indicated in Eq. (\ref{queue density2}) is not only determined by the outflow rate but also by the backward wave propagation.

The queue length $ L_{q}(k)$ and free-flow length $ L_{f}(k)$ of the link at $k$ can now be determined by:
\begin{equation}
     L_{q}(k) = \frac{ N_{i}^{qu}(k) -  N_{i}^{out}(k)}{ \rho_{i}^{q}(k)}
     \label{q length}
\end{equation}
\begin{equation}
    L_{f}(k) = L_{i} -  L_{q}(k)
    \label{f length}
\end{equation}
where the cumulative queue inflow $ N_{i}^{qu}(k)$ and outflow $ N_{i}^{out}(k)$ for the link need to be further determined in ways described in the sequel. 

\begin{figure}[H]
\captionsetup{font={small}}
\centering
\includegraphics[width=3.5in]{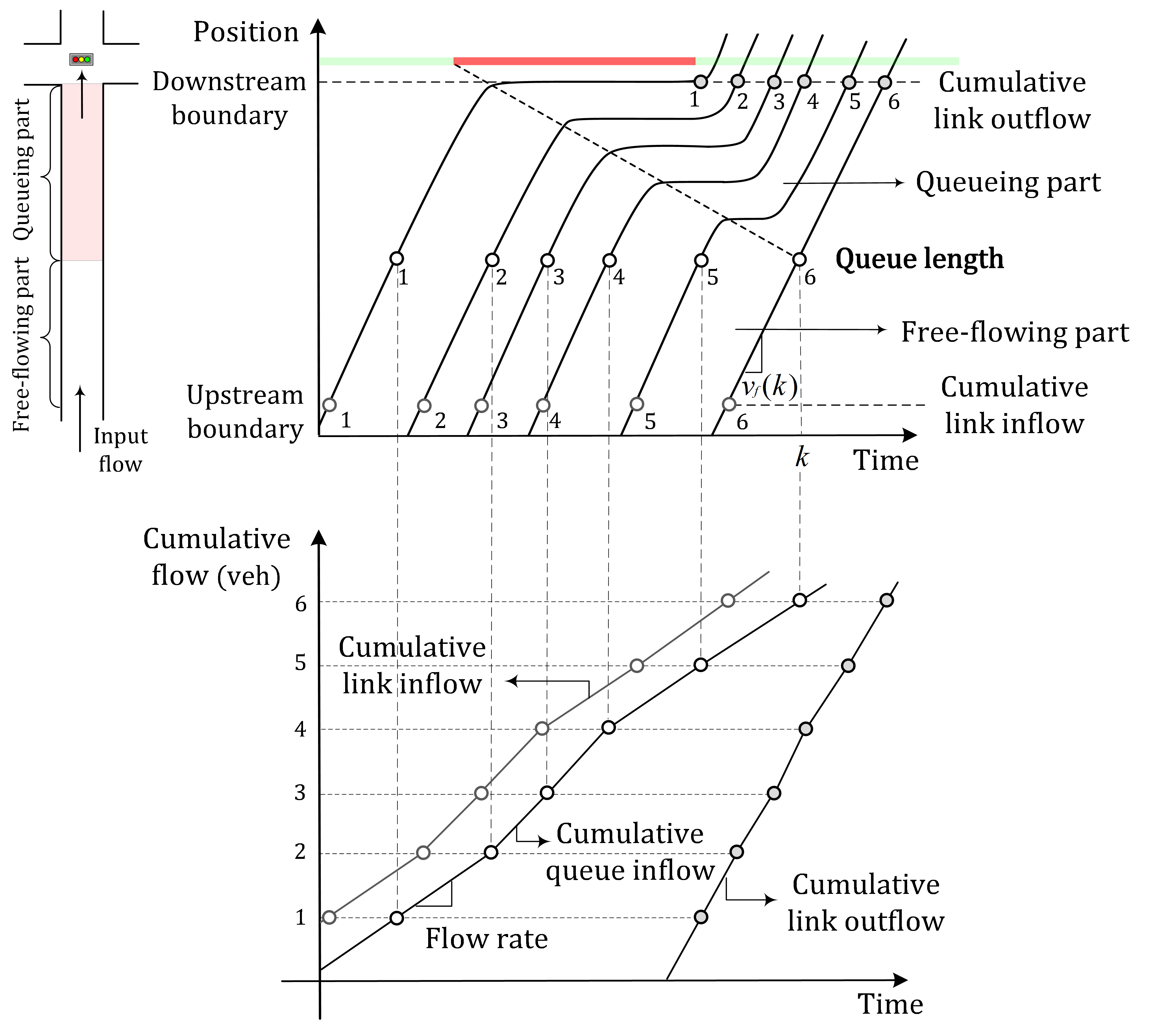}
\caption{{Illustration of vehicle trajectories (top) and corresponding cumulative flow dynamics (bottom) at a certain sampling time step $k$ on a turn link, where $v_f(k)$ is the link free-flow speed at $k$; the flow dynamics include cumulative link inflow, cumulative queue inflow and cumulative link outflow that are determined by the proposed link model.  }}
\label{framework}
\end{figure}

%% file: Node_model.tex
\begin{figure}[H]
\captionsetup{font={small}}
\centering
\includegraphics[width=3.5in]{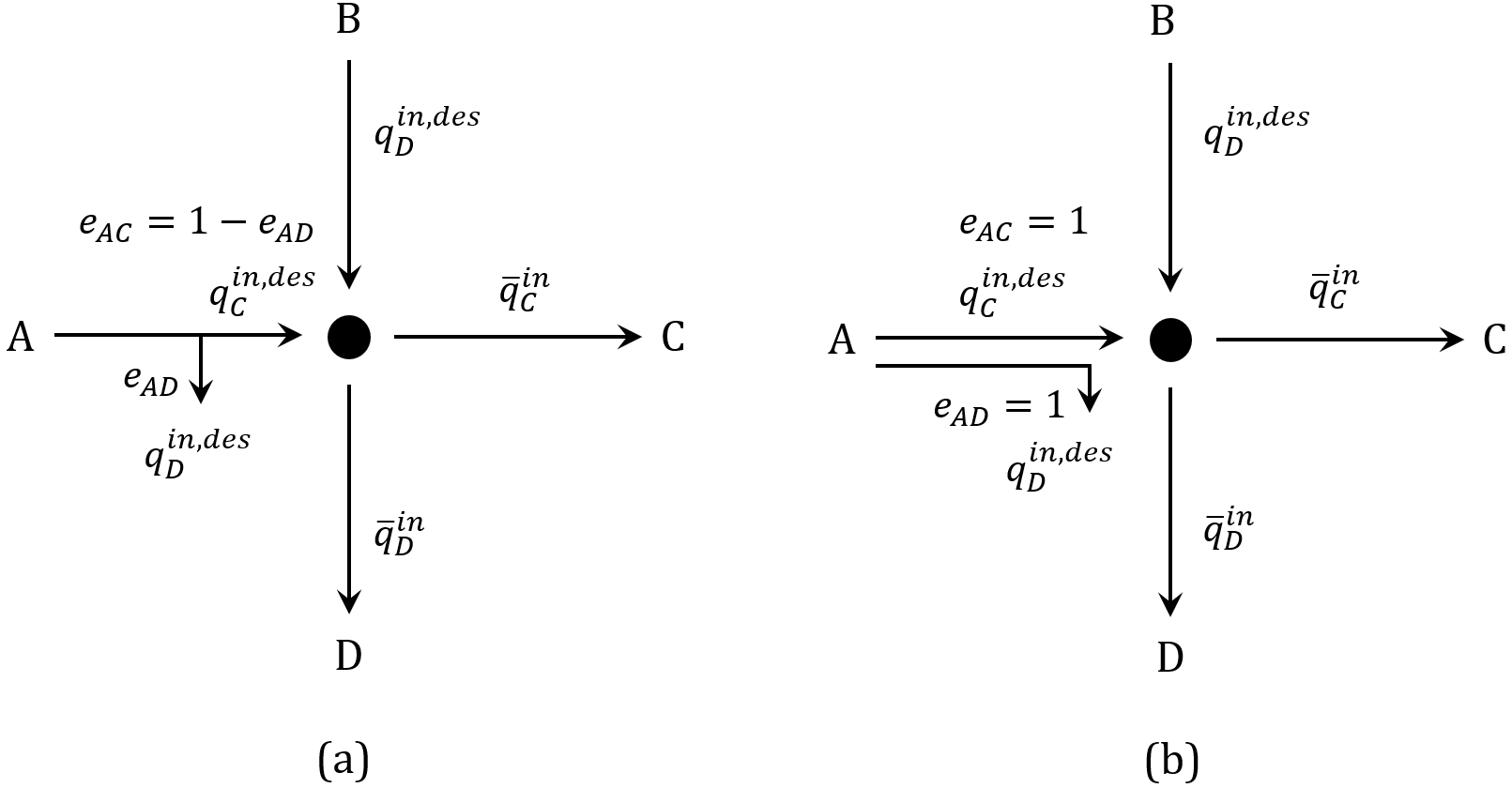}
\caption{{Illustration of nodes at the intersection and the critical notations used in the node model. (a) node's incoming link with lane sharing; (b) node's incoming link without lane sharing, where A and B are node's incoming links; C and D are node's outgoing links; $e_{i,j}$ is the turning rate from link $i$ to $j$; $q_i^{in,des}$ is the desired inflow rate of link $i$; $\bar q_i^{in}$ is the allowed maximum inflow rate of link $i$; Note that $q_i^{in,des}$ and $\bar q_i^{in}$ are time-dependent and the time index $k$ is omitted from notations.}}
\label{node}
\end{figure}

In the link-based models, the links are connected through the nodes, thus the transition of traffic flow among adjacent RSs can be achieved by the node model. Due to the limited receiving capacity of the downstream RSs, the actual number of vehicles that can travel from the node’s incoming RSs to the outgoing RSs is further determined by the node model. The nodes at the intersection and the critical notations used are as shown in Figure \ref{node}. For the node between a common link and turn-links as shown in Figure \ref{LTM_structure}, it can be regarded as the node with lane sharing presented in Figure \ref{node}(a). The sequel illustrates the proposed node model. It should be noted that we limited the scope of signal control impacts at the intersections to include the green time fractions ${b}_{i}(k) \in (0,1]$, for various turning movements, which constrains the outflows of the node's incoming links as shown in Eq. (\ref{max_out}). 

% \begin{equation}
% \label{green}
%     b_{i}(k)=\frac{q_{out}(k)}{q_{sat}(k)}
% \end{equation}
% where $q_{out}(k)$ represent the realized outflow rate at $k$.

For a node's outgoing link $i$, in the case that the desired link inflow rate $ q_{i}^{in,des} (k)$ is greater than its allowed maximum value $ \bar {q}_{i}^{in} (k)$ as calculated by Eq. (\ref{in_max}), the outflows of its upstream feeding links need to be decreased due to the downstream supply constraint. Using $ \beta_{i}(k)$ to indicate the extent of supply constraint at $k$:

\begin{equation}
     \beta_{i}(k) = \mathit{\min}\left( {1,\frac{ \bar {q}_{i}^{in}(k)}{ q_{i}^{in,des}(k)}} \right)
    \label{decreased index}
\end{equation}
where $ q_{i}^{in,des}(k)$ is calculated by:
\begin{equation}
 q_{i}^{in,des}(k)=\left( N_{i}^{in,des} (k+1)- N_{i}^{in} (k) \right)/\Delta t
\end{equation}
and $ N_{i}^{in,des}(k)$ is the desired cumulative link inflow that can be given by:
\begin{equation}
\begin{split}
     N_i^{in,des}(k)&= N_i^{in}(k-1)+{\sum\limits_{j \in \boldsymbol{I}_{i}^{in}}e_{j,i}( N_j^{out,des}(k)- N_j^{out}(k-1))}\\& = N_i^{in}(k-1)+{\sum\limits_{j \in \boldsymbol{I}_{i}^{in}}e_{j,i}\left(\mathit{\min}\left( {N}_{j}^{out}(k) + {q}_{sat}^{out}(k){b}_{j}(k),\bar{N}_{j}^{out}\left( {k} \right) \right)- N_j^{out}(k-1)\right)}
    \end{split}
    \label{desired in}
\end{equation}
where $e_{j,i}$ represents the turning rate from link $j$ to $i$; $\boldsymbol{I}_{i}^{in}$ represents the set of the feeding links of $i$ ; $ b_{j}(k)$ represents the corresponding green time fraction of $j$ at $k$. To avoid the green times between the conflict phases overlapping, the assignment of the green time fraction should satisfy $
{\sum_{ c \in \boldsymbol{p}^{c}}{ b_{c}(k)}} \leq 1$, where $\boldsymbol {p}^c$ represents the set of signal phases whose moving directions conflict with each other.

{Therefore, it can be found when $ q_{i}^{in,des}(k)=0$ or $ \bar {q}_{i}^{in}(k)> q_{i}^{in,des}(k)$, $ \beta_{i}(k)=1$, otherwise $0\le  \beta_{i}(k)<1$. If $ \beta_{i}(k)<1$, the outflow of the node’s incoming links and the inflow of the node’s outgoing links need to be adjusted. Hence, the task of the node model is to distribute the supply over their corresponding feeding links, and an integrative procedure is designed as provided in Algorithm \ref{algotithm1} based on \cite{jabari2016node} and \cite{tampere2011generic}, where the outflows of the feeding links can be reduced by Eq. (\ref{out_max}) based on the supply that does not related to the demand change of the feeding link, and the potential violations of invariance principle (IP) caused by the demand proportional distribution can be overcome. The IP states that the flow solutions for a node model should remain invariant to increases in supplies (or demands) when restricted by demands (or supplies). This principle should be satisfied for intersection models \citep{daganzo1995cell,lebacque2005first, jabari2016node,tampere2011generic}.}

\RestyleAlgo{ruled}
\begin{algorithm}
\small
\caption{{Node’s outflow and inflow adjustments}}\label{algotithm1}
{{\bf Inputs}: cumulative link inflow $ N_i^{in} (k)$ at $k$, cumulative inflow limit $ \bar N_i^{in} (k+1)$, desired cumulative link inflow $ N_i^{in, des} (k+1)$, and maximum cumulative outflow
$\bar {N}_{i}^{out}(k+1)$ at $k+1$ which are determined by the link model; node’s incoming links $\boldsymbol{I}^{in}$ and outgoing links $\boldsymbol{I}^{out}$; vehicle turning rates within the turn-links inner\\
{\bf Outputs}: actual cumulative outflows of $\boldsymbol{I}^{in}$ and actual cumulative inflows of $\boldsymbol{I}^{out}$ \\
{\bf Step 1}: Find the node's outgoing link with the smallest decreased index $ \beta_{i}(k)$ and its feeding links\\
Given ${i\in \boldsymbol I^{out}}$, $ \bar N_i^{in} (k+1)$, $ N_i^{in, des} (k+1)$, and $ N_i^{in} (k)$ calculate the decreased index $ \beta_{i}(k)$ for all node's outgoing links $\boldsymbol I^{out}$ by Eq. (\ref{decreased index}). \\
{\eIf{$ \beta_{i}(k) < 1$}{Find the link index $\xi$ corresponding to the smallest decreased index $\xi=arg \min_{i\in \boldsymbol I^{out}}  \beta_{i}(k) $\\Find all the links $\boldsymbol  I_{\xi}^{in}$ feeding link $\xi$, which means to identify the link $j$ that has a non-negative turning rate $e_{j,\xi}>0$ towards link $\xi$}{All $ \beta_{i}(k)$ values are equal to 1, and continue to {\bf Step 5}}}}
{{\bf Step 2}: Decrease the maximum cumulative outflows of the feeding links $\boldsymbol  I_{\xi}^{in}$\\
{
\For{ all $j \in \boldsymbol  I_{\xi}^{in}$}{Let $s_i= \bar N_i^{in} (k+1) $ represent the supply of $i$, then calculate the available supply of $j$: 
\begin{equation}
s_j=\min_{e_{j,i}>0} \frac{s_i}{e_{j,i}}
\label{supply}
\end{equation}
\\
}
Update the maximum cumulative outflow of $j$:
\begin{equation}
     \bar N_j^{out} (k+1)\longleftarrow \min\left( \bar N_j^{out} (k+1), s_j\right)
    \label{out_max} 
\end{equation}
    $j\longleftarrow j+1$}}

{{\bf Step 3}: Update the maximum cumulative inflow and the desired cumulative inflow of $i$\\
Now, the maximum outflows of the links in $\boldsymbol  I_{\xi}^{in}$ are determined, hence, their corresponding downstream supply represented by the link maximum cumulative inflow $\bar N_i^{in}(k+1)$ can be reduced with these values:
\begin{equation}
     \bar N_i^{in} (k+1)\longleftarrow  \bar N_i^{in} (k+1)-{\sum\limits_{j \in \boldsymbol I_{\xi}^{in}}e_{j,i}\left( \bar N_j^{out} (k+1)- N_j^{out} (k) \right)}
     \label{max_update}
\end{equation}
Also, update desired cumulative inflow $ N_i^{in,des}(k+1)$:
\begin{equation}
     N_i^{in, des} (k+1)\longleftarrow  N_i^{in,des} (k+1)-{\sum\limits_{j \in \boldsymbol  I_{\xi}^{in}}e_{j,i}\left( \bar N_j^{out} (k+1)- N_j^{out} (k) \right)}
    \label{des_update}
\end{equation}}
\\
{{\bf Step 4}: Set the turning rates to zero due to they have been involved in determining $ \bar N_i^{in} (k+1)$ and $ N_i^{in, des} (k+1)$, implying their effect is already present in the cumulative flows 
\begin{equation}
   e_{j,i}=0, ~\forall j \in \boldsymbol  I_{\xi}^{in}, i\in \boldsymbol I^{out}
\end{equation}
Return to {\bf Step 1}.}
\\
{{\bf Step 5}: Estimate the actual cumulative outflow of the node's incoming link
\begin{equation}
    N_j^{out} (k+1)= \bar N_j^{out} (k+1),~ j \in \boldsymbol I^{in}
   \label{actual out}
\end{equation}}
\\
{According to Eq. (\ref{actual out}), the actual cumulative inflow of the node's outgoing link can be obtained:
\begin{equation}
    N_i^{in} (k+1)= N_i^{in} (k)+{\sum\limits_{j \in \boldsymbol{I}_{i}^{in}}e_{j,i}\left( N_j^{out} (k+1)- N_j^{out} (k) \right)}, ~i\in \boldsymbol I^{out}
\end{equation}}
\end{algorithm}

{The main aim of Algorithm \ref{algotithm1} is to identify the node's outgoing link $\xi$ with the smallest $ \beta_{i}(k)$ that represents the strongest supply constraint on the demand \citep{van2016urban}, and then reduce the outflow of the upstream links feeding this link based on Eq. (\ref{out_max}). Note that if a shared lane exists on the node's incoming link, as depicted in Figure \ref{node}(a), this procedure will reduce the outflows of all turning directions of the feeding link, implying that some supply becomes available for other links. For example, we assumed that Link C in Figure \ref{node}(a) has insufficient supply at $k$, hence the desired outflow of its upstream feeding Link A should be reduced using the node model as summarised in Algorithm \ref{algotithm1}. Obviously, this process will simultaneously reduce the right-turn outflow of Link A due to the lane sharing, which would make some supply free for Link B. Consequently, the outflow of Link B might be increased if there is great demand. Thus, to avoid hold-back traffic, after reducing the outflow of a certain node's incoming link, $ \beta_{i}(k)$ for all node's outgoing links will be updated accordingly as shown in Step 4, potentially increasing or remaining unchanged. This iterative process continues until all $ \beta_{i}(k)$ values are equal to 1, indicating that at this point, all outflows do not exceed downstream supply. Besides, the node flows obtained by Algorithm \ref{algotithm1} are invariant to increases in supply when constrained by demand and vice versa. It is an invariant holding-free algorithm that satisfies the IP and can provide a holding-free solution. The detailed proof is provided in the Appendix B.}

{For the node's incoming link without lane sharing, as depicted in Figure \ref{node}(b), setting the turning rate within the turn-link to 1 suffices, as the demand towards each destination link separately, and the impacts of turning rates are already reflected in the outflow of the upstream common link. We will next turn to prove that Algorithm \ref{algotithm1} is indeed an invariant holding-free solution algorithm.}
\label{node model}

{The complete algorithm for the proposed LQM, incorporating both link and node models, is outlined in Algorithm \ref{algotithm2}.}

\RestyleAlgo{ruled}
\begin{algorithm}[H]
\renewcommand{\thealgocf}{2}
\small
\caption{{Proposed LQM including link and node models}}\label{algotithm2}
{{\bf Inputs}: link parameters; traffic demands; vehicle turning rates; green time fractions\\
{\bf Outputs}: queue lengths, cumulative link inflows, queue inflows and outflows\\
{\bf Step 1}: Estimate queue inflow and queue length of each link\\
Calculate the queue density $\rho_i^{cr}(k)$ of link $i$ at $k$ based on Eq. (\ref{queue density2}), where $i \in \boldsymbol{I}^{in} \cup \boldsymbol{I}^{out}$, $k \in [0,T]$;\\
Calculate the corresponding queue length $L_q(k)$ and free-flowing length $L_f(k)$ based on Eqs. (\ref{q length}) and (\ref{f length}), respectively.\\
Update the queue inflow of $i$ based on Eq. (\ref{N_qu_LT3});\\
{\bf Step 2}: Determine the desired inflow for each link by Eq. (\ref{desired in})}

{{\bf Step 3}: Determine the desired outflow for each link\\
The desired outflow for $i$ at $k$ is defined as the maximum cumulative link outflow as elaborated in Section 2.2.4. This can be estimated by
\begin{equation}
\bar {N}_{i}^{out}\left( {k} \right) = \mathit{\min}\left( {N}_{i}^{out}(k) + {q}_{sat}(k){b}_{i}(k),{N}_{i}^{out,des}\left( {k} \right) \right)
\end{equation}
\\
{\bf Step 4}: Utilize the developed node model to compute the actual inflows and outflows for each link by employing Algorithm 1.}

{Update time step $k=k+1$, if $k<T$, return to {\bf Step 1}.
\\
{\bf Step 5}: Output queue lengths, cumulative link inflows, queue inflows and outflows}
\end{algorithm}

%% file: Simulation_experiments.tex
\subsection{Scenario 1: Simulation experiments at a single signalized intersection}

{Since both the free-flow speed and queue transmission significantly influence the flow propagation results. In the following Section 3.1.1, we initially set all link free-flow speeds as constants to concentrate on analyzing the effects of queue transmission on the typical link-based model. Subsequently, in Section 3.1.2, the free-flow speed of a specific link will be made time-varying to further validate the model's performance under TFS.}
\subsubsection{Model performance under fixed free-flow speed}
\label{Basic settings}

\textbf{(a) Experimental set-up}

\begin{figure}[H]
\captionsetup{font={small}}
\centering  
{
\includegraphics[width=2.6in]{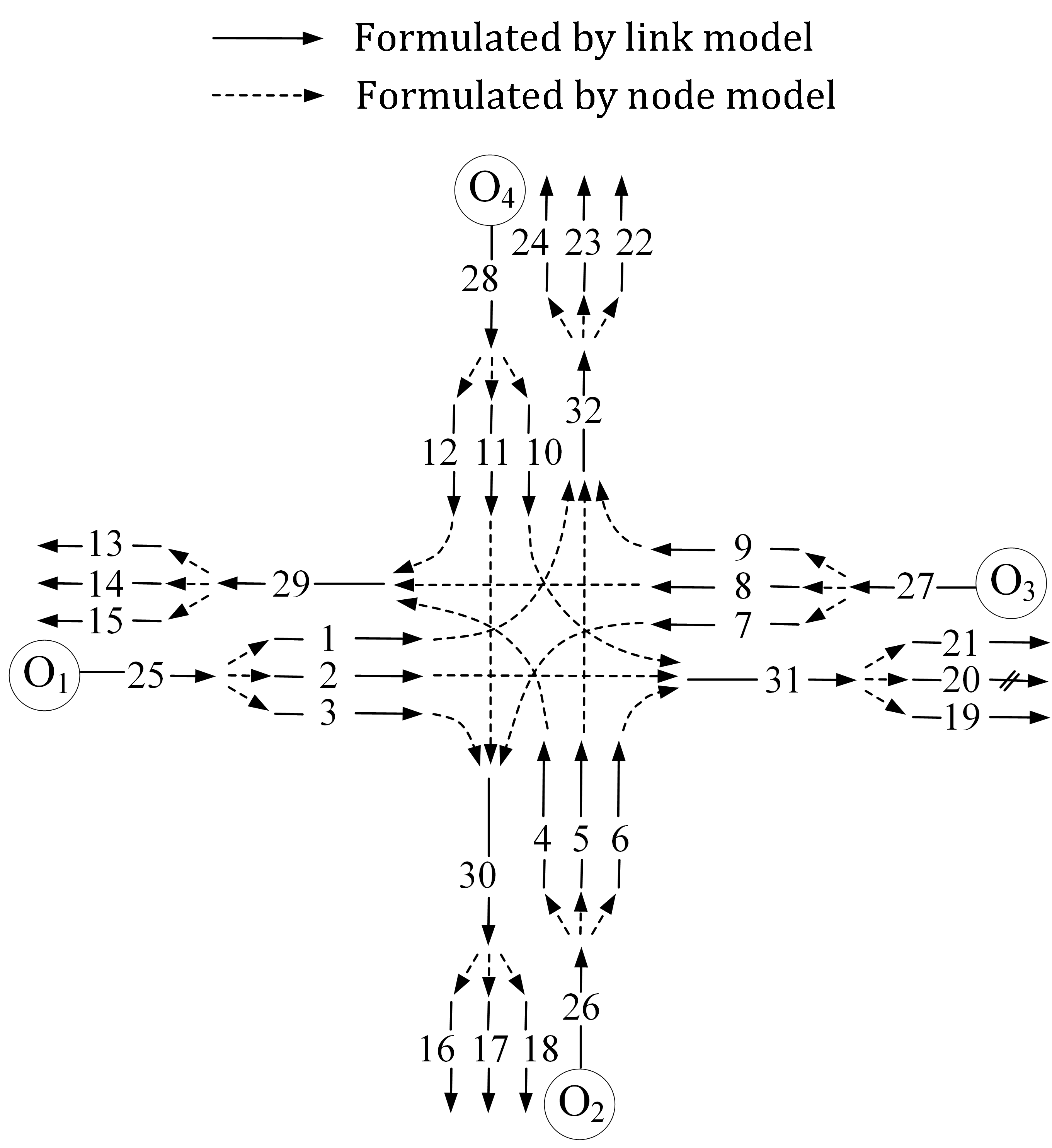}}
\caption{{Representation of the signalized intersection depicted in LQM, where LQM assigns a common link and turn links to each road,  and Links 1-24 are the turn links, each with a length of 100m, while the remaining links represent the common links, each with a length of 500m; Link 20 is set to a bottleneck; $\mathrm O_1$, $\mathrm O_2$, $\mathrm O_3$, $\mathrm O_4$ are the input demands.}}
\label{intersection}
\end{figure}

\begin{figure}[H]
\captionsetup{font={small}}
\centering
\includegraphics[width=2.6in]{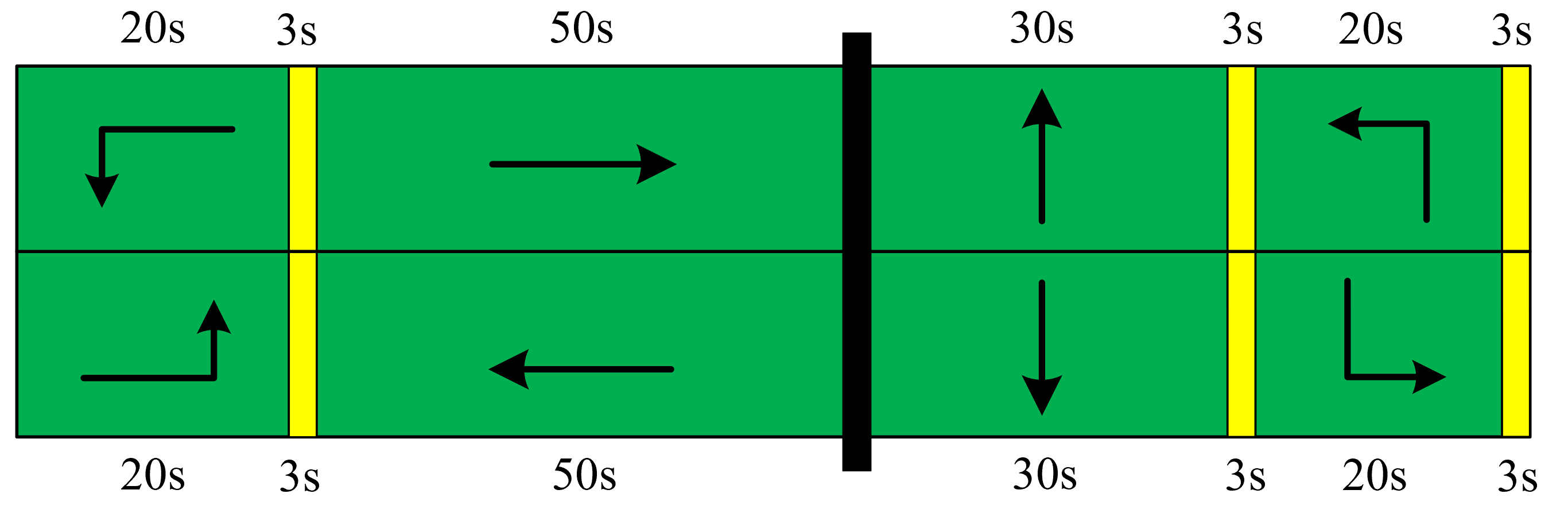}
\caption{{Signal timing plan}}
\label{Signal timing plan}
\end{figure}

% \noindent\textbf{(b) Simulation results}
{Simulation experiments with a standard four-arm intersection are conducted to assess the effectiveness of the proposed LQM with TFS. The simulated signalized intersection consists of four incoming RSs, each with a length of 600m and divided into a common link and three downstream turn links, as shown in Figure \ref{intersection}. Tables \ref{Road parameters} and \ref{turning rate} provide information on link related parameters and vehicle turning rates. Note that the free-flow speed for the turn links is smaller than the common links due to the physical characteristics of turning vehicles. The four-phase signal timing plan implemented at the intersection is shown in Figure \ref{Signal timing plan}. We assumed a fixed signal timing signal plan, thus the green time fractions in Table \ref{turning rate} are constant values. Besides, to validate the proposed LQM for representing the queue changes and spillbacks, we set the green time fraction of Link 20, which is the through direction of a node's outgoing link, to 0.1. This adjustment aimed to initiate queue spillback from Link 20.}

{The link-based model proposed for signalized intersections in \cite{van2018hierarchical} serves as the baseline model (LM) for comparative analysis. LM operates at the segment level, treating roads with multiple turning directions as a unified, homogeneous link. Consequently, in LM, the distinct turn links depicted in Figure \ref{intersection} are aggregated into a single link, which is also commonly adopted by other studies \citep{yperman2005link,himpe2016efficient}. This aggregation results in the turn-level flow operation differences, e.g., green time, speed, capacity, that cannot be reflected by their models. Moreover, the LM only focuses on the formulation of link inflow and outflow at road boundaries. The impacts of turn-level queue dynamics on both link inflow and outflow are disregarded, which cannot capture the queue length change. A more comprehensive comparison will be presented in subsequent discussions. }

% \begin{figure}[H]
% \captionsetup{font={small}}
% \centering
% \includegraphics[width=4.9in]{Figure/intersection_f2.png}
% \caption{Visualization of the signalized intersection modeled in LTM: (a) LQM integrates a common link and multiple turn links for each road segment, where Links 1-24 are the turn links and others are the common links; Link 20 is set to a bottleneck;
% (b) LM allocates homogeneous links for each road segment, where Link 14 is set to a bottleneck; $\mathrm O_1$, $\mathrm O_2$, $\mathrm O_3$, $\mathrm O_4$ are the input demands.}
% \label{intersection}
% \end{figure}

\begin{table}[H]
\captionsetup{font={small}}
\footnotesize
\renewcommand\arraystretch{1}
\caption{{Link parameters \label{Road parameters}}}
%\begin{tabular}{cccp{10cm}}
\centering
\begin{tabular}{m{3.9cm}<{\centering} m{3.85cm}<{\centering} m{3.65cm}<{\centering} }
\hline
\makecell[l]{Parameter}&Left-turn and right-turn links& Through and common links\\
\hline
\makecell[l]{Free-flow speed (m/s)}&4&11\\
% \makecell[l]{Saturation flow rate (veh/h)}&1400&1400\\
\makecell[l]{Backward wave speed (km/h)}&20&20\\
\makecell[l]{Jam density (veh/km)}&100&100\\
\hline
\end{tabular}
\end{table}

\begin{table}[H]
\captionsetup{font={small}}
\footnotesize
\caption{{Vehicle turning rates $e_{j,i}$ and green time fractions $b_R$\label{turning rate}}}
\centering
    \begin{tabular}{m{1.3cm}<{\centering} m{0.6cm}<{\centering} m{0.6cm}<{\centering} m{0.6cm}<{\centering} m{0.6cm}<{\centering} m{0.6cm}<{\centering} m{0.6cm}<{\centering}} 
    \hline
         \multirow{2}*{Direction}&  \multicolumn{2}{c}{Through}&  \multicolumn{2}{c}{Left turn}&  \multicolumn{2}{c}{Right turn}\\ 
         \cline{2-7}
         &  $e_{j,i}$&  $b_R$&  $e_{j,i}$&  $b_R$&  $e_{j,i}$& $b_R$\\ 
         \hline
         West& 0.6 &0.38  &0.3  &0.15  &0.1  &1 \\ 
         South& 0.4 &0.23  & 0.1 &0.15  &0.5  &1 \\ 
         East& 0.6 &0.38  &0.3  &0.15  & 0.1 &1 \\ 
         North&0.5  &0.23  &0.4  &0.15  &0.1  &1 \\ 
         \hline
    \end{tabular}
\end{table}

% \begin{table}[H]
% \captionsetup{font={small}}
% \footnotesize
% \renewcommand\arraystretch{1}
% \caption{Link parameters\label{Road parameters}}
% %\begin{tabular}{cccp{10cm}}
% \centering
% \begin{tabular}{m{3.7cm}<{\centering} m{0.7cm}<{\centering} m{0.7cm}<{\centering} m{0.7cm}<{\centering} m{0.7cm}<{\centering} m{0.7cm}<{\centering} m{0.7cm}<{\centering} m{0.7cm}<{\centering} m{0.7cm}<{\centering}}
% \hline
% Parameter&RS 1&RS 2&RS 3&RS 4&RS 5&RS 6&RS 7&RS 8\\
% \hline
% Saturation flow rate (veh/h)&1000&1800&800&950&950&1100&1500&900\\
% Shockwave speed (km/h)&20&20&20&20&20&20&20&20\\
% Jam density (veh/km)&150&150&150&150&150&150&150&150\\
% \hline
% \end{tabular}
% \end{table}

% \begin{figure}
% \captionsetup{font={small}}
% \centering  
% \subfigure[Eastern direction]{
% \includegraphics[width=5.7cm,height = 4.9cm]{Figure/demandw.png}}~\subfigure[Southern direction]{
% \includegraphics[width=5.7cm,height = 4.9cm]{Figure/demands.png}}\\
% \subfigure[Eastern direction]{
% \includegraphics[width=5.7cm,height = 4.9cm]{Figure/demande.png}}~\subfigure[Northern direction]{
% \includegraphics[width=5.7cm,height = 4.9cm]{Figure/demandn.png}}
% \caption{Traffic demand of each incoming link}
% \label{demand}
% \end{figure}

\begin{figure}[H]
\captionsetup{font={small}}
\centering
\includegraphics[width=2.4in]{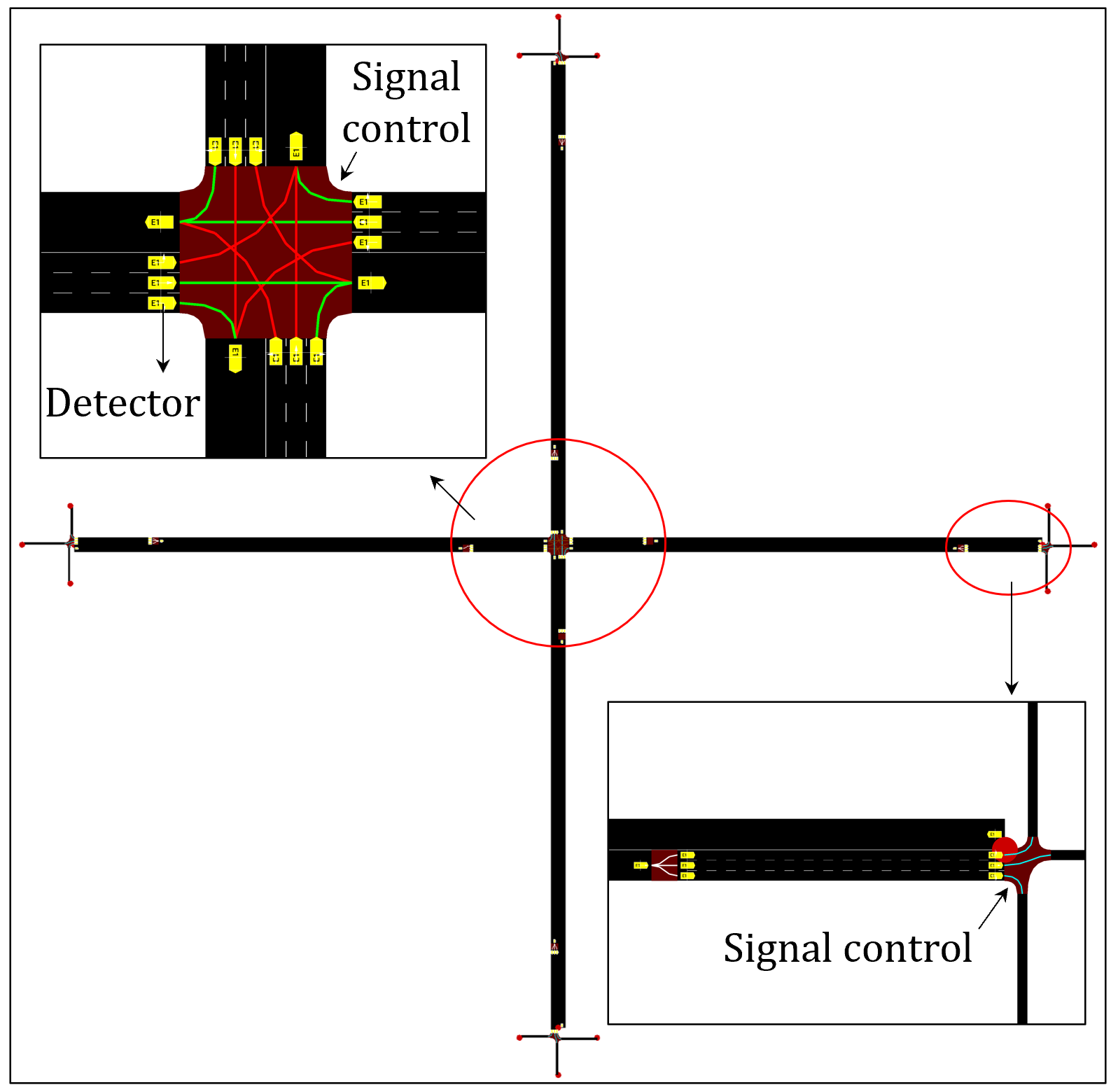}
\caption{{Illustration of simulated signalized intersection in SUMO}}
\label{s intersection}
\end{figure}

{SUMO, an open-source microscopic traffic simulation tool for managing and controlling urban traffic is used to compare model performance of  the LM and LQM and to verify our LQM model. To ensure a fair comparison, as shown in Figure \ref{s intersection}, the same simulation settings and inputs for both link-based models, including road length, free-flow speed and signal control plan are the same as those in SUMO. The node's outgoing links, e.g., Links 19, 20, 21,  are also signalized as presented in Figure \ref{s intersection}. }

\begin{figure}[H]
\captionsetup{font={small}}
\centering
\includegraphics[width=2.65in]{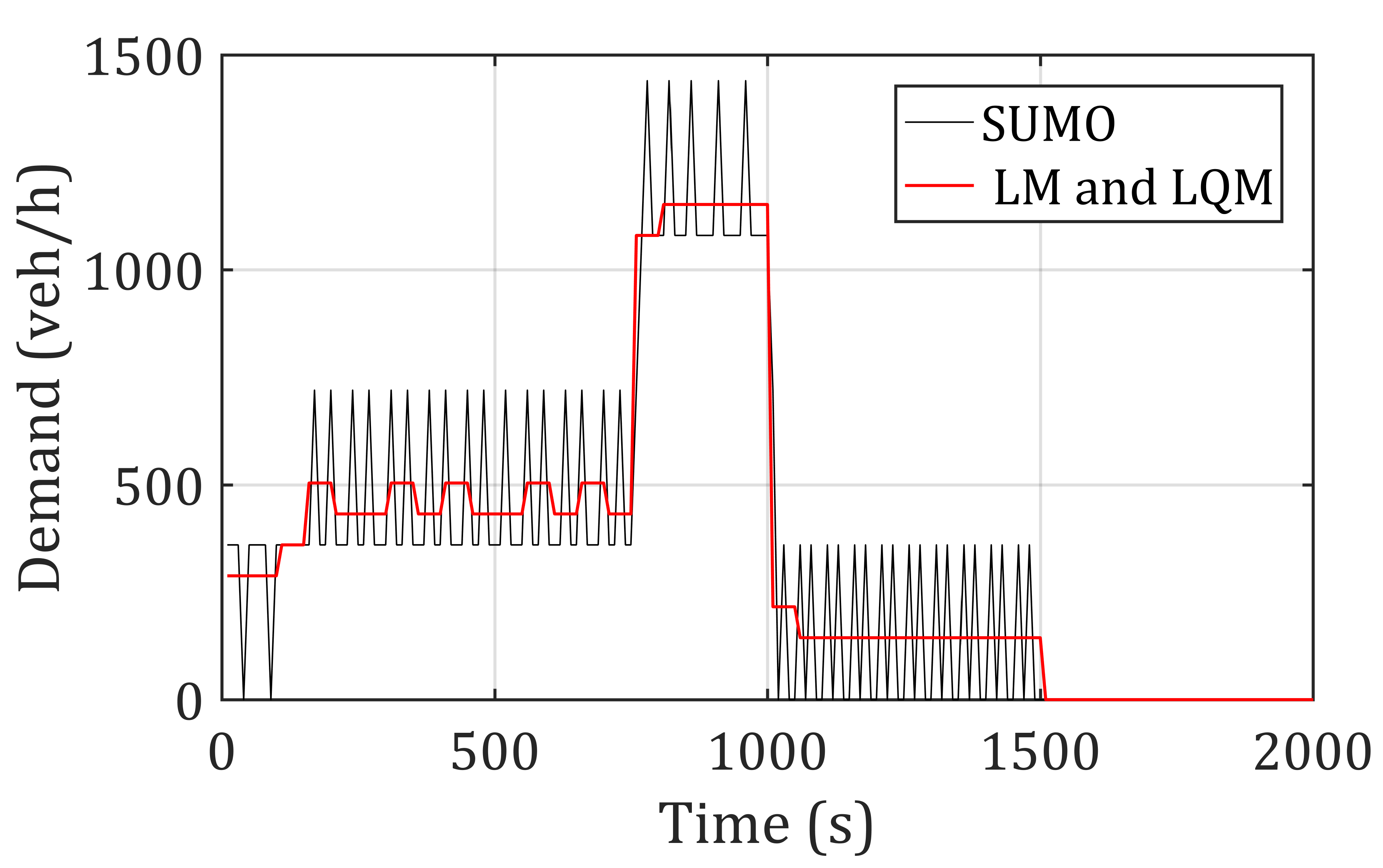}
\caption{{Input traffic demands of the simulated signalized intersection}}
\label{demand}
\end{figure}

{Figure \ref{demand} provides the time-varying input demands during the simulation process, where the demands of SUMO are collected by setting the detectors on the link as shown in \ref{s intersection}, and the demand sampling interval is 10s. The first 750s is the warm-up stage, after which the input demand is increased to simulate the traffic peak period until the 1000s. After the 1000s, the demand gradually decreases and returns to the initial stage. From the 1000s to the 1500s, the input demand remains at zero. Due to the significant oscillations in the SUMO demand curve, averaging every five demand values (meaning taking a 50s interval, e.g., 5$\times$10=50s) as the input demand for LM and LQM. Both the LM and LQM, along with SUMO, were simulated for 2000s, where the sampling interval of LM and LQM is set to 10s.}

\noindent
\textbf{(b) Simulation results}

\begin{figure}[H]
\captionsetup{font={small}}
\centering  
\subfigure[Cumulative flow of L20 by LM]{
\includegraphics[width=5cm,height = 4.1cm]{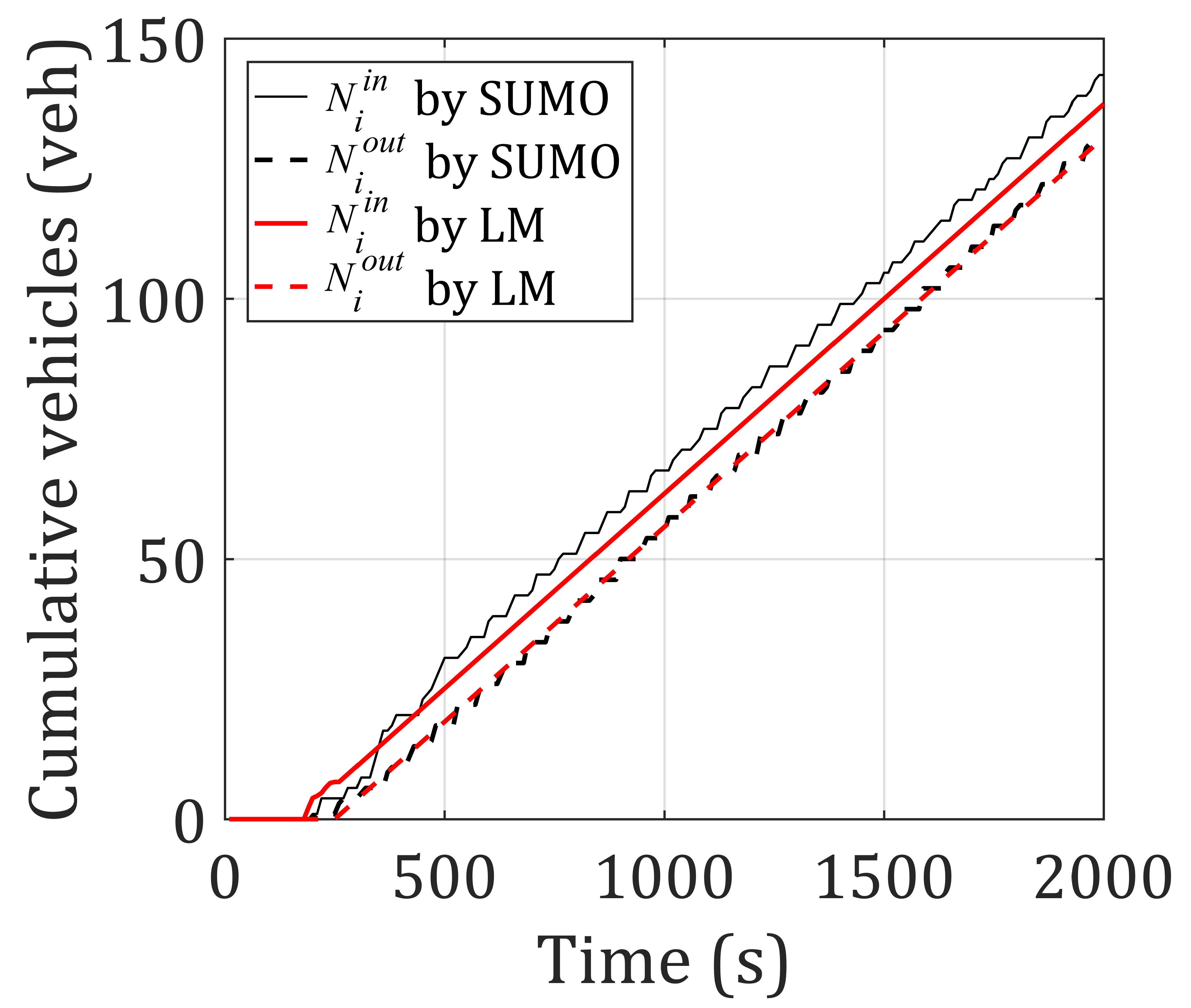}}\subfigure[Cumulative flow of L20 by LQM]{
\includegraphics[width=5cm,height = 4.1cm]{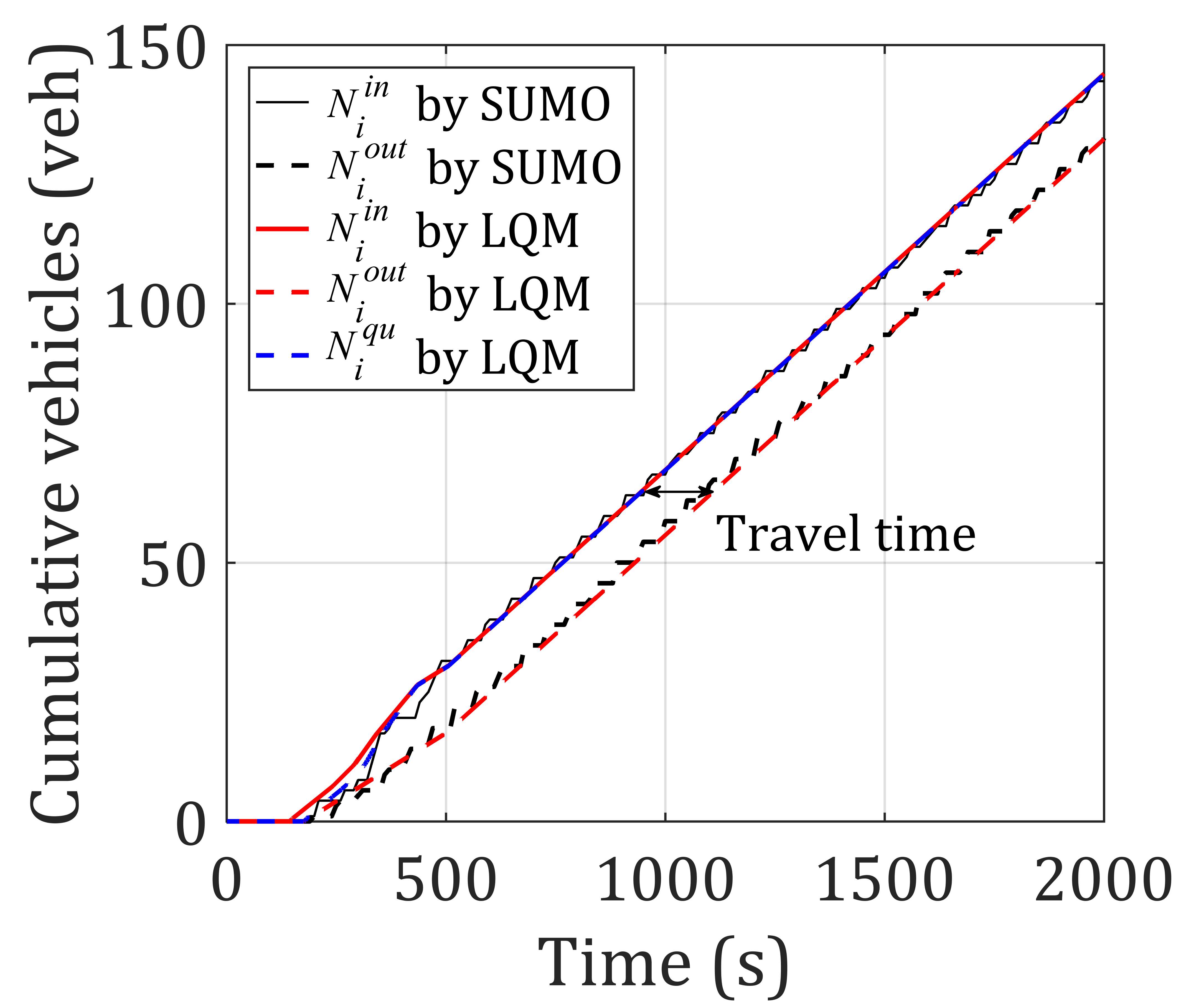}}\subfigure[Queue length of L20 by LQM]{
\includegraphics[width=5cm,height = 4.1cm]{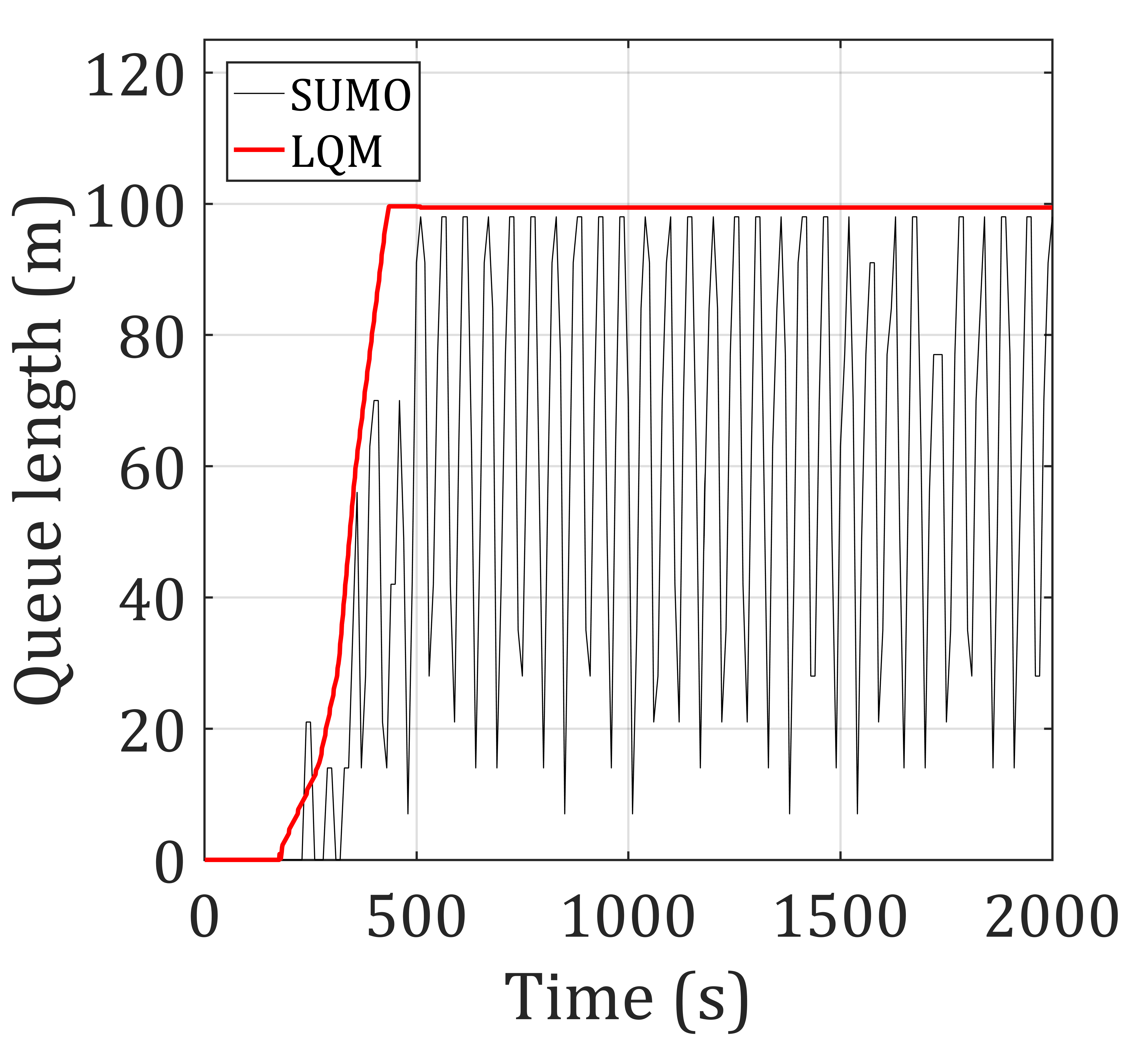}}\\
\subfigure[Cumulative flow of L31 by LM]{
\includegraphics[width=5cm,height = 4.1cm]{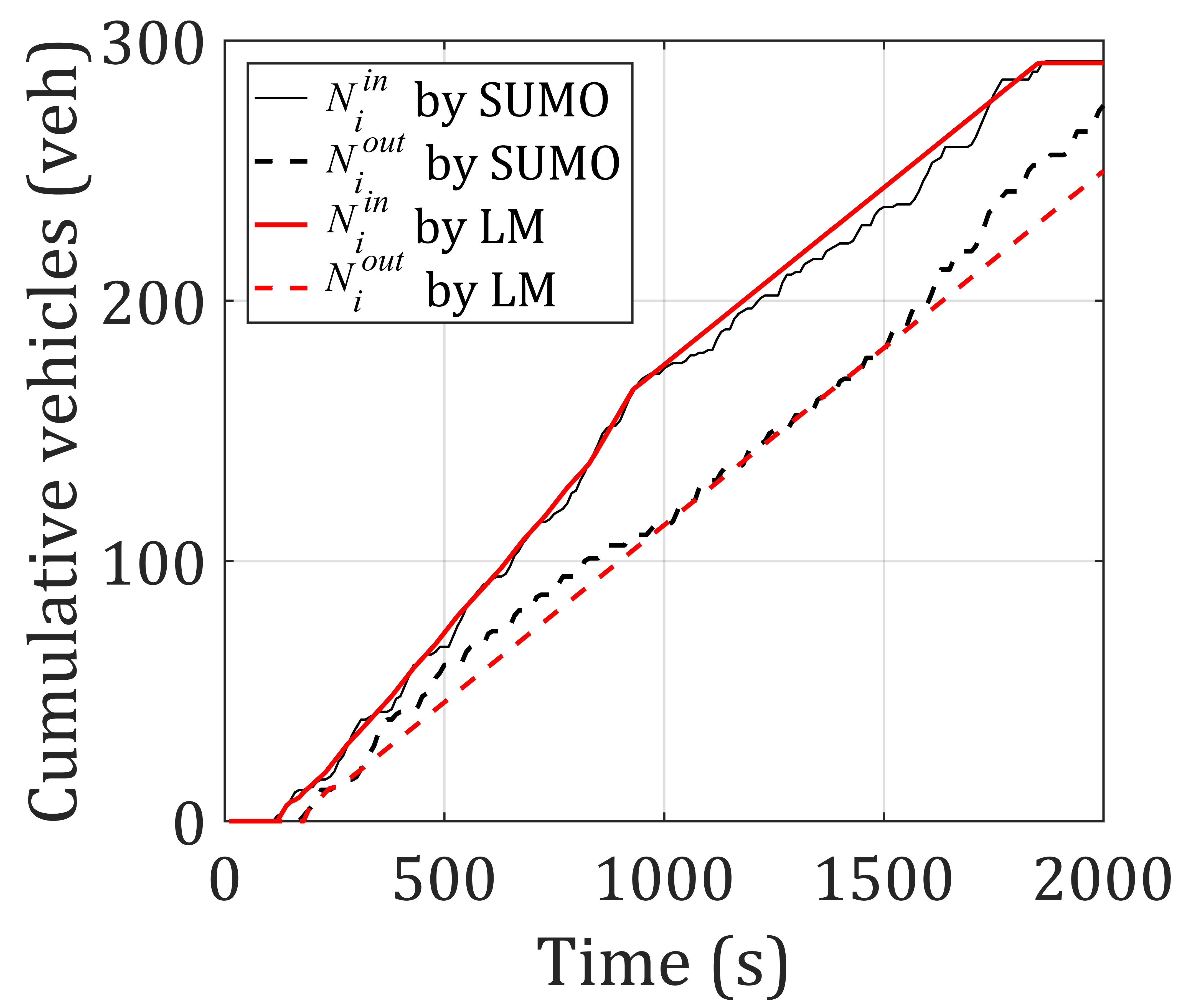}}\subfigure[Cumulative flow of L31 by LQM]{
\includegraphics[width=4.9cm,height = 4.1cm]{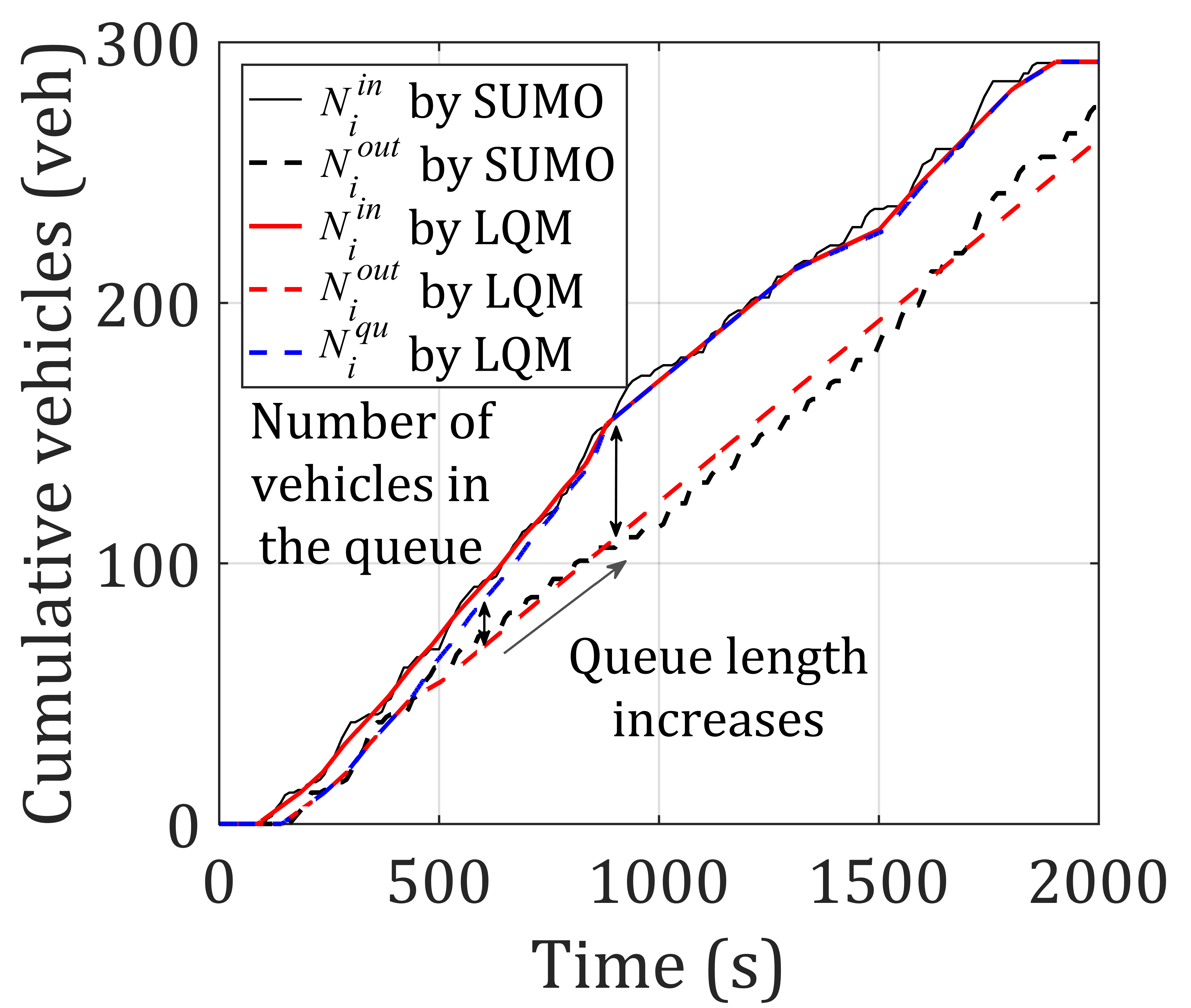}}\subfigure[Queue length of L31 by LQM]{
\includegraphics[width=4.9cm,height = 4.1cm]{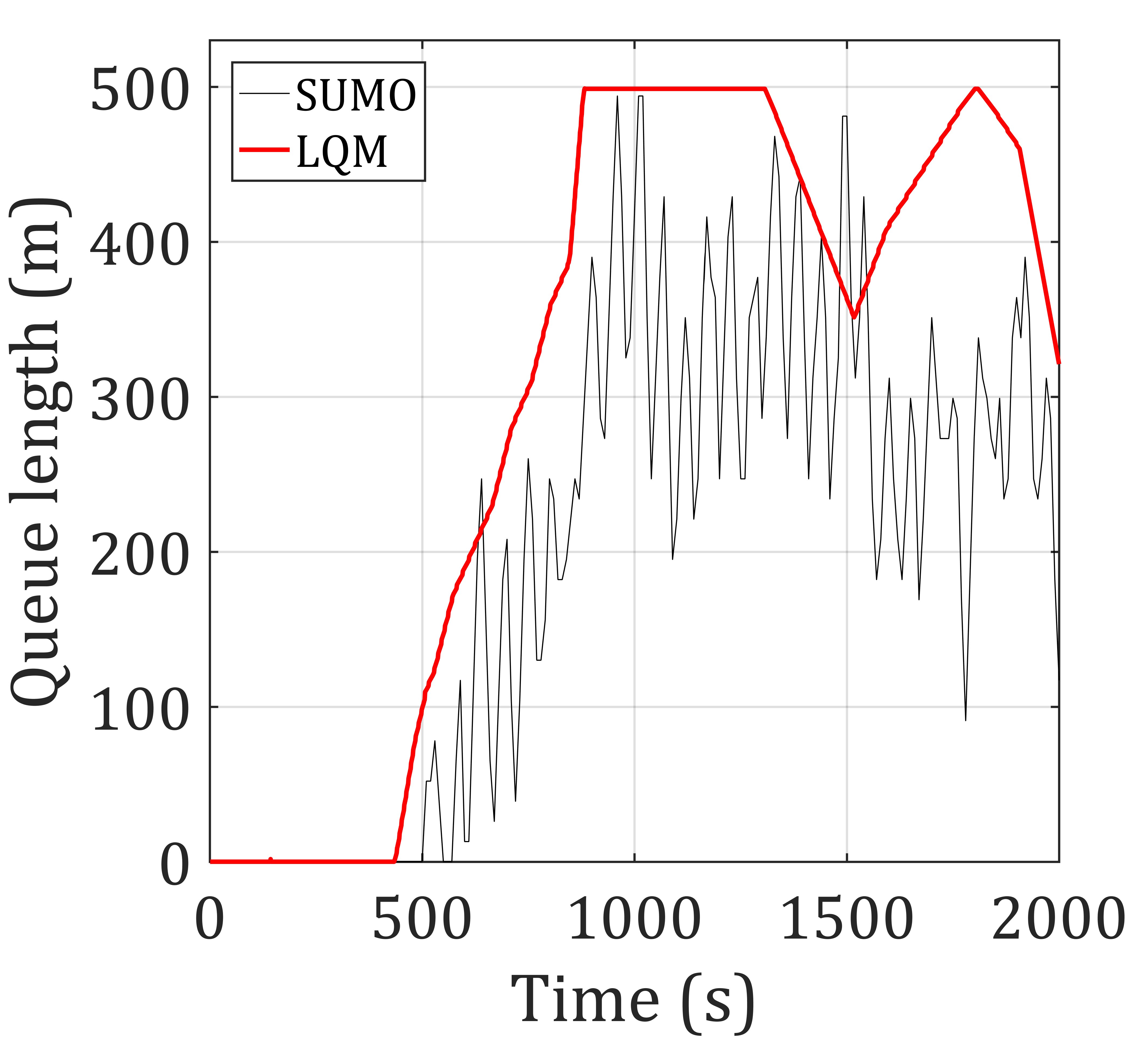}}
\subfigure[Flow rate (veh/s) of L20 by LQM]{
\includegraphics[width=6.4cm,height = 4.3cm]{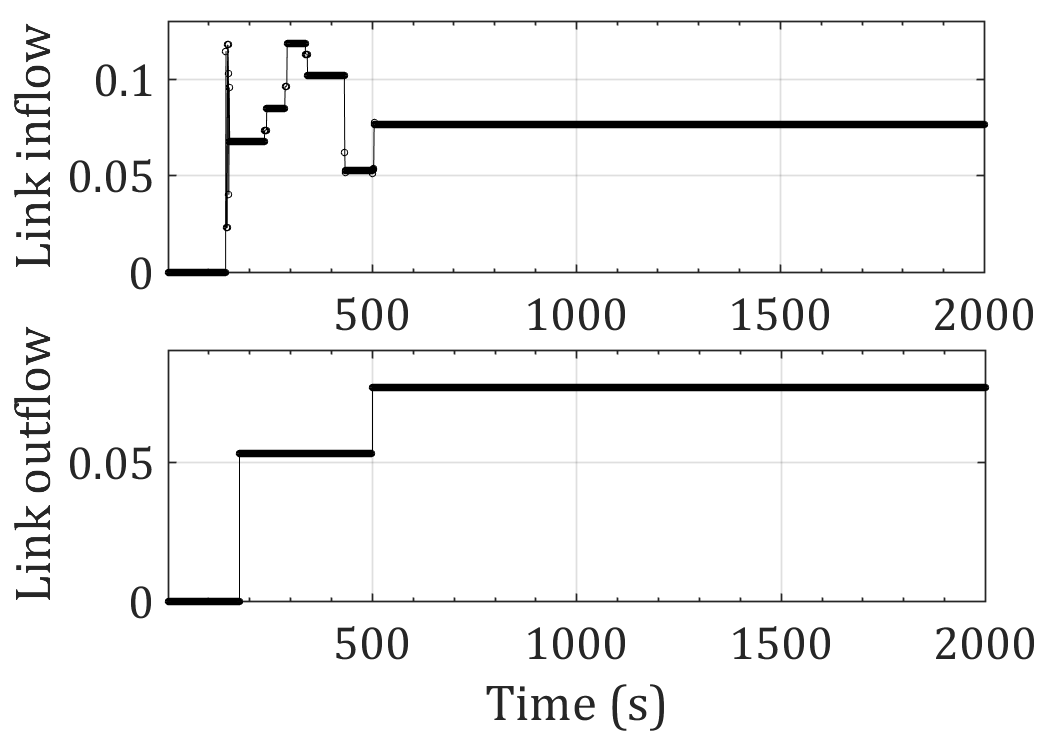}}~~~\subfigure[Flow rate (veh/s) of L31 by LQM]{
\includegraphics[width=6.4cm,height = 4.3cm]{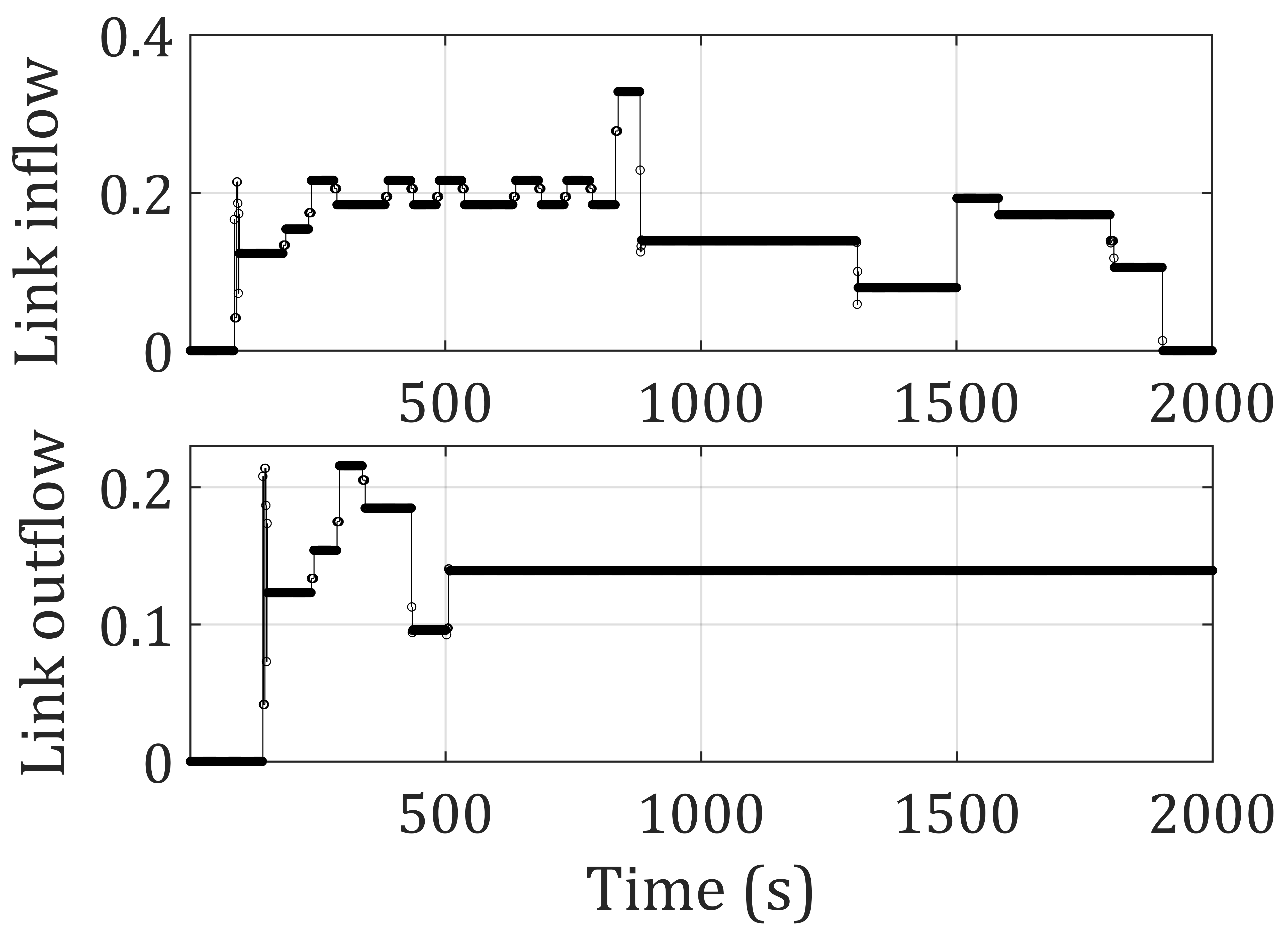}} \caption{{Comparison results of the bottleneck flows simulated by SUMO, LM and LQM, where L20 is the bottleneck link 20; L31 is the corresponding upstream common link 31.}}
\label{bottleneck flows}
\end{figure}

% The simulation results of LM, LQM and SUMO are quantified by using the cumulative link inflows, outflows, and flow rates. To verify the dynamic queue inflow and the LQM's capability of capturing turn-level spillback, queue length changes are compared with those obtained from SUMO at each time step.
{The simulation results of LM, LQM and SUMO are quantified by using the cumulative link inflow/outflow, and corresponding flow rate, along with queue length at each time step. The results of the bottleneck link and its upstream common link are first presented in Figure \ref{bottleneck flows}. Note that the cumulative queue inflow in LQM is estimated based on the link inflow that cannot be directly captured by SUMO. Therefore, only the queue inflow by LQM is presented in Figure \ref{bottleneck flows}. As LM cannot capture detailed queue length changes within the link, only queue lengths from LQM and SUMO are compared. }

{As shown in Figure \ref{bottleneck flows}, the simulation results of the bottleneck turn link 20 and its upstream common link 31 by LM and LQM are quite similar. Both LM and LQM can match the simulation results by SUMO, which indicates that these two models are able to reproduce the flow propagation at the bottleneck with severe congestion. However, our proposed LQM can provide more information about flow propagation compared to LM, e.g., detailed queue length changes as shown in Figure \ref{bottleneck flows}(c) and (f), owing to its integrated queue transmission features. Although the frequent oscillations in SUMO's queue length changes caused by green/red phase transitions cannot be captured by LQM, the changing trends of queue length and maximum values observed in SUMO can be accurately reproduced by the proposed LQM. This capability is crucial for a macroscopic flow model intended for traffic control, such as MPC \citep{zhou2016two,van2016urban}, as it requires fewer computational resources without relying on simulations and complex calibration of vehicle driving behaviors in macroscopic models, e.g., SUMO. Taking Link 31 as an example, as shown in Figure \ref{bottleneck flows}(e), the vertical distance between the cumulative queue inflow and link outflow curves indicates the number of vehicles in the queue. The horizontal distance between the cumulative link inflow and outflow curves represents travel time. When no queuing vehicles are present at the initial simulation stage, e.g., 0 to 400s shown in Figure \ref{bottleneck flows}(f), the cumulative queue inflow curve overlaps with the cumulative link outflow curve, and the travel time is equal to the free-flow travel time. This indicates that incoming queueing vehicles directly form an outflow and leave the link. As the link inflow increases, the cumulative queue inflow curve and the cumulative link inflow curve become closer, indicating an increased queue length on the link. }

{Note that the link outflow rate has some frequent fluctuations before 500s as shown in Figure \ref{bottleneck flows}(h). These fluctuations stem from corresponding fluctuations in its link inflow rates. The fluctuations of link inflow rates, in turn, originate from changes in input demands, illustrated in Figure \ref{demand}, where the demand doesn't maintain a constant value before the 750s. Actually, any oscillations in the link's input flow can cause outflow oscillations when the link is at free-flow conditions. After about 500s, these fluctuations dissipate due to a substantial queue length, stabilizing the link outflow at the maximum allowed outflow rate. At this moment, the inflow is primarily governed by the outflow rates rather than the input demands. Consequently, the link outflow leads to link inflow rates toward constant values. }

\begin{figure}[H]
\captionsetup{font={small}}
\centering  
\subfigure[Cumulative flow of L2 by LM]{
\includegraphics[width=5cm,height = 4.1cm]{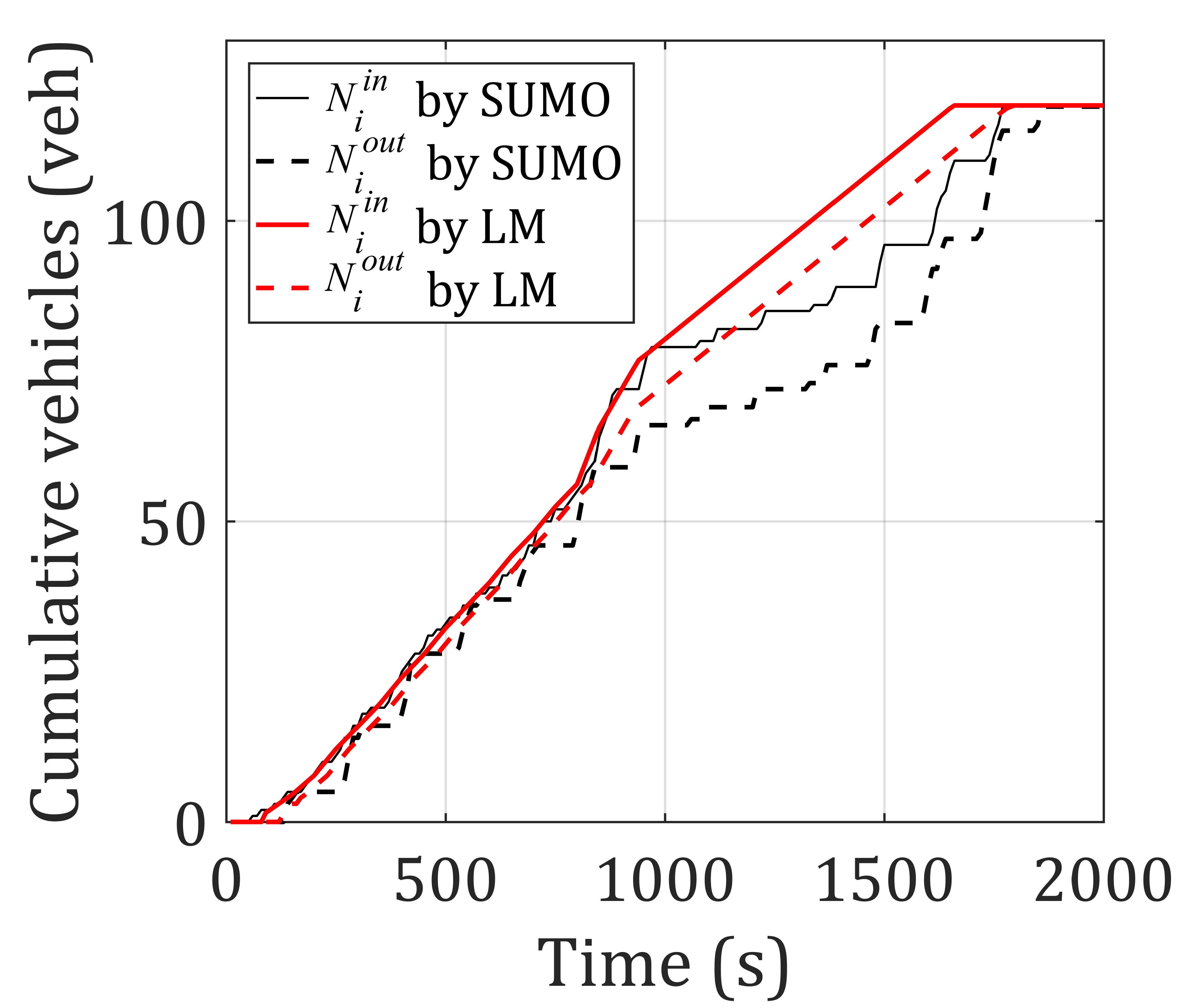}} \subfigure[Cumulative flow of L2 by LQM]{
\includegraphics[width=5cm,height = 4.1cm]{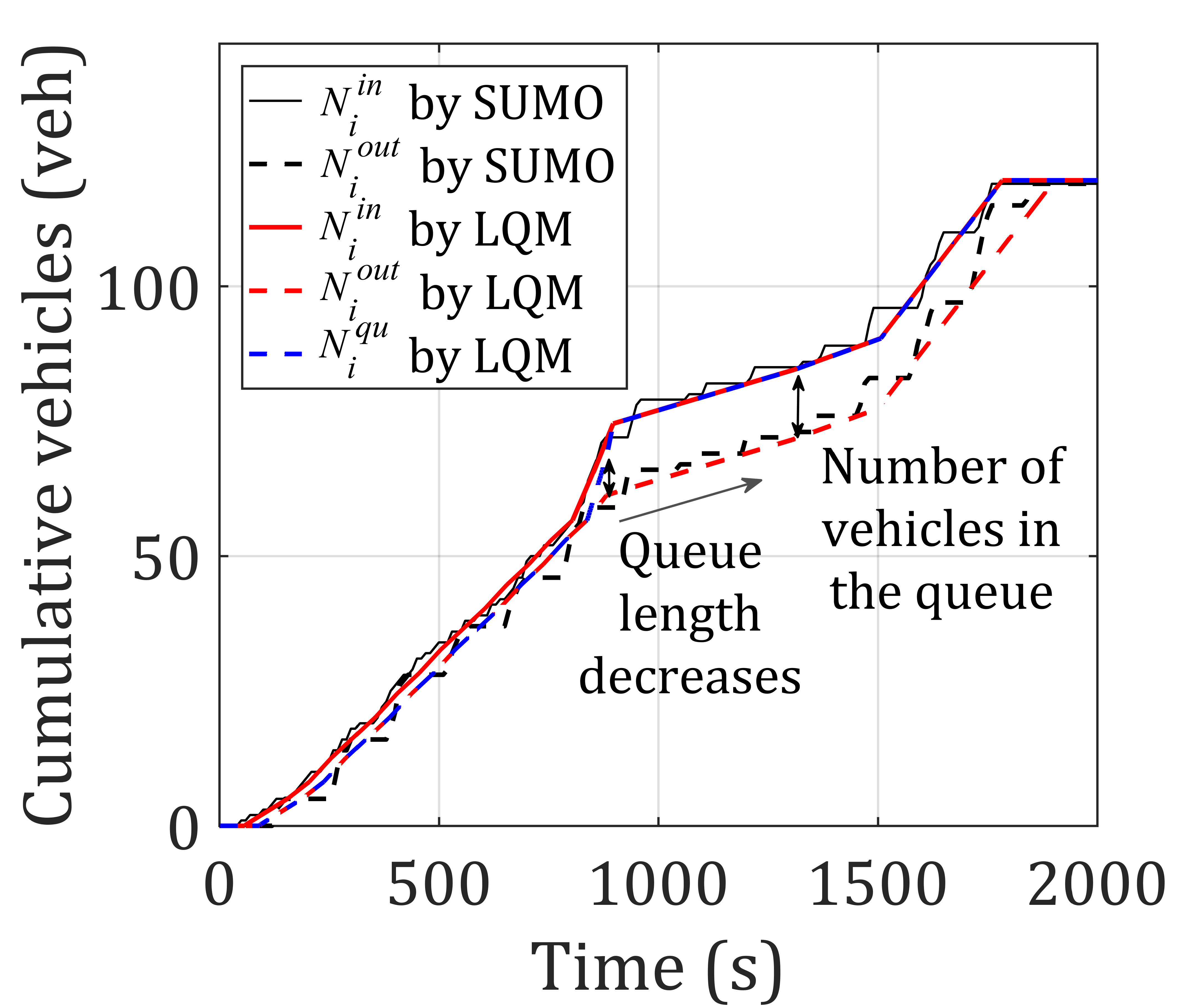}}\subfigure[Queue length of L2 by LQM]{
\includegraphics[width=4.9cm,height = 4.1cm]{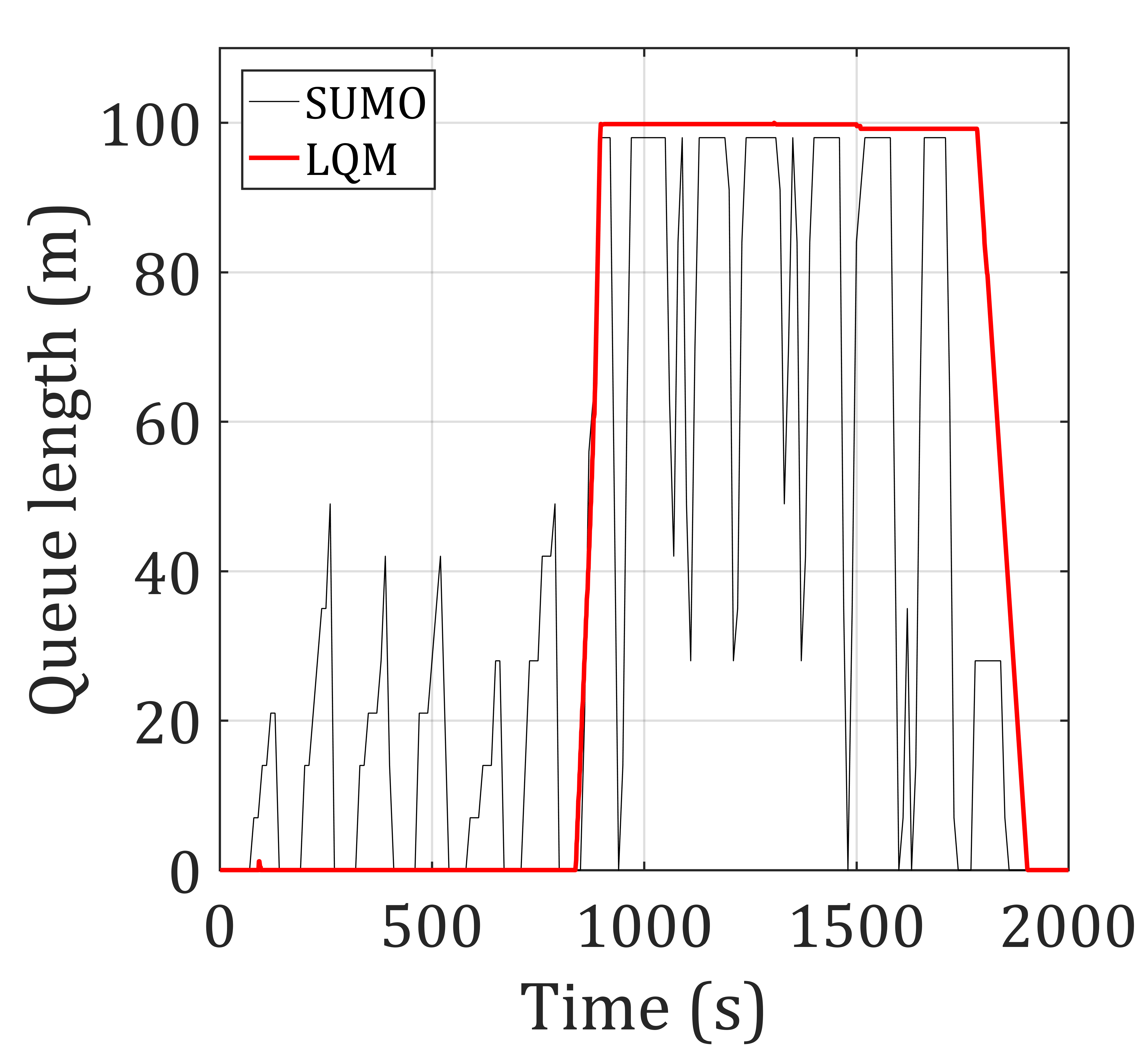}}\\
% \caption{Comparison of the upstream through direction related to bottleneck spillback from LM, LQM and SUMO simulations, where L2 is the turn link 2; L25 is the corresponding upstream common link 25.}
% \label{b l flow}
% \end{figure}

% \begin{figure}[H]\ContinuedFloat
% \captionsetup{font={small}}
% \centering  
\subfigure[Cumulative flow of L25 by LM]{
\includegraphics[width=5cm,height = 4.1cm]{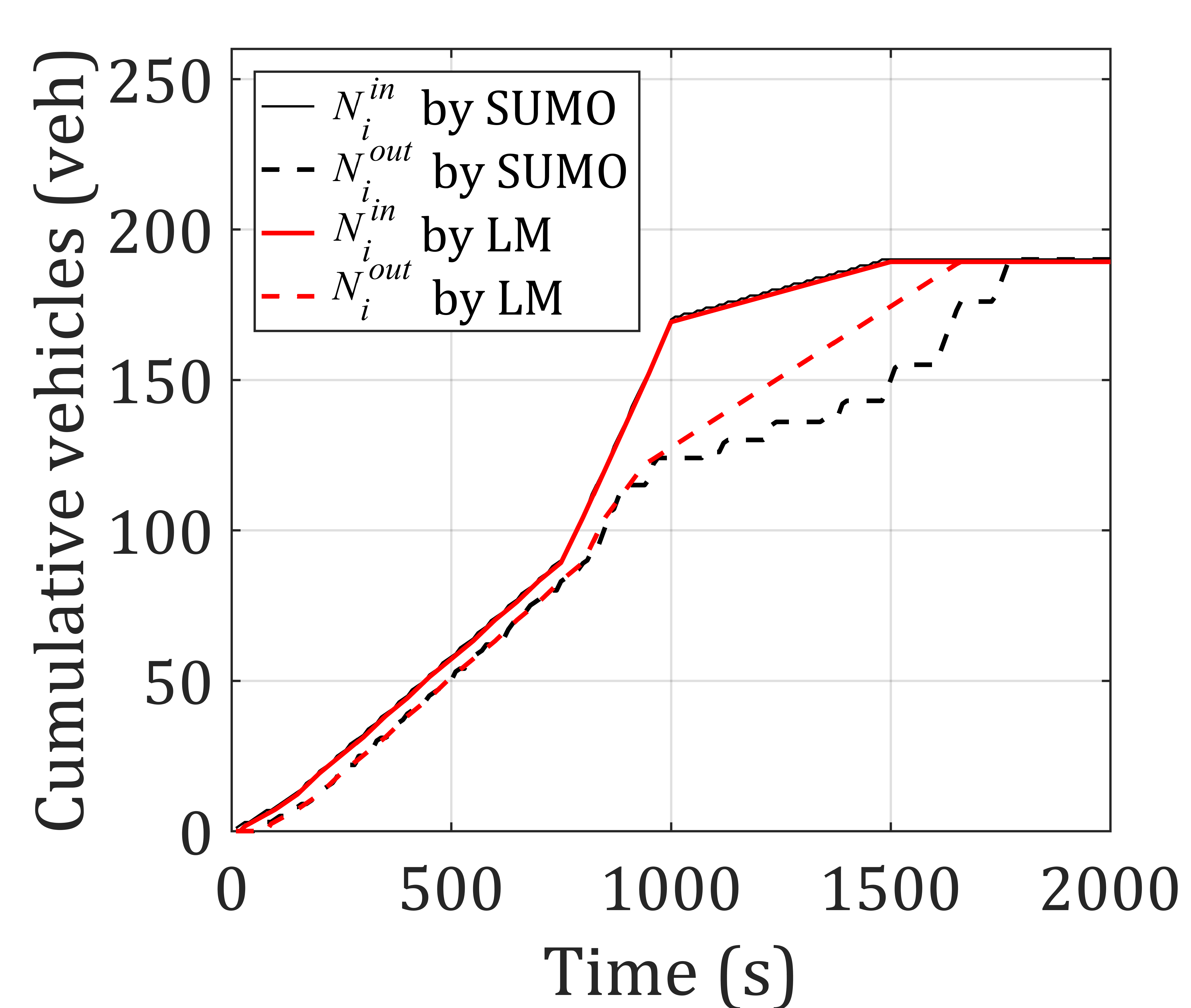}}\subfigure[Cumulative flow of L25 by LQM]{
\includegraphics[width=5cm,height = 4.1cm]{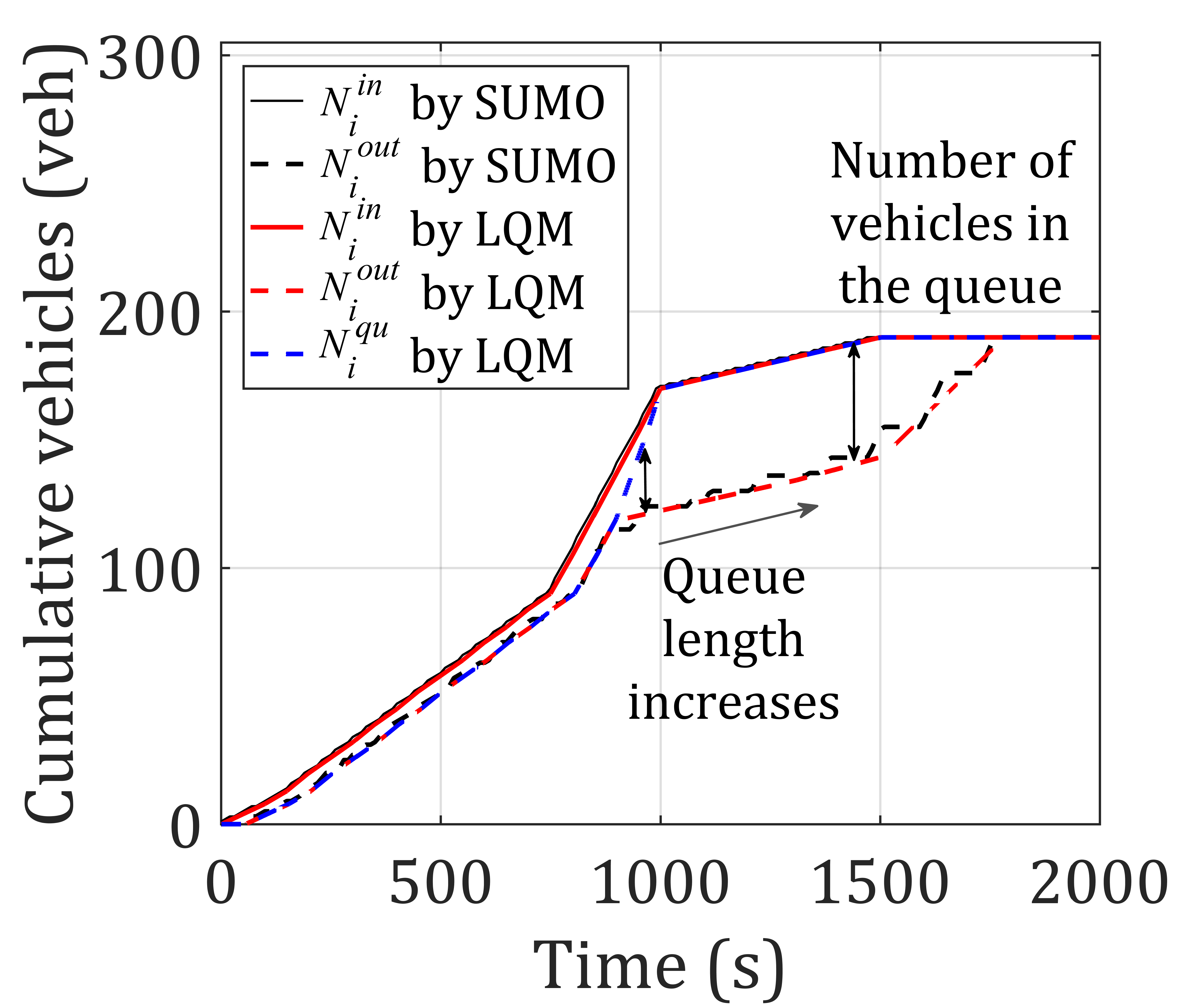}}\subfigure[Queue length of L25 by LQM]{
\includegraphics[width=4.9cm,height = 4.1cm]{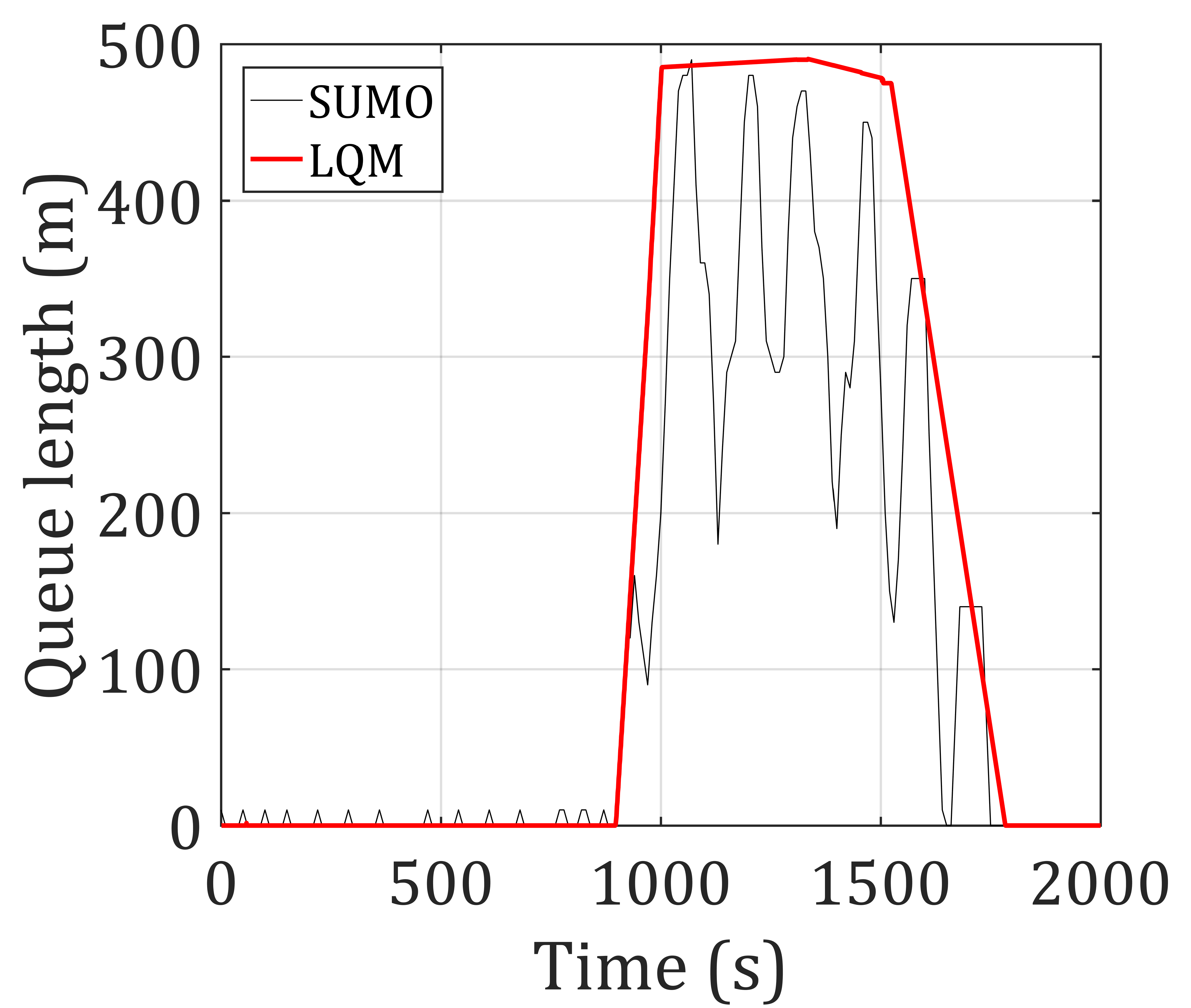}}
\subfigure[Flow rate (m/s) of L2 by LQM]{
\includegraphics[width=6.4cm,height = 4.3cm]{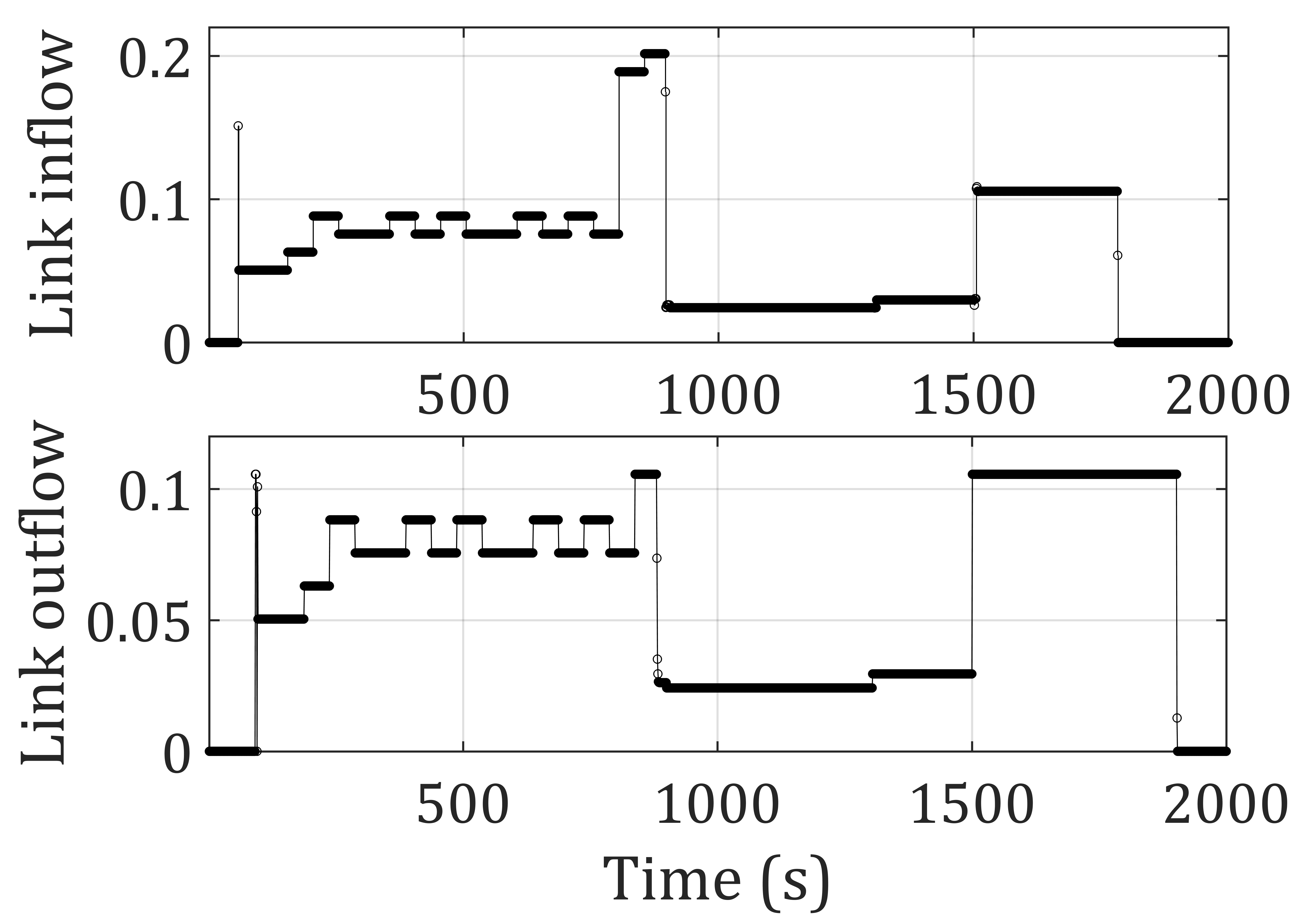}}~~~\subfigure[Flow rate (m/s) of L25 by LQM]{
\includegraphics[width=6.4cm,height = 4.3cm]{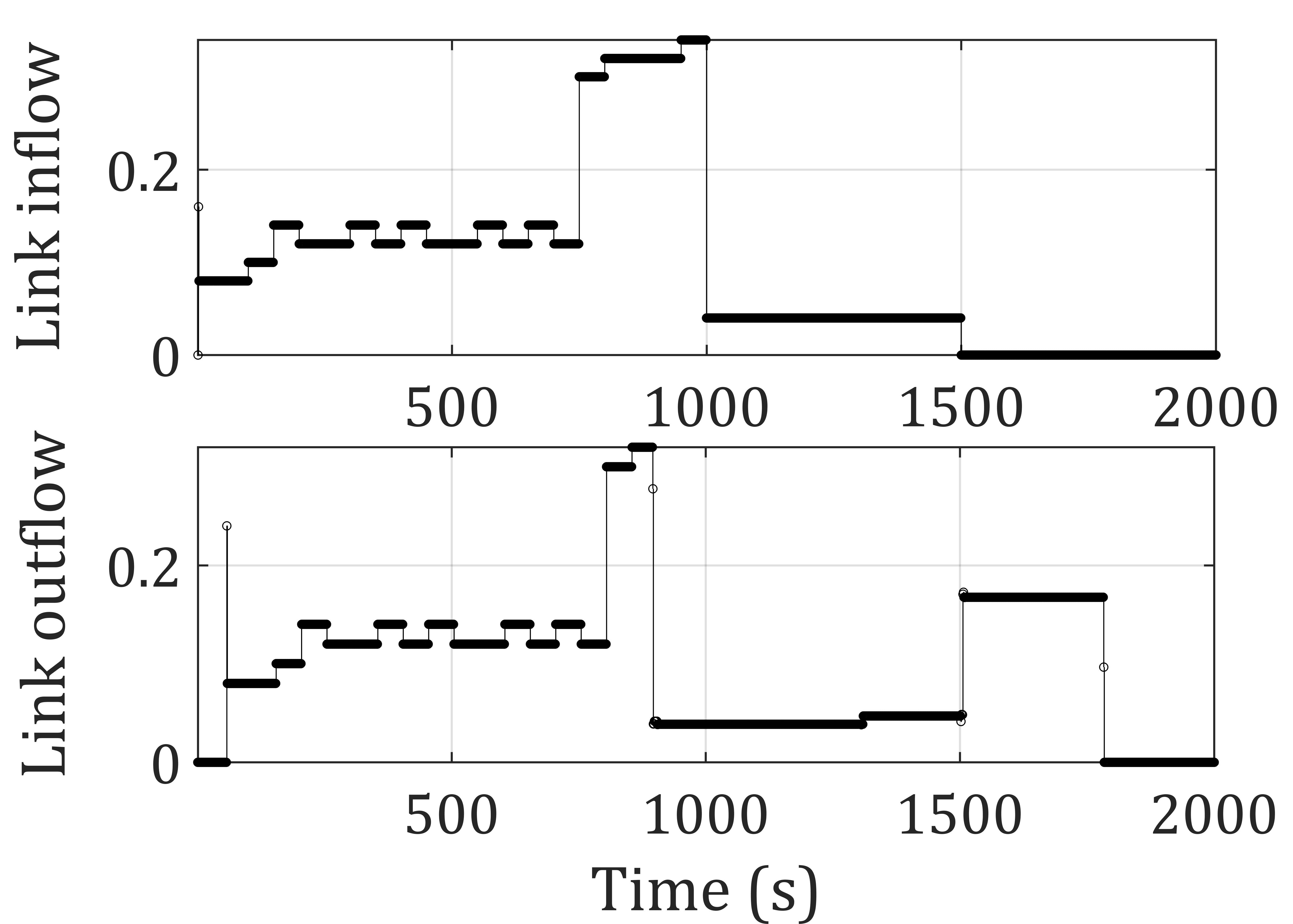}}\\
\caption{{Comparison of the upstream through direction related to bottleneck spillback from LM, LQM and SUMO simulations, where L2 is the turn link 2; L25 is the corresponding upstream common link 25.}}
\label{b t flow}
\end{figure}

{From the final queue length results depicted in Figure \ref{bottleneck flows}(c) and (f), we can find that the spillback initially emerges on the bottleneck-located turn link 20 around 480s, consequently causing the queue length of its upstream common link 31 to increase from 480s. At approximately 900s, the spillback occurs on Link 31. To further analyze the flow propagation, the simulation results of Link 31's upstream through direction are shown in Figure \ref{b t flow}. The results for other incoming directions including left-turn and right-turn are similar to that of Figure \ref{b t flow} and thus supplemented in Appendix C.}
% encompassing through, left-turn, and right-turn incoming directions, are illustrated in Figures \ref{b t flow}, \ref{b l flow}, and \ref{b r flow}, correspondingly, where the common links 25, 26, 28 are also the origin links.

\begin{figure}[H]
\captionsetup{font={small}}
\centering  
\subfigure[West left turn (Link 1)]{
\includegraphics[width=5.6cm,height = 4.8cm]{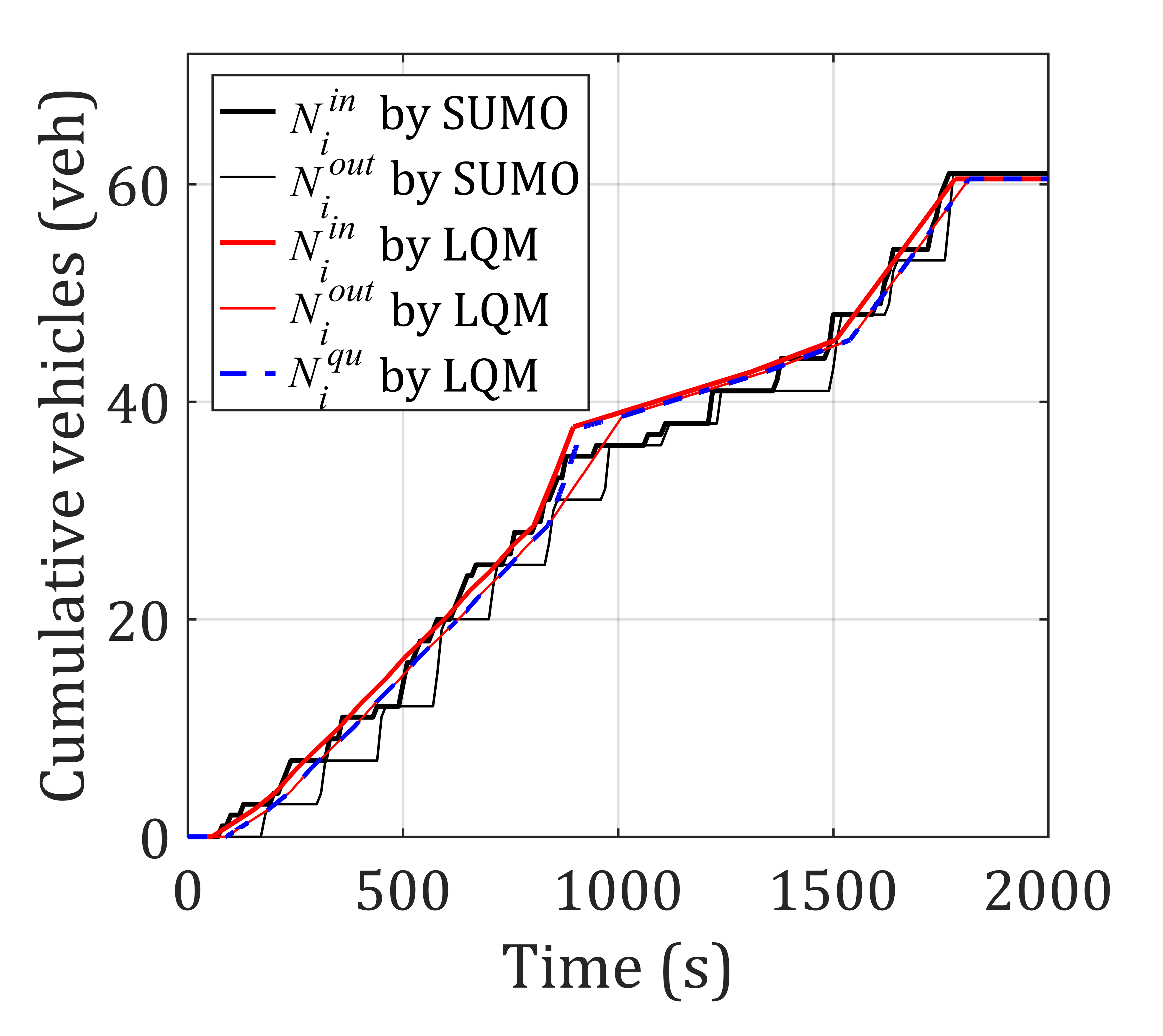}}~~\subfigure[West right turn (Link 3)]{
\includegraphics[width=5.6cm,height = 4.8cm]{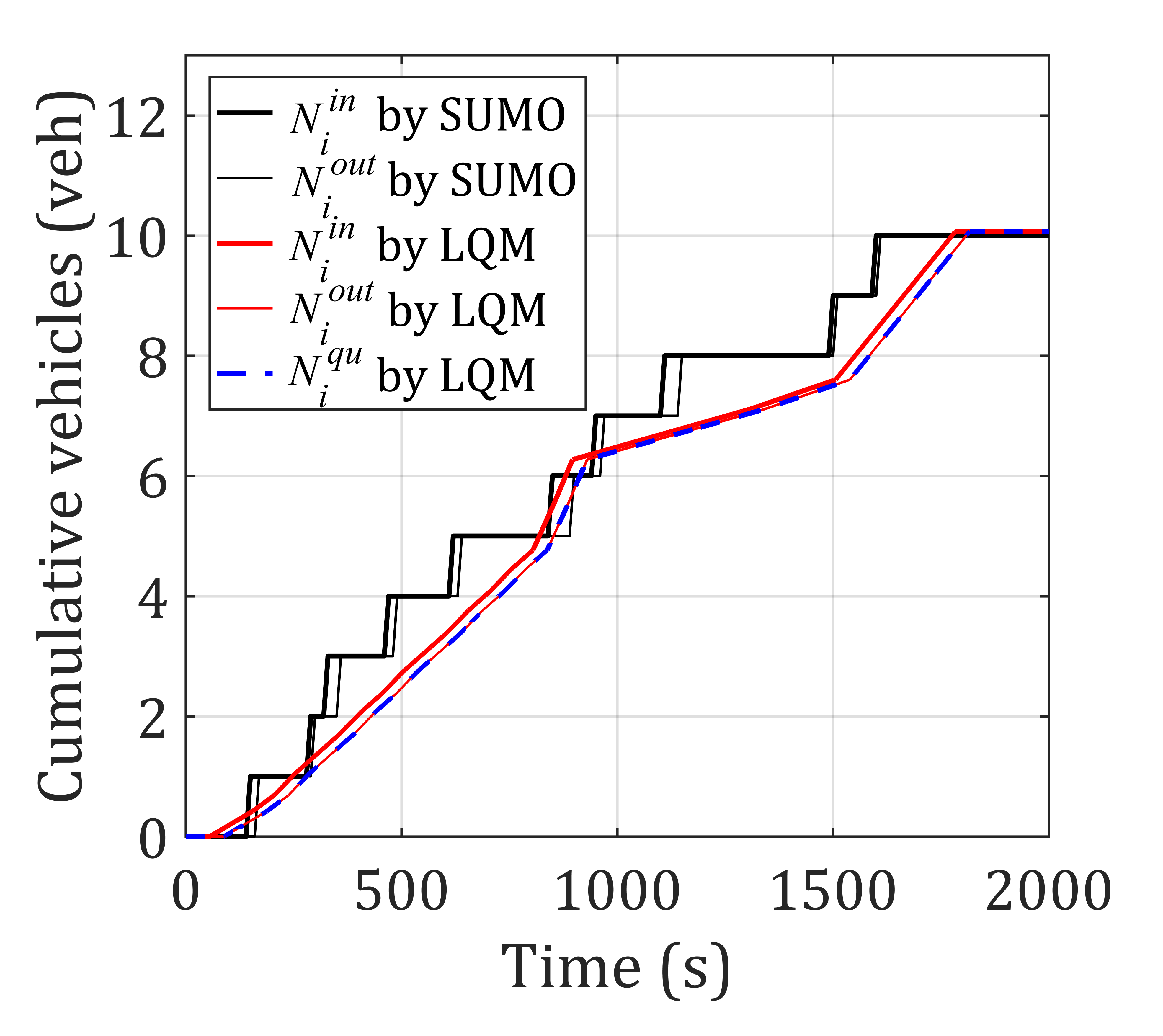}}
% \caption{Comparison results of LQM and SUMO for other upstream turning direction flows related to bottleneck spillback}
% \label{b l flow}
% \end{figure}

% \begin{figure}[H]\ContinuedFloat
% \captionsetup{font={small}}
% \centering  
\subfigure[South left turn (Link 4)]{
\includegraphics[width=5.6cm,height = 4.8cm]{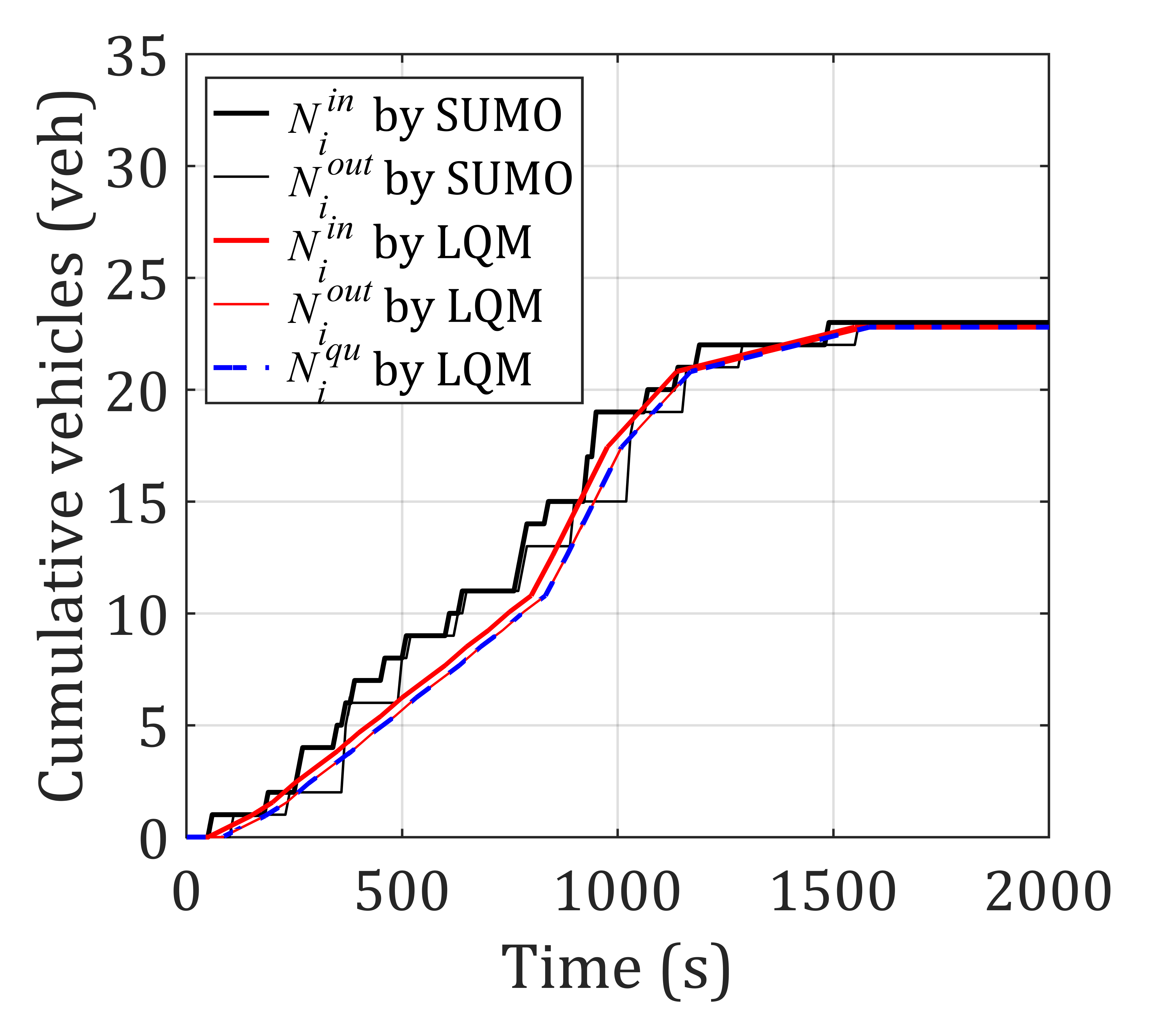}}~\subfigure[South through (Link 5)]{
\includegraphics[width=5.6cm,height = 4.8cm]{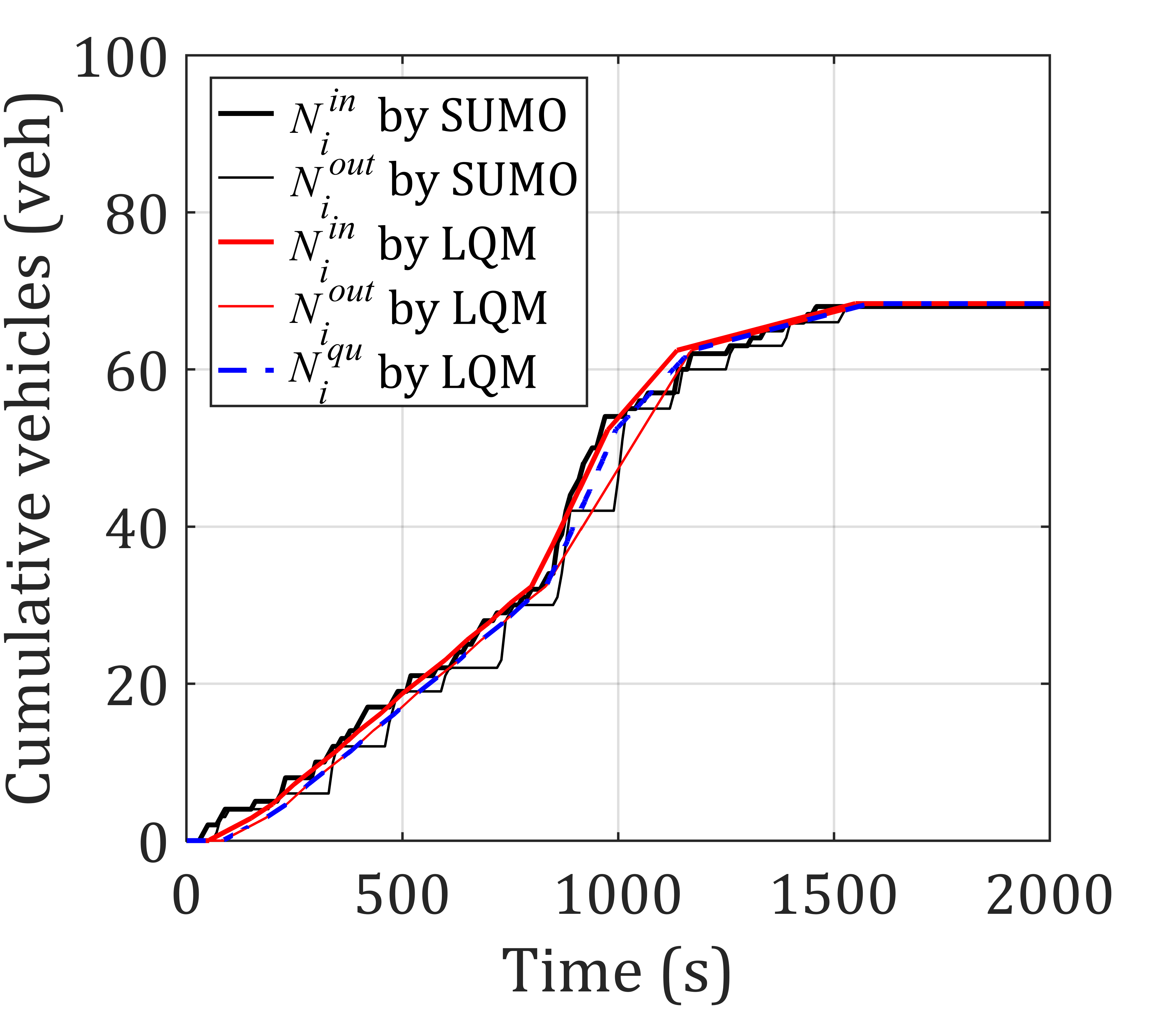}}\\
\subfigure[North through (Link 11)]{
\includegraphics[width=5.6cm,height = 4.8cm]{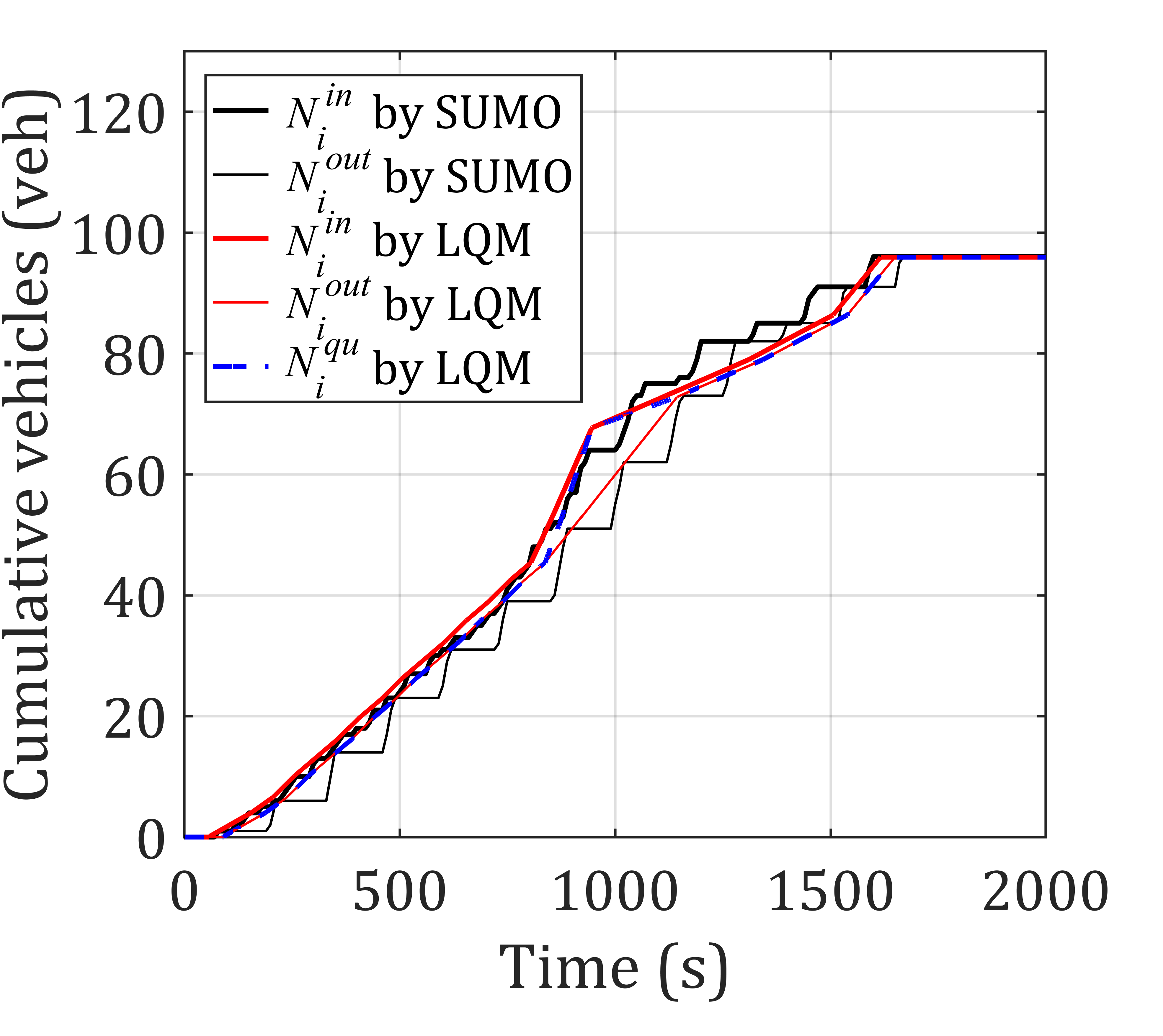}}~\subfigure[North right turn (Link 12)]{
\includegraphics[width=5.6cm,height = 4.8cm]{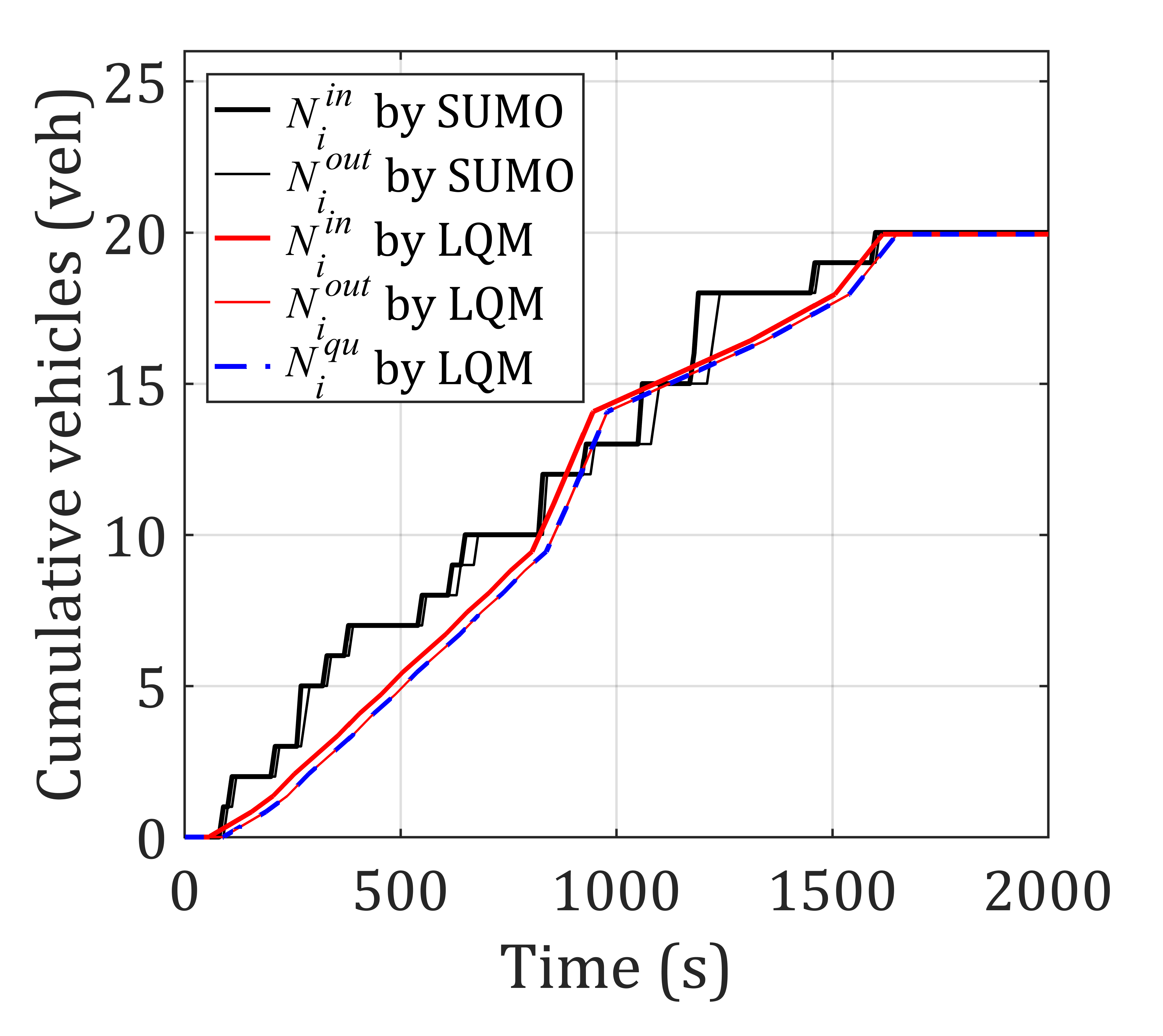}}
\caption{{Comparison results of LQM and SUMO for other upstream turning direction flows related to bottleneck spillback}}
\label{other turns}
\end{figure}

{As depicted in Figure \ref{b t flow}, under free-flow conditions (e.g., 0-800s), the performance of LM and LQM exhibits similarity. However, as the inflow increases and queue lengths emerge, the distinctions in simulation results between LM and LQM become increasingly pronounced. Although the input flows of the origin links under LM and LQM are all the same and consistent with SUMO, e.g., as indicated in Figure \ref{b t flow}(d) and (e), the flow propagation results are quite different. This is mainly caused by the fact that the intersections in LQM and SUMO are both constructed considering the actual turn-level flow operation by involving dedicated turn links, while LM directly assumes a uniform and homogeneous link for different turn movements. Hence, once a turning direction of a link has a downstream supply limit, the outflows of all turning directions of that link may decrease simultaneously, resulting in potential traffic holdback. Figure \ref{other turns} supplements the flow propagation results for other bottleneck-related upstream turning directions of the node's incoming links. The results simulated by LQM present a notable proximity to SUMO, indicating that LQM effectively represents the intersection's turn-level flow dynamics.}

{By analyzing Figures \ref{bottleneck flows} and \ref{b t flow}, it can be observed that the changing trends of cumulative flows and queue length obtained from LQM and SUMO are quite similar despite the differences at each time step. The differences at each time step arises because SUMO operates as a vehicle-based microscopic simulation, whereas LQM is a link-based macroscopic simulation that formulates flow dynamics based on specific links. Consequently, LQM does not precisely capture individual vehicle movements with the same level of detail as SUMO, such as stop-and-go traffic, which may result in some discrepancies in the queue lengths and travel times observed at each time step. To quantify the difference between the LM, LQM and SUMO more specific, an error metric $
\varepsilon\left( {\boldsymbol{x},\boldsymbol{\psi}} \right)$ for a link is defined below:}

{\begin{equation}
\label{error}
    \varepsilon\left( {\boldsymbol{x},\boldsymbol{\psi}} \right) = \sqrt{\frac{1}{T}{\sum_{k = 1}^{T}\left( x_{k} - \psi_{k} \right)^2}}
\end{equation}
where $\boldsymbol{x} = \left\lbrack x_{1},\ldots,x_{T} \right\rbrack$ represents simulation results by LM or LQM over simulation period $T$; $\boldsymbol{\psi} = \left\lbrack \psi_{1},\ldots,\psi_{T} \right\rbrack$ represents simulation results by SUMO over $T$. It can be found that $\varepsilon\left( {\boldsymbol{x},\boldsymbol{\psi}} \right) \ge 0$. $\varepsilon\left( {\boldsymbol{x},\boldsymbol{\psi}} \right) = 0$ indicates that the simulation results by the LQM (LM) and SUMO match perfectly. The greater the $\varepsilon\left( {\boldsymbol{x},\boldsymbol{\psi}} \right)$ is, the greater the simulation differences between LQM (LM) and SUMO. The statistical simulation differences for all links at the intersection as shown in Figure \ref{intersection} measured by $\varepsilon\left( {\boldsymbol{x},\boldsymbol{\psi}} \right) $ are provided in Tables \ref{difference for in} and \ref{difference for out}, and the summarized results are provided in Table \ref{Statistical results}.}
{It can be seen that the proposed LQM and SUMO produced similar simulation results for the incoming and outgoing links, with an average difference of only 2.45 vehicles. This indicates that the simulation difference between LQM and SUMO at each time step can be eliminated over a time span. Additionally, the average difference can be reduced from 6.7 vehicles with LM to 2.45 vehicles with the proposed LQM. Thus,  the proposed LQM outperforms LM in simulating flow propagation at signalized intersections with multiple turn movements.}
% Tables \ref{difference for in}, \ref{difference for out} and \ref{Statistical results} show that LQM presents smaller simulation differences with SUMO in terms of LI and LO when compared to LM. Notably, the range of average differences in queue lengths between LQM and SUMO is substantial, varying from 5.2m to 140.96m. This variation arises because SUMO operates as a vehicle-based microscopic simulation, whereas LTM is a link-based macroscopic simulation that formulates flow dynamics based on specific links. Consequently, LTM does not precisely capture individual vehicle movements with the same level of detail as SUMO, such as stop-and-go traffic. These differences may lead to disparities in flow propagation and queue lengths at each time step as shown in Figure \ref{bottleneck flows}, resulting in potentially large average difference. Nevertheless, the changing trends and maximum values remain quite similar, as presented and discussed earlier. 

\begin{table}[H]
\captionsetup{font={small}}
\footnotesize
\renewcommand\arraystretch{1}
\caption{{Simulation difference $\varepsilon\left( {\boldsymbol{x},\boldsymbol{\psi}} \right) $ between LM and LQM compared to SUMO for incoming links\label{difference for in}}}
%\begin{tabular}{cccp{10cm}}
\centering
\begin{tabular}{ccccccccccc}
\hline
\multirow{2}{*}{West}&\multicolumn{2}{c}{Link 25}&\multicolumn{2}{c}{Link 1}&\multicolumn{2}{c}{Link 2}&\multicolumn{2}{c}{Link 3}&\multicolumn{2}{c}{Average}\\
&LM&LQM&LM&LQM&LM&LQM&LM&LQM&LM&LQM\\ \hline
$N_i^{in}$&1.42&1.4&5.91&1.62&9.3&1.61&2.1&0.77&4.68&1.35\\
$N_i^{out}$ &9.62&2.57&6.42&2.31&10.59&2.77&1.53&0.82&7.04&2.12\\
% QL (m)&-&103.26&-&10.59&-&33.3&- &6.68&-&38.46\\
\hline
\multirow{2}{*}{South}&\multicolumn{2}{c}{Link 26}&\multicolumn{2}{c}{Link 4}&\multicolumn{2}{c}{Link 5}&\multicolumn{2}{c}{Link 6}&\multicolumn{2}{c}{Average}\\
&LM&LQM&LM&LQM&LM&LQM&LM&LQM&LM&LQM\\ \hline
$N_i^{in}$&1.46&1.44&6.13&1.16&7.2&1.26&10.38&1.78&6.29&1.41\\
$N_i^{out}$&15.2&2.74&9.61&1.14&6.79&1.91&11.18&2.6&10.7&2.1\\
% QL (m)&-&17.9&-&5.2&-&23.02&- &19.75&-&16.47\\
\hline
\multirow{2}{*}{East}&\multicolumn{2}{c}{Link 27}&\multicolumn{2}{c}{Link 7}&\multicolumn{2}{c}{Link 8}&\multicolumn{2}{c}{Link 9}&\multicolumn{2}{c}{Average}\\
&LM&LQM&LM&LQM&LM&LQM&LM&LQM&LM&LQM\\ \hline
$N_i^{in}$&1.45&1.44&5.39&2.21&6.37&3.86&2.37&1.04&3.9&2.14\\
$N_i^{out}$&6.31&4.91&3.11&1.62&8.13&4.36&2.86&1.11&5.1&3\\
% QL (m)&-&20.41&-&15.8&-&10.44&- &18.72&-&16.34\\
\hline
\multirow{2}{*}{North}&\multicolumn{2}{c}{Link 28}&\multicolumn{2}{c}{Link 10}&\multicolumn{2}{c}{Link 11}&\multicolumn{2}{c}{Link 12}&\multicolumn{2}{c}{Average}\\
&LM&LQM&LM&LQM&LM&LQM&LM&LQM&LM&LQM\\ \hline
$N_i^{in}$&1.48&1.42&2.31&1.17&4.26&2.43&4.12&1.36&3.04&1.6\\
$N_i^{out}$&5.36&3.72&5.67&1.72&8.72&3.29&3.16&1.48&5.73&2.56\\
% QL (m)&-&69.39&-&26.16&-&30.21&- &5.43&-&32.8\\
\hline
\end{tabular}
\end{table}

\begin{table}[H]
\centering
\captionsetup{font={small}}
\footnotesize
\renewcommand\arraystretch{1}
\caption{{Simulation difference $\varepsilon\left( {\boldsymbol{x},\boldsymbol{\psi}} \right) $ between LM and LQM compared to SUMO for outgoing links\label{difference for out}}}
%\begin{tabular}{cccp{10cm}}
\begin{tabular}{ccccccccccc}
\hline
\multirow{2}{*}{West}&\multicolumn{2}{c}{Link 29}&\multicolumn{2}{c}{Link 13}&\multicolumn{2}{c}{Link 14}&\multicolumn{2}{c}{Link 15}&\multicolumn{2}{c}{Average}\\
&LM&LQM&LM&LQM&LM&LQM&LM&LQM&LM&LQM\\ \hline
$N_i^{in}$&7.63&4.59&5.31&2.27&3.03&1.91&4.95&2.61&5.23&2.85\\
$N_i^{out}$&9.61&4.72&5.92&2.18&4.98&1.77&3.1&2.51&5.9&2.8\\
% QL (m)&-&15.01&-&13.44&-&15.61&-&12.8&-&15.52\\
\hline
\multirow{2}{*}{South}&\multicolumn{2}{c}{Link 30}&\multicolumn{2}{c}{Link 16}&\multicolumn{2}{c}{Link 17}&\multicolumn{2}{c}{Link 18}&\multicolumn{2}{c}{Average}\\
&LM&LQM&LM&LQM&LM&LQM&LM&LQM&LM&LQM\\ \hline
$N_i^{in}$&9.62&3.14&9.1&1&9.72&2.7&8.93&1.55&9.34&2.1\\
$N_i^{out}$&10.75&3.28&11.42&0.9&10.63&2.3&7.5&1.69&10.08&2.05\\
% QL (m)&-&18.32&-&12.71&-&15.12&-&10.89&-&14.26\\
\hline
\multirow{2}{*}{East}&\multicolumn{2}{c}{Link 31}&\multicolumn{2}{c}{Link 19}&\multicolumn{2}{c}{Link 20}&\multicolumn{2}{c}{Link 21}&\multicolumn{2}{c}{Average}\\
&LM&LQM&LM&LQM&LM&LQM&LM&LQM&LM&LQM\\ \hline
$N_i^{in}$&4.09&3.26&4.9&1.33&5&1.6&12.03&6.19&6.51&3.01\\
$N_i^{out}$&9.23&6.54&8.44&1.27&2.3&1.36&14.82 &6.11&8.7&3.82\\
% QL (m)&-&140.96&-&20.51&-&42.45&-&16.21&-&55\\
\hline
\multirow{2}{*}{North}&\multicolumn{2}{c}{Link 32}&\multicolumn{2}{c}{Link 22}&\multicolumn{2}{c}{Link 23}&\multicolumn{2}{c}{Link 24}&\multicolumn{2}{c}{Average}\\
&LM&LQM&LM&LQM&LM&LQM&LM&LQM&LM&LQM\\ \hline
$N_i^{in}$&6.11&3.14&5.06&2.74&9.21&3.63& 5.92&3.23&6.58&3.18\\
$N_i^{out}$&8.62&3.34&8.14&2.89&10.08&3.13&6.3 &2.93&8.23&3.07\\
% QL (m)&-&18.12&-&14.57&-&16.51&-&16.02&-&16.31\\
\hline
\end{tabular}
\end{table}

\begin{table}[H]
\captionsetup{font={small}}
\footnotesize
\renewcommand\arraystretch{1}
    \centering
     \caption{{Summarized results of the simulation differences between LM and LQM}}
    \begin{tabular}{ccc}
    \hline
         Parameter&  LM& LQM\\
         \hline
         $N_i^{in}$&  5.7& 2.21\\
         $N_i^{out}$&  7.69& 2.69\\
         Avarage&  6.7& 2.45\\
         \hline
    \end{tabular}
    \label{Statistical results}
\end{table}

\subsubsection{Model performance under time-varying free-flow speed}
\begin{figure}[H]
\captionsetup{font={small}}
\centering
\includegraphics[width=2.75in]{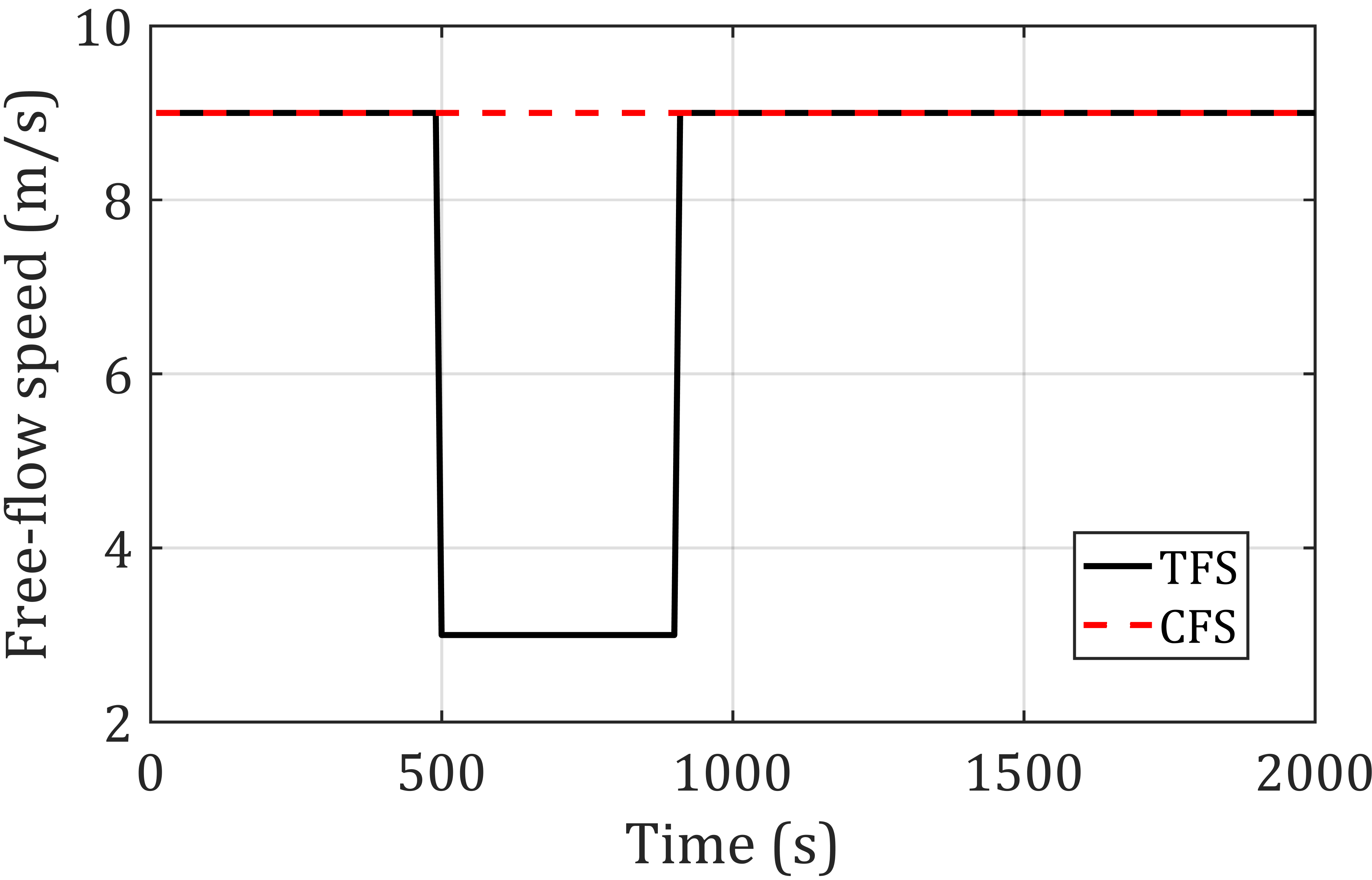}
\caption{{Free-flow speed of Link 25 (TFS: time-varying free-flow speed; CFS: constant free-flow speed) }}
\label{Link25 speed}
\end{figure}

\begin{figure}[H]
\captionsetup{font={small}}
\centering  
\subfigure[Cumulative flow with CFS]{
\includegraphics[width=5.7cm,height = 4.9cm]{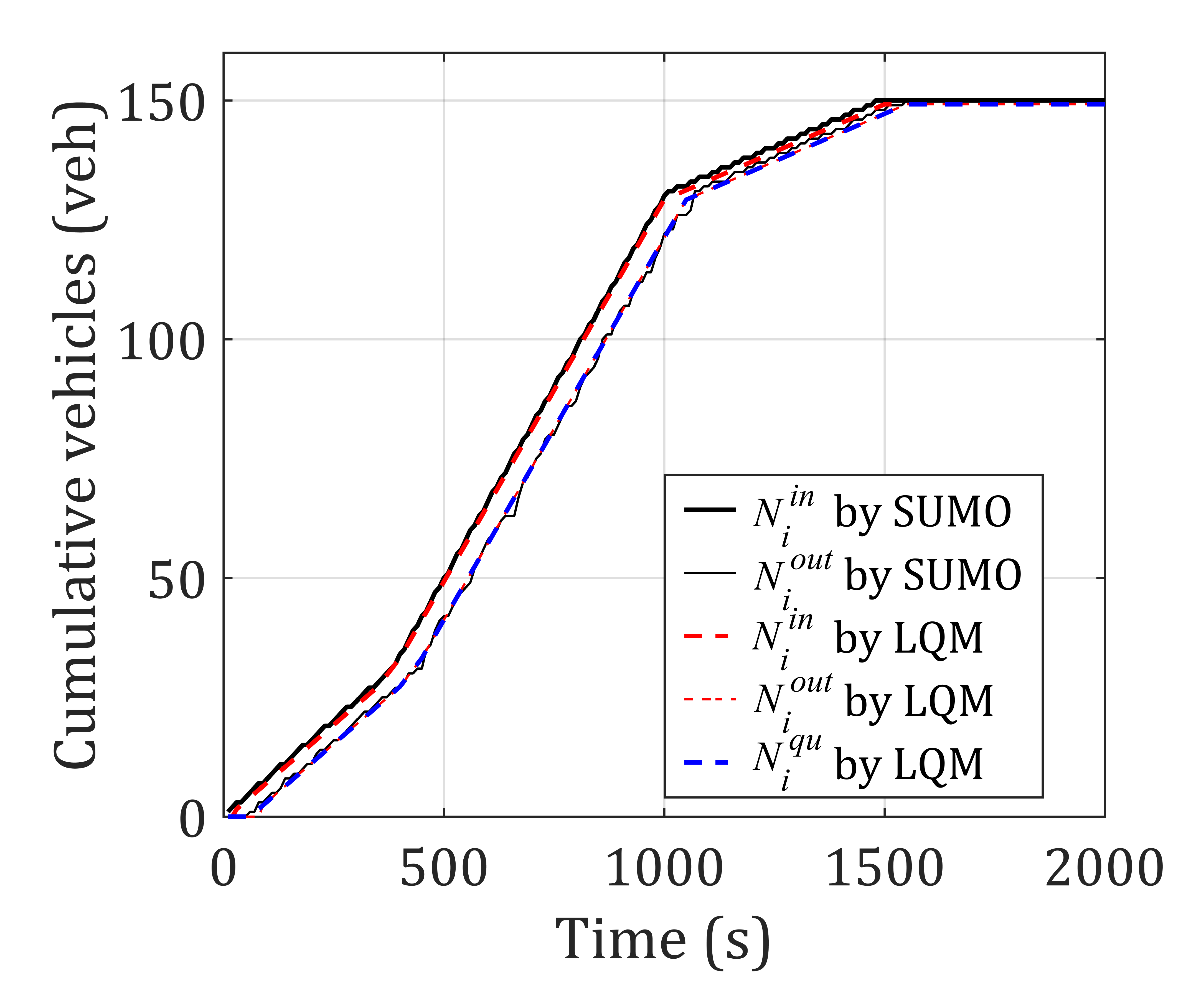}}~~\subfigure[Cumulative flow with TFS]{
\includegraphics[width=5.7cm,height = 4.9cm]{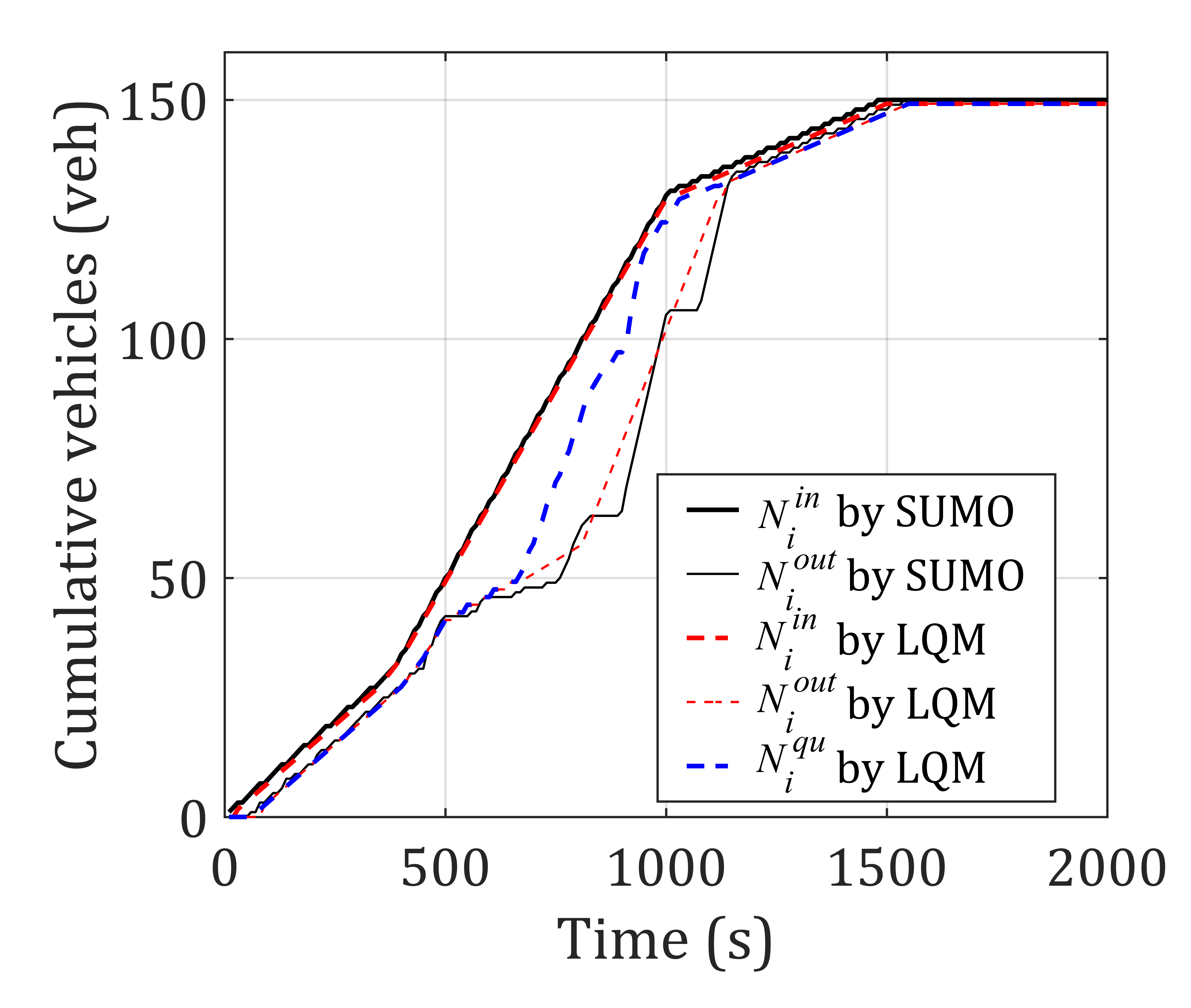}}
\subfigure[Link outflow rate by LQM]{
\includegraphics[width=5.7cm,height = 4.9cm]{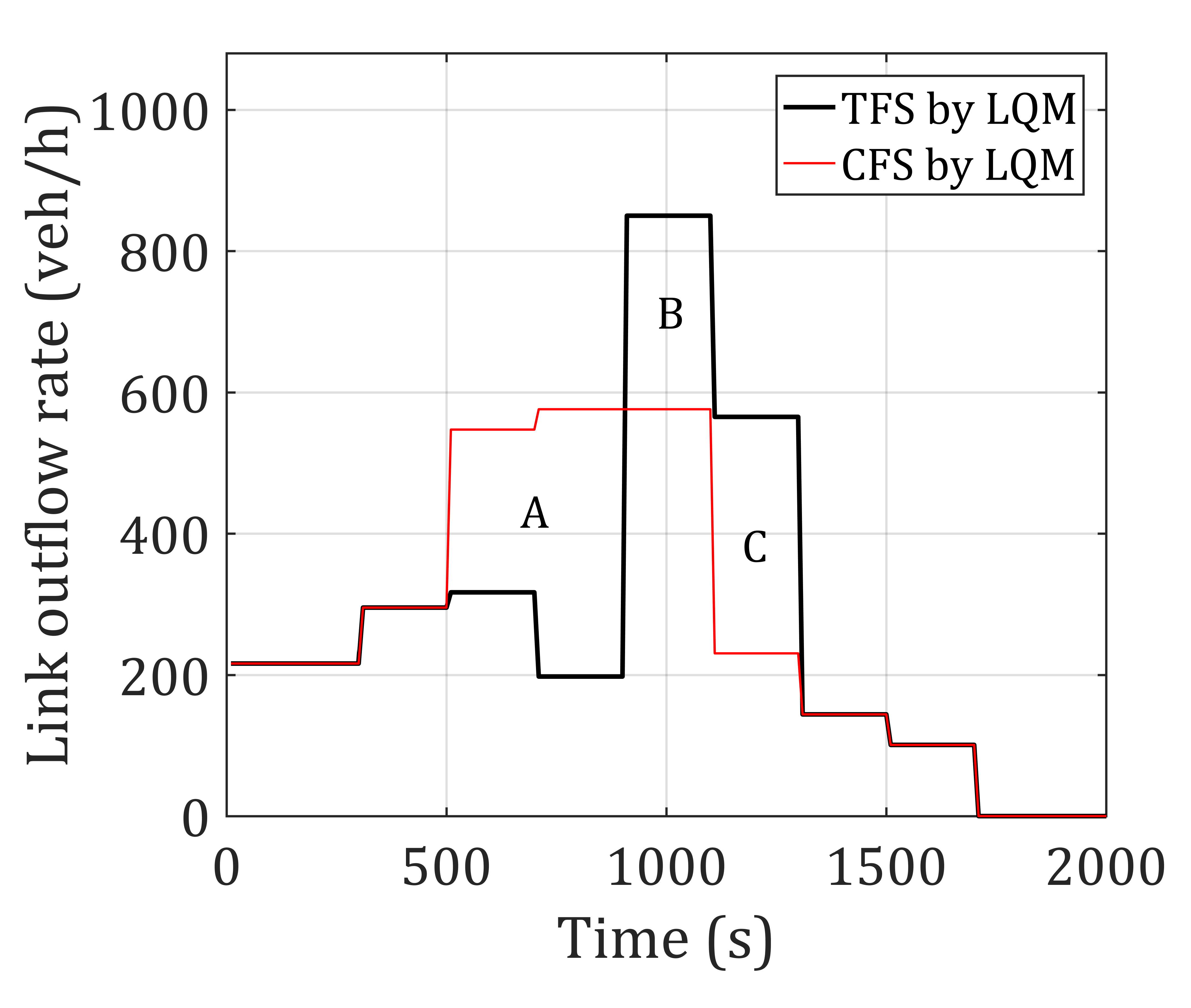}}~~\subfigure[Link outflow rate by SUMO]{
\includegraphics[width=5.7cm,height = 4.9cm]{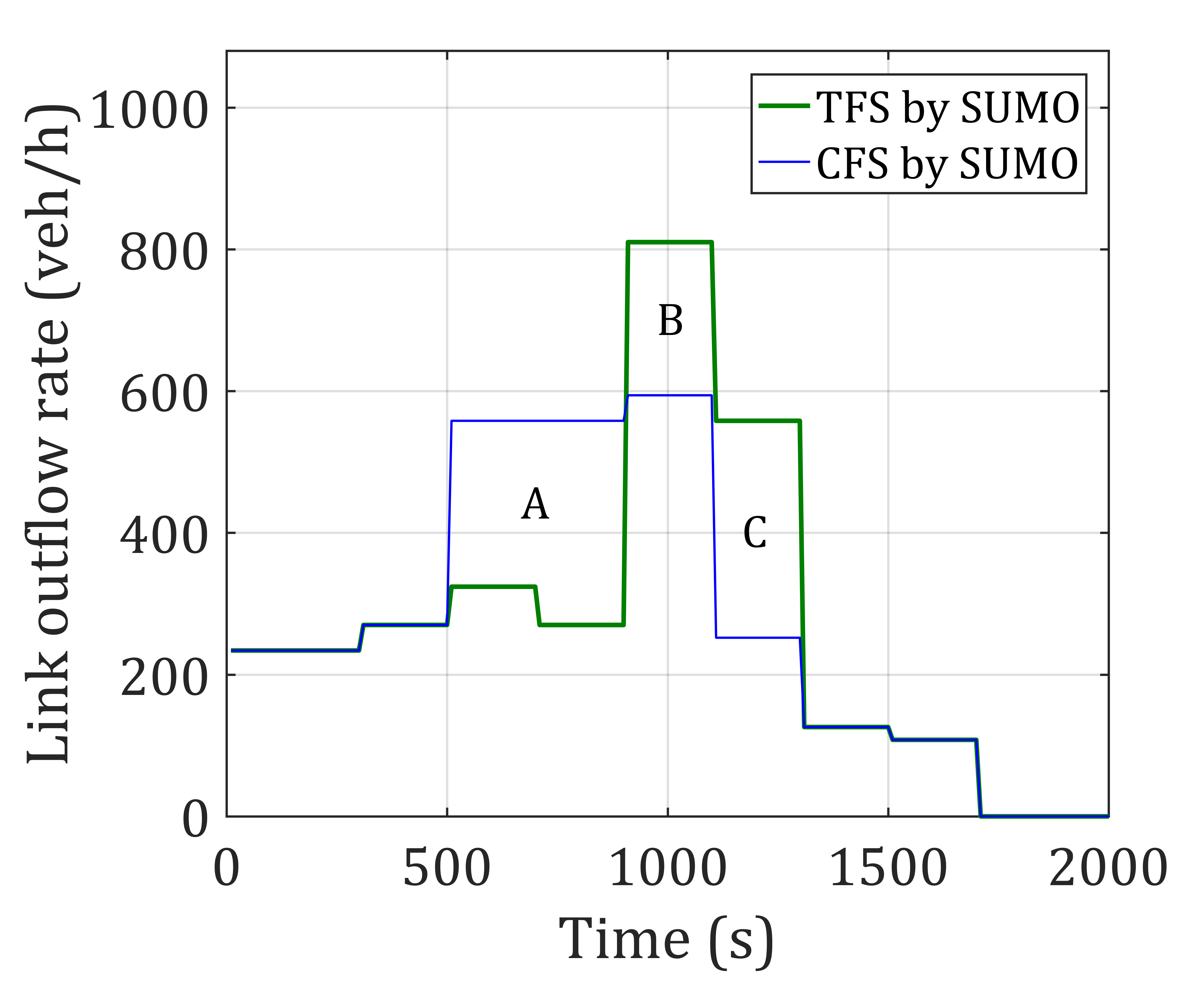}}
\caption{{Comparison of flow propagation of Link 25 between LQM and SUMO under TFS and CFS}}
\label{Link25 flow}
\end{figure}

{The aforementioned experiments verified the impacts of involving queue transmission on the typical link-based model. To further verify the performance of the proposed LQM under TFS, we then increase the green time fraction of the bottleneck Link 20 to 0.38, which is consistent with other west through directions as shown in Table \ref{Road parameters}. Consequently, the fixed bottleneck at Link 20 was eliminated. Subsequently, we assumed that there was a moving bottleneck, e.g., a leading CAV impacting platoon operation, that appeared on Link 25 in Figure \ref{intersection} from 500s to 900s, which caused the free-flow speed of Link 25 to reduce in this period as shown in Figure \ref{Link25 speed}. The cumulative flows and their corresponding outflow rates of Link 25 under such TFS are provided in Figure \ref{Link25 flow}. Due to the 10s intervals in data collection using SUMO, the flow rates present great oscillations in SUMO. Hence, the outflow rates in Figure \ref{Link25 flow} are averaged every 10 values to provide a clearer representation. Note that since the LM assumed the free-flow speed to be a constant value, which is less capable of simulating TFS, thus only the simulation results of LQM and SUMO are compared. It can be found that the proposed LQM can still match the simulation results by SUMO under TFS. A noticeable reduction in the link outflow rate is observed when the free-flow speed decreases to 3m/s at 500 seconds. Subsequently, at 900 seconds, as the free-flow speed recovers to 9m/s, the link outflow rate experiences a significant increase due to the accumulation of vehicles on the link during the 500s to 900s period. Adhering to flow conservation principles, the area of region A equals the sum of areas of regions B and C. }

{It should be noted that the above simulation was carried out by MATLAB on a laptop with an Intel i7-8550U CPU 1.80 GHz and 16GB RAM. The computational time for simulating 2000s is about 16s, which equals an average of 0.008s per step, obviously smaller than the sampling interval of 10s. Therefore, despite incorporating turn-level queue transmission and TFS, the proposed LQM is still an efficient simulation tool.}
\subsection{ Scenario 2: Simulation experiments in a real-world network}
\begin{figure}[H]
\captionsetup{font={small}}
\centering  
\subfigure[Overview of the network]{
\includegraphics[width=2.3in]{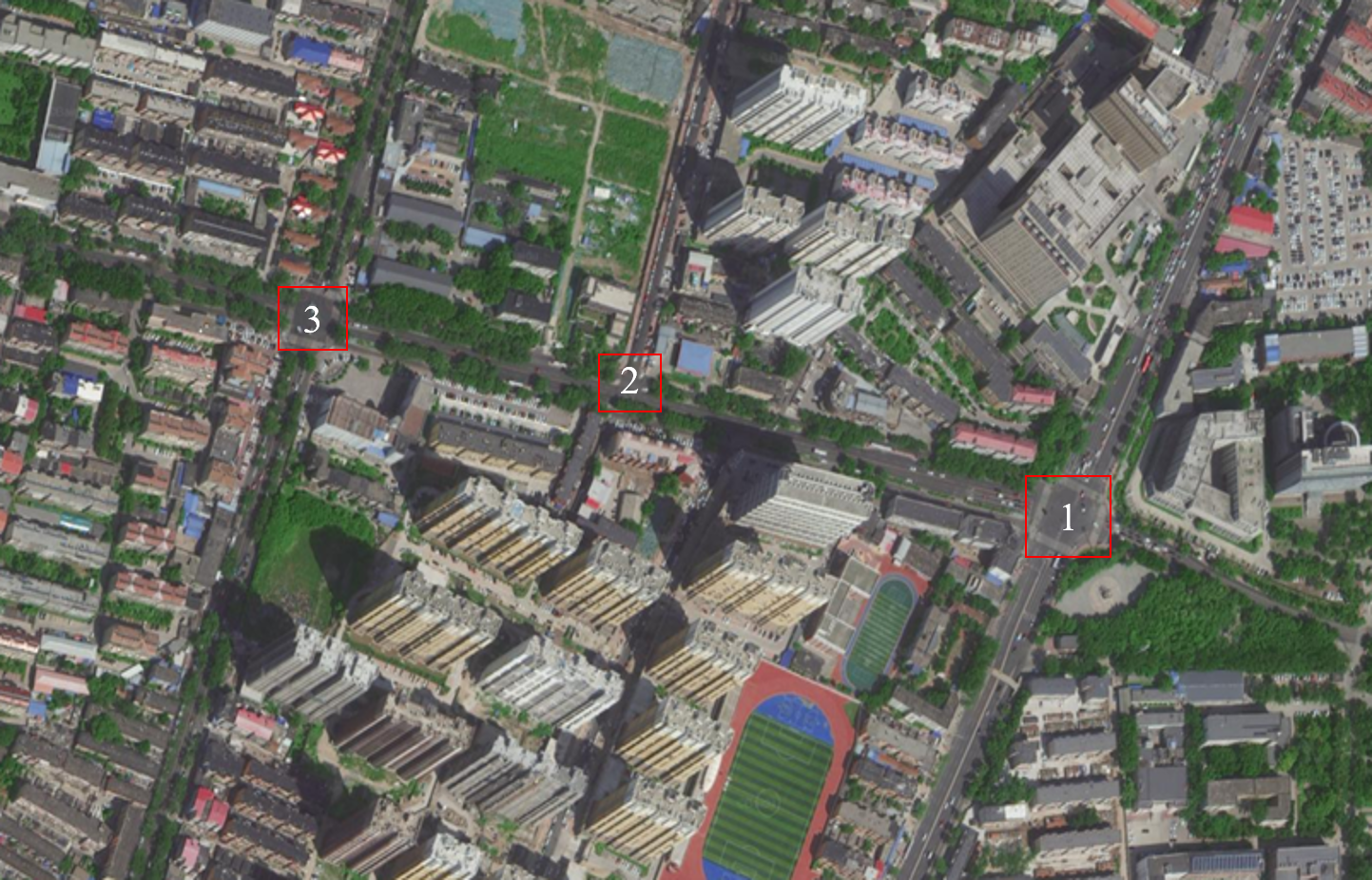}}~~~\subfigure[Simulation model in SUMO]{
\includegraphics[width=2.3in]{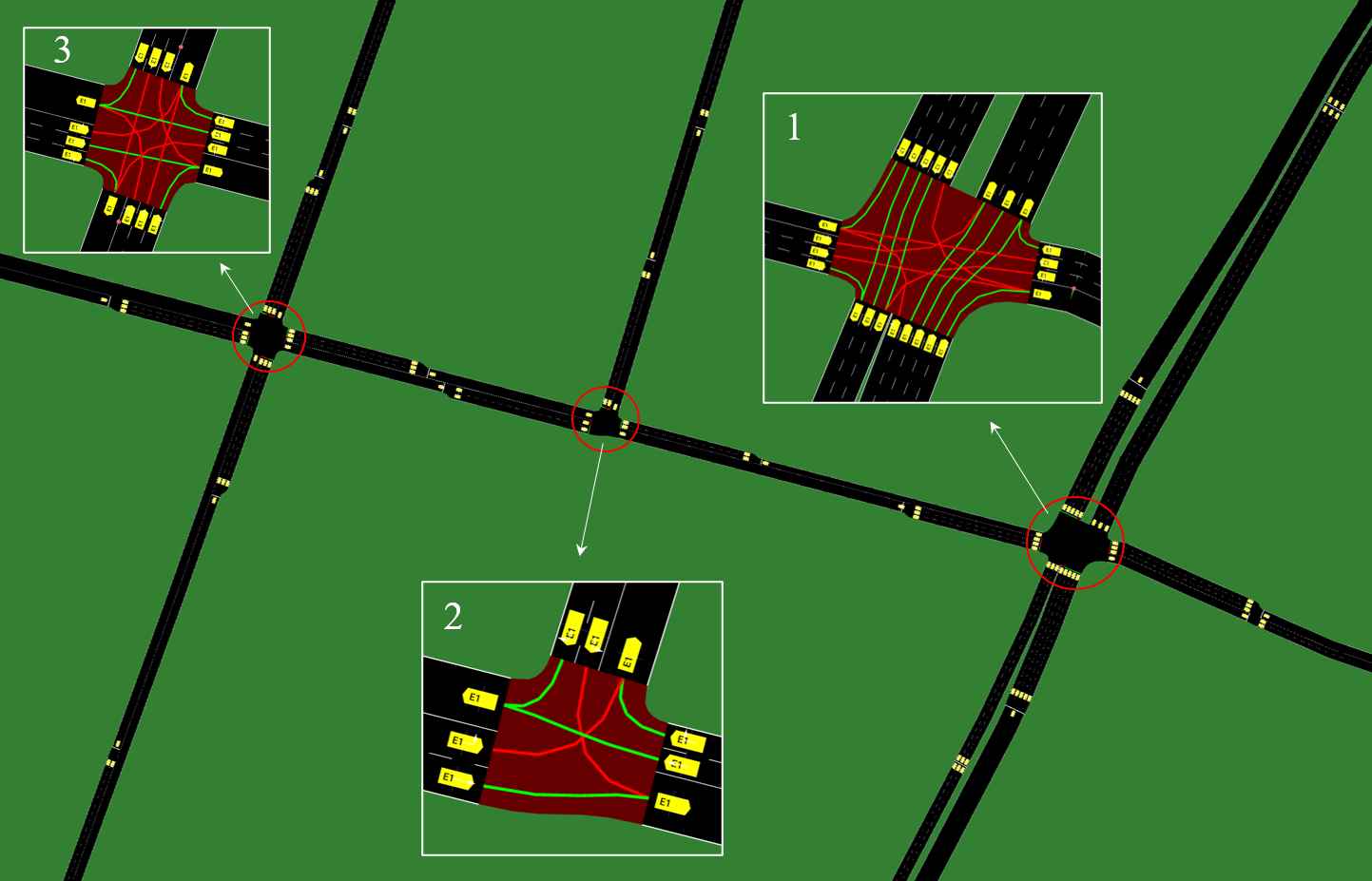}}
\caption{{Layout of the target network}}
\label{net}
\end{figure}

\begin{figure}[H]
\captionsetup{font={small}}
\centering
\includegraphics[width=4.8in]{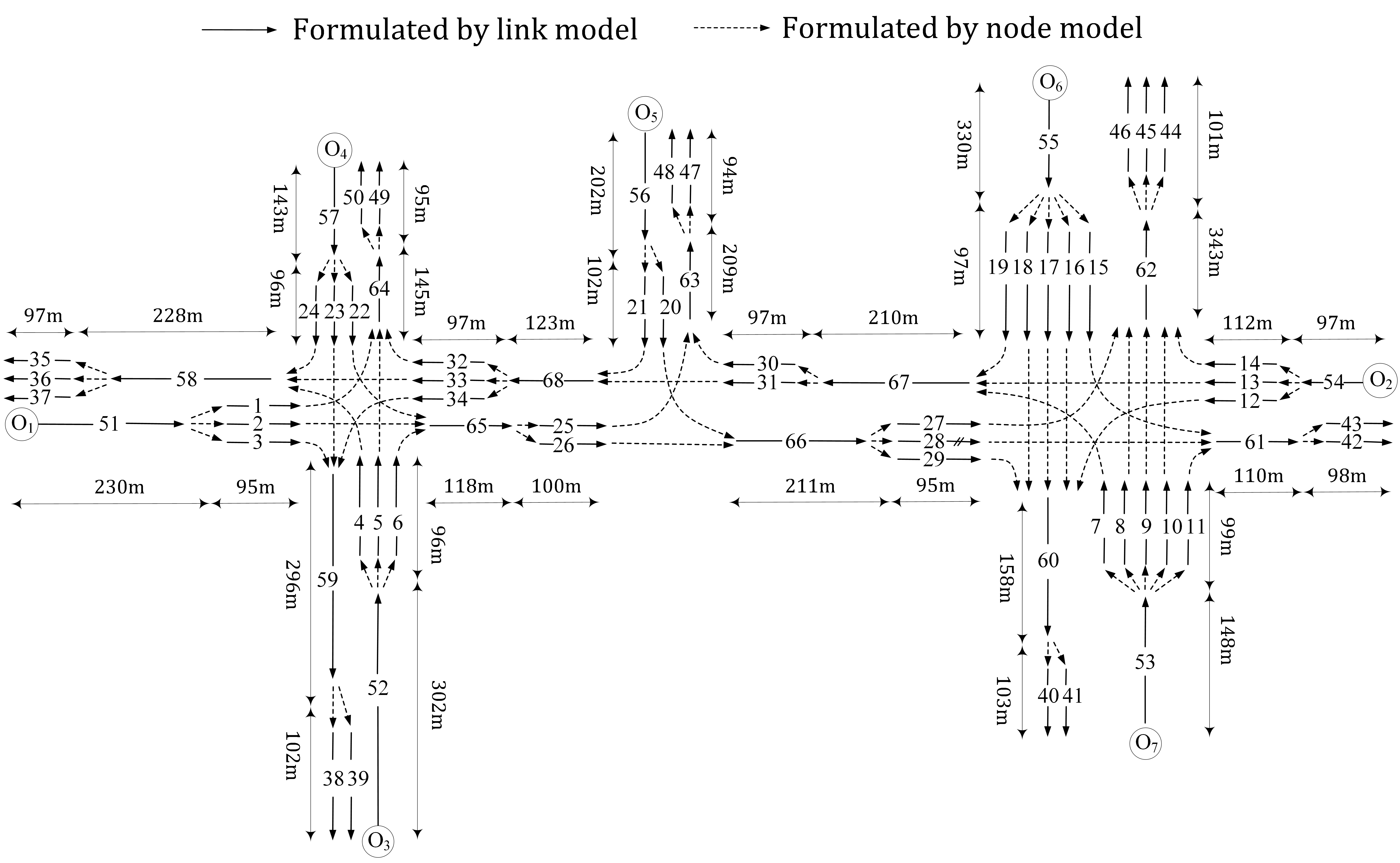}
\caption{{Lane configuration of the network, where $\mathrm O_1$ to $\mathrm O_7$  are the input demands. }}
\label{net LTM}
\end{figure}

\begin{figure}[H]
\captionsetup{font={small}}
\centering  
\includegraphics[width=4.8in]{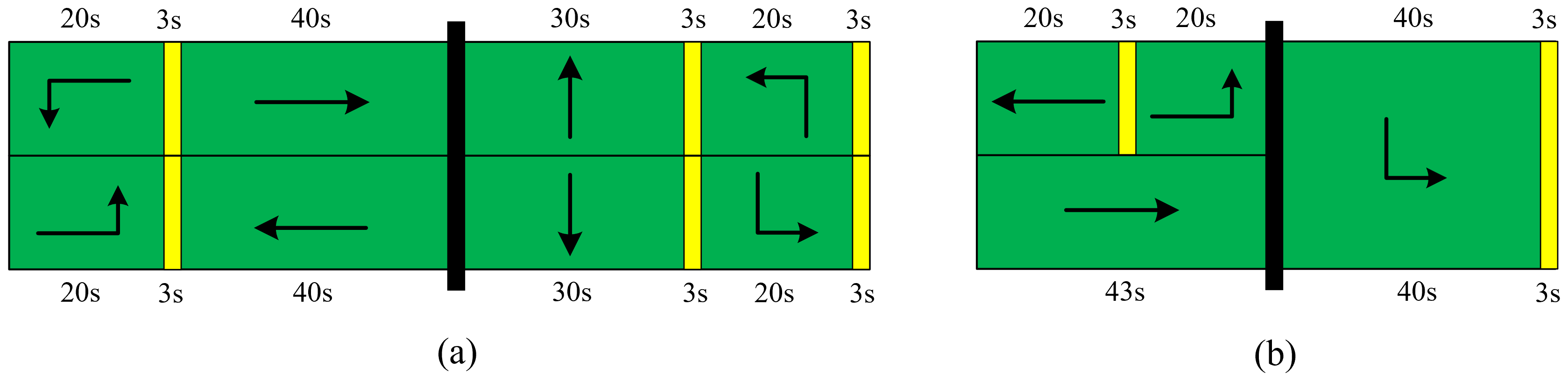}
\caption{{Signal timing plans of the network: (a) Intersections 1 and 3; (b) Intersection 2.}}
\label{net signal}
\end{figure}

\begin{figure}[H]
\captionsetup{font={small}}
\centering
\includegraphics[width=2.7in]{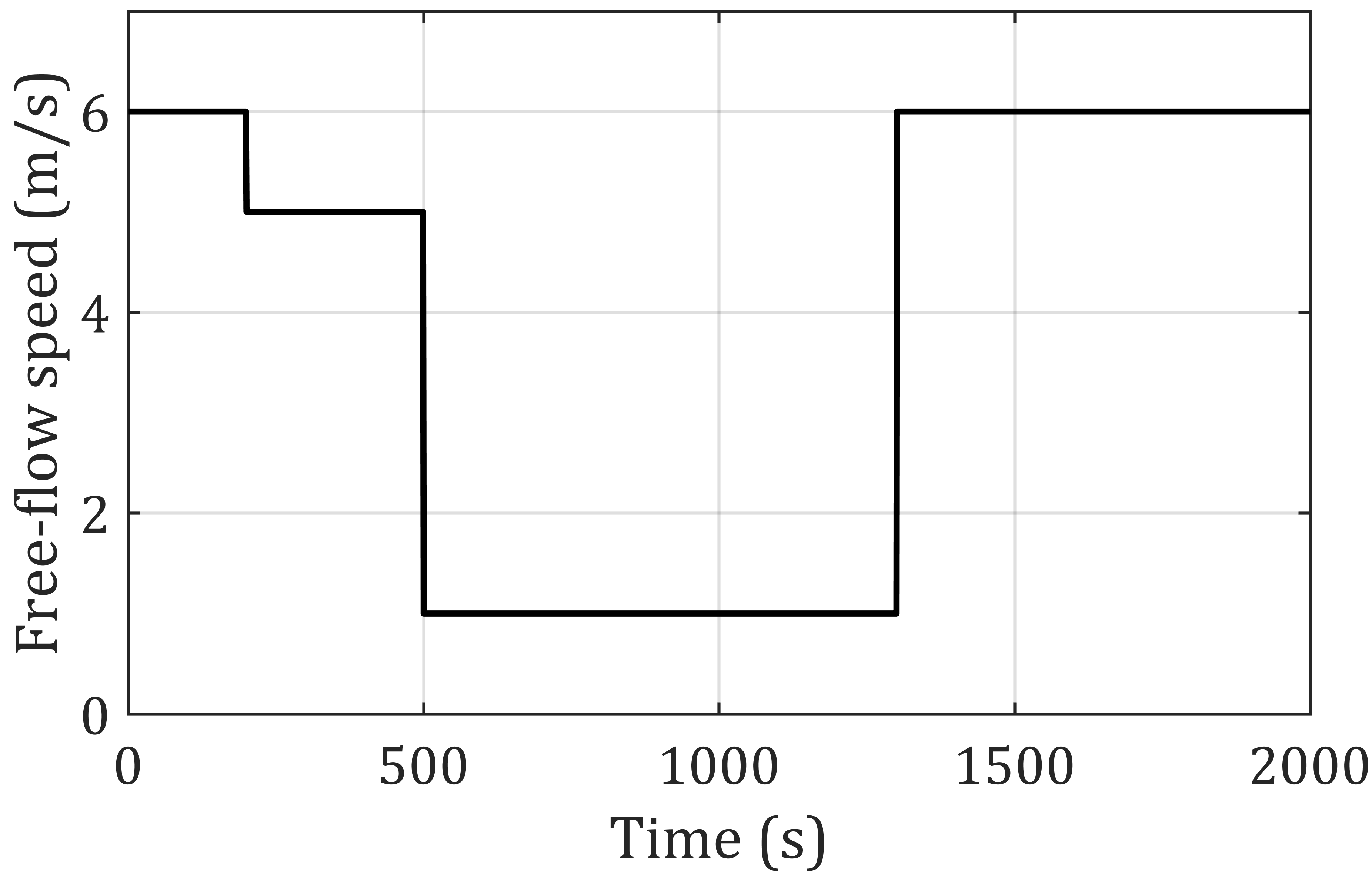}
\caption{{Time-varying free-flow speed of Link 28 in the network}}
\label{net fs}
\end{figure}

\begin{figure}[H]
\captionsetup{font={small}}
\centering  
\subfigure[West and east]{
\includegraphics[width=5.8cm,height = 5cm]{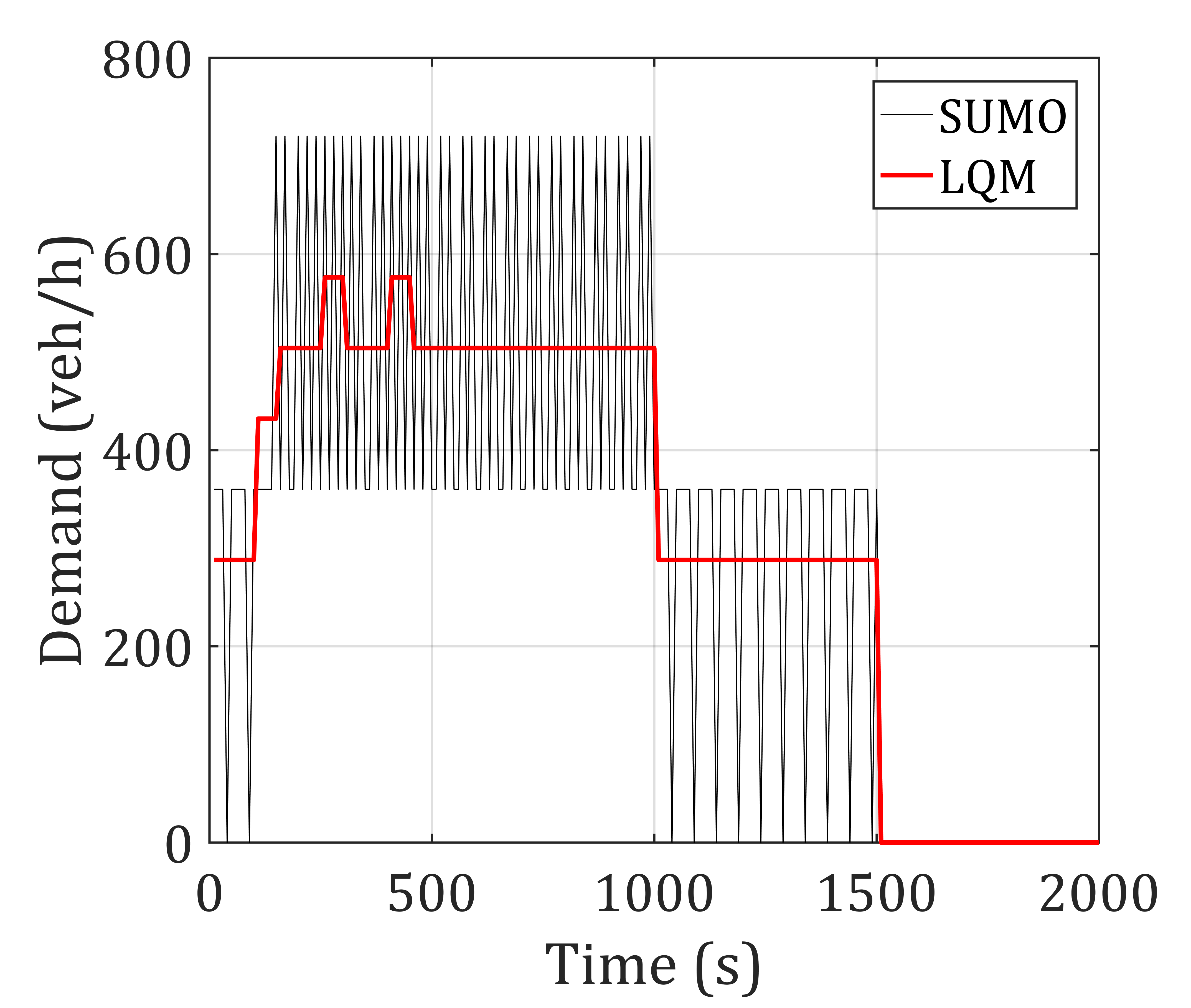}}~~\subfigure[North and south]{
\includegraphics[width=5.8cm,height = 5cm]{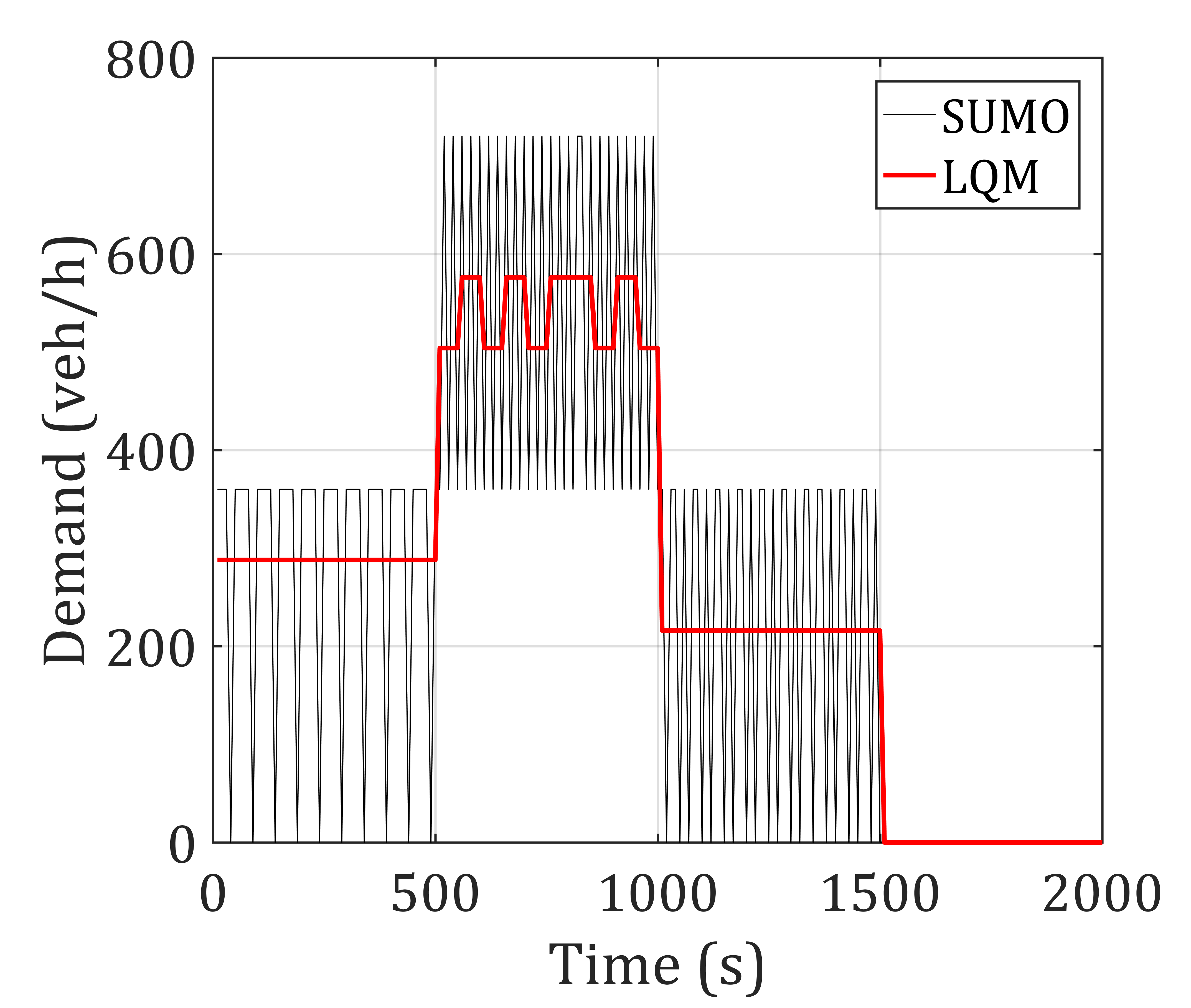}}
\caption{{Traffic demands of the network: (a) $\mathrm O_1$ and $\mathrm O_2$; (b) $\mathrm O_3$ to $\mathrm O_7$.}}
\label{net demand}
\end{figure}

 {To further verify the performance of the proposed LQM in real-life network structures, we expand the simulation scale to a network with three consecutive signalized intersections in Baoding City, China. An overview of the network is visually depicted in Figure \ref{net}(a). The network comprises 50 turn links and 18 common links, each exhibiting varied lengths as illustrated in Figure \ref{net LTM}. Detailed information regarding signal phases and green time durations at these three intersections is provided in Figure \ref{net signal}. Tables \ref{net Road parameters} and \ref{net turning rate} present link-related parameters and the assumed vehicle turning rates within the network. The simulation incorporates TFS on Link 28 to simulate a bottleneck that appears on the network, leading to a variable free-flow speed of Link 28 as specified in Figure \ref{net fs}. As a benchmark model for comparison, we employ SUMO, which accounts for dynamic changes in queue build-up and dissipation, is also capable of accommodating TFS. The simulation model in SUMO is outlined in Figure \ref{net}(b).The input demands for LQM are derived by averaging the input demands in SUMO every 50s due to demand oscillation, as depicted in Figure \ref{net demand}.}

\begin{table}[H]
\captionsetup{font={small}}
\footnotesize
\renewcommand\arraystretch{1}
\caption{{Network ink parameters}\label{net Road parameters}}
%\begin{tabular}{cccp{10cm}}
\centering
\begin{tabular}{m{3.9cm}<{\centering} m{3.85cm}<{\centering} m{3.65cm}<{\centering} }
\hline
\makecell[l]{Parameter}&Left-turn and right-turn links&Through and common links\\
\hline
\makecell[l]{Free-flow speed (m/s)}&6 &11 (except Link 28)\\
% \makecell[l]{Saturation flow rate (veh/h)}&1400&1400\\
\makecell[l]{Backward wave speed (km/h)}&20&20\\
\makecell[l]{Jam density (veh/km)}&100&100\\
\hline
\end{tabular}
\end{table}

\begin{table}[H]
\captionsetup{font={small}}
\footnotesize
\caption{{Vehicle turning rates $e_{j,i}$ and green time fractions $b_R$ in the network\label{net turning rate}}}
\centering
    \begin{tabular}{m{1.3cm}<{\centering} m{0.6cm}<{\centering} m{0.6cm}<{\centering} m{0.6cm}<{\centering} m{0.6cm}<{\centering} m{0.6cm}<{\centering} m{0.6cm}<{\centering} <{\centering} m{0.6cm}<{\centering} m{0.6cm}<{\centering} m{0.6cm}<{\centering} m{0.6cm}<{\centering} m{0.6cm}<{\centering} m{0.6cm}<{\centering}} 
    \hline
    Intersection& \multicolumn{6}{c}{1 and 3}& \multicolumn{6}{c}{2}\\ \hline
         \multirow{2}*{Direction}&  \multicolumn{2}{c}{Through}&  \multicolumn{2}{c}{Left turn}&  \multicolumn{2}{c}{Right turn}&\multicolumn{2}{c}{Through}&  \multicolumn{2}{c}{Left turn}&  \multicolumn{2}{c}{Right turn}\\ 
         \cline{2-13}
         &  $e_{j,i}$&  $b_R$&  $e_{j,i}$&  $b_R$&  $e_{j,i}$& $b_R$&  $e_{j,i}$&  $b_R$&  $e_{j,i}$&  $b_R$&  $e_{j,i}$& $b_R$\\ 
         \hline
         West& 0.6 &0.33  &0.3  &0.16  &0.1  &1 & 0.8& 0.48& 0.2&0.22& 0& 0\\ 
         South& 0.6 &0.25  & 0.3 &0.16  &0.1  &1&0 &0 &0 &0 &0 &0\\ 
         East& 0.6 &0.33  &0.3  &0.16  & 0.1 &1 &0.6 &0.22 &0 &0 &0.4 &1 \\ 
         North&0.6  &0.25  &0.3  &0.16  &0.1  &1 & 0&0&0.6&0.45& 0.4&1\\ 
         \hline
    \end{tabular}
\end{table}

{We quantify the differences in terms of the cumulative flows, and corresponding flow rates, along with queue lengths at each time step between LQM and SUMO. The differences, $\varepsilon\left( {\boldsymbol{x},\boldsymbol{\psi}} \right) $, for all the links in the entire network are summarized in Table \ref{net error}. Figure \ref{net flows} illustrates the results for two representative paths in the network: Path 28-66-26-65-2 and Path 28-66-20. These paths are particularly relevant to bottleneck congestion, offering  insights into queuing phenomena. The simulation period is 2000s and the sampling time interval of LQM is 10s. }

\begin{table}[H]
\captionsetup{font={small}}
\footnotesize
\renewcommand\arraystretch{1}
\caption{{Simulation difference $\varepsilon\left( {\boldsymbol{x},\boldsymbol{\psi}} \right) $ between LQM and SUMO for the entire network}}
    \centering
    \begin{tabular}{ccc} 
    \hline
         Parameter& Average $\varepsilon\left( {\boldsymbol{x},\boldsymbol{\psi}} \right) $  & Standard deviation of $\varepsilon\left( {\boldsymbol{x},\boldsymbol{\psi}} \right) $\\ 
         \hline
         \makecell[l]{Cumulative link inflow}&  2.31veh&  1.31\\ 
         \makecell[l]{Cumulative link outflow}&  2.99veh&  2.16\\ 
         \makecell[l]{Queue length}&  43.05m&  28.24\\ 
         \hline
    \end{tabular}
    \label{net error}
\end{table}

\begin{figure}[H]
\captionsetup{font={small}}
\centering  
\subfigure[Cumulative flow of L28]{
\includegraphics[width=4.8cm,height = 4.1cm]{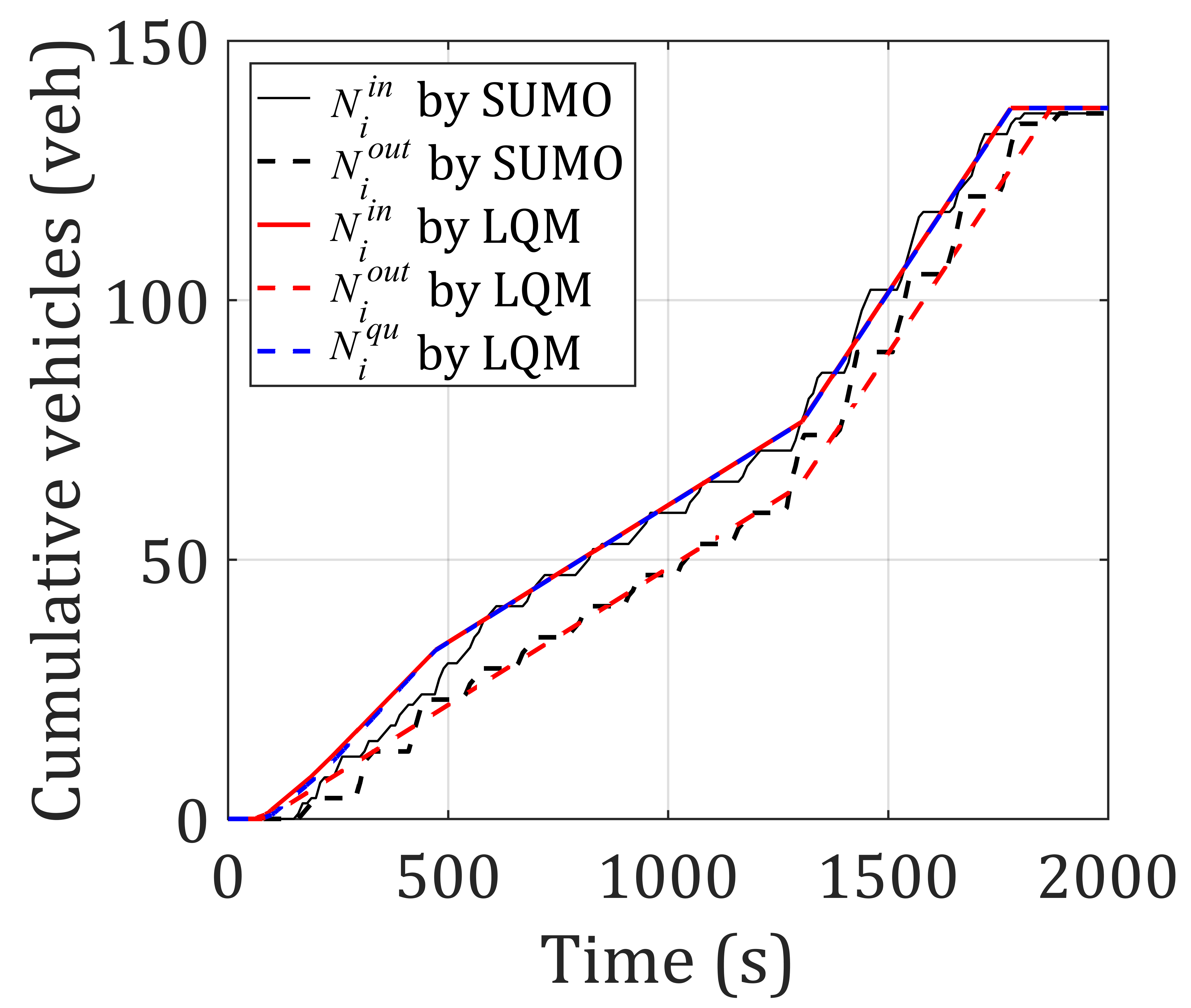}}\subfigure[Flow rate (veh/s) of L28 by LQM]{
\includegraphics[width=4.9cm,height = 4.1cm]{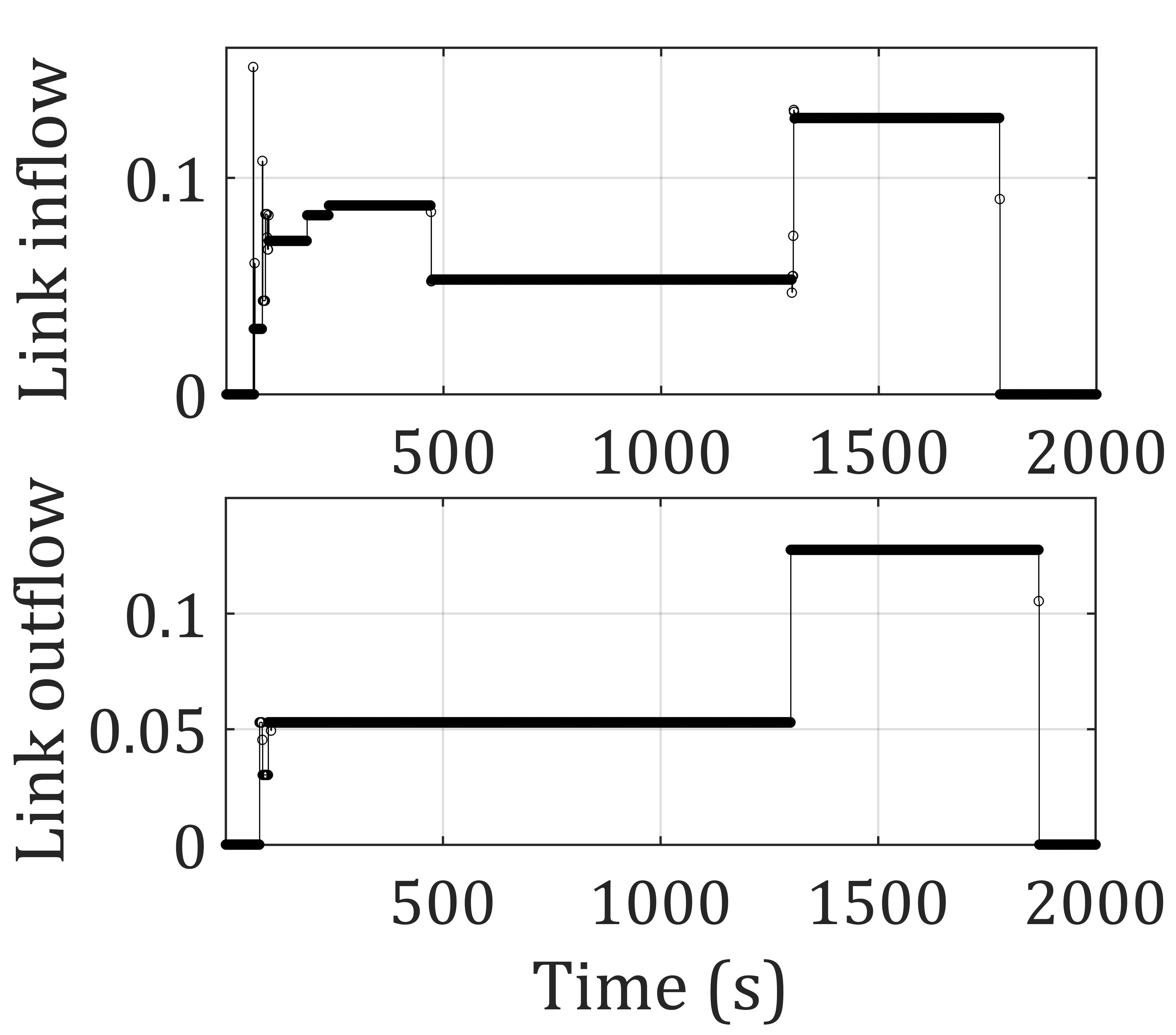} }\subfigure[Queue length of L28]{
\includegraphics[width=4.8cm,height = 4.1cm]{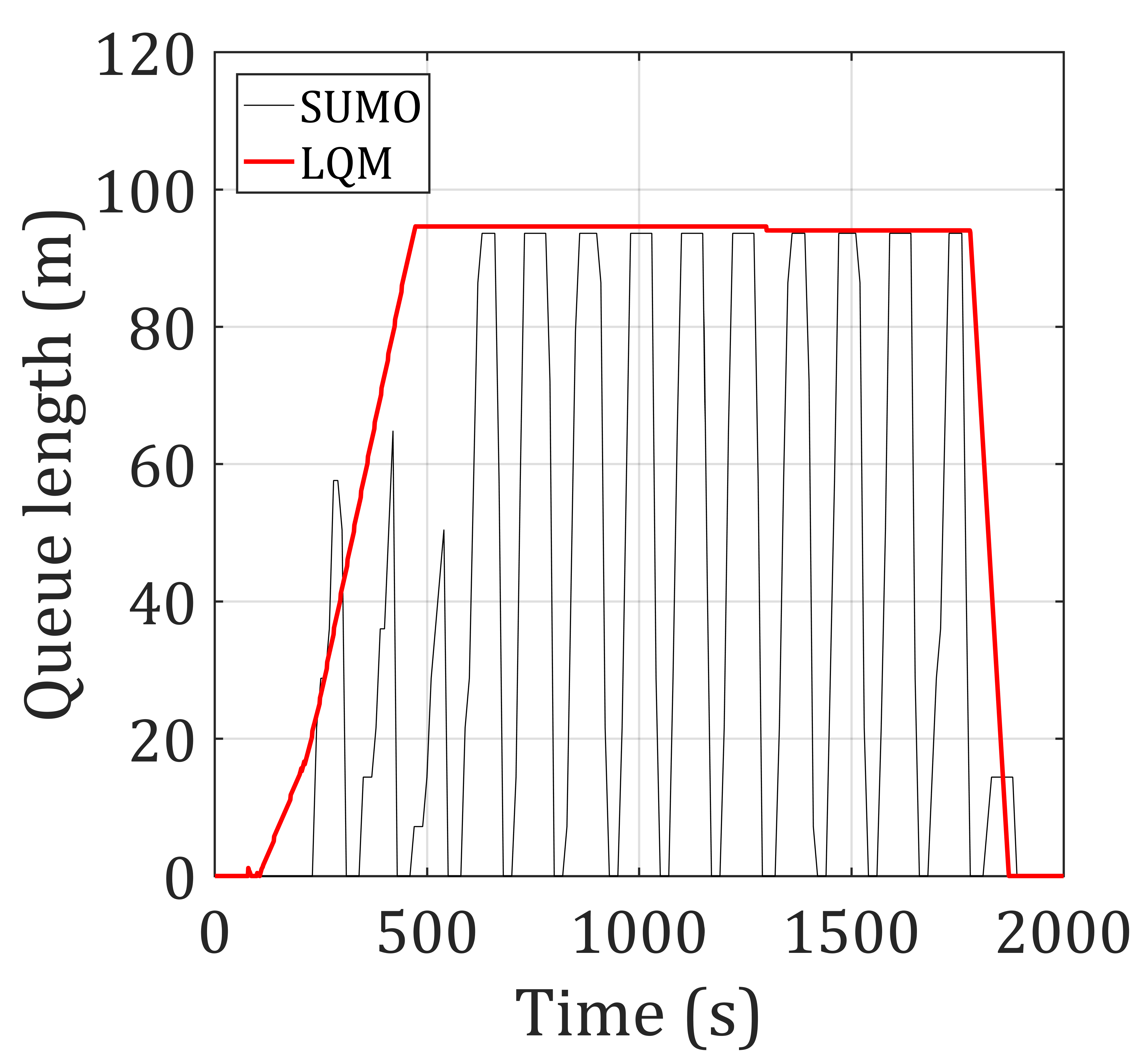}}\\
\subfigure[Cumulative flow of L66]{
\includegraphics[width=4.8cm,height = 4.1cm]{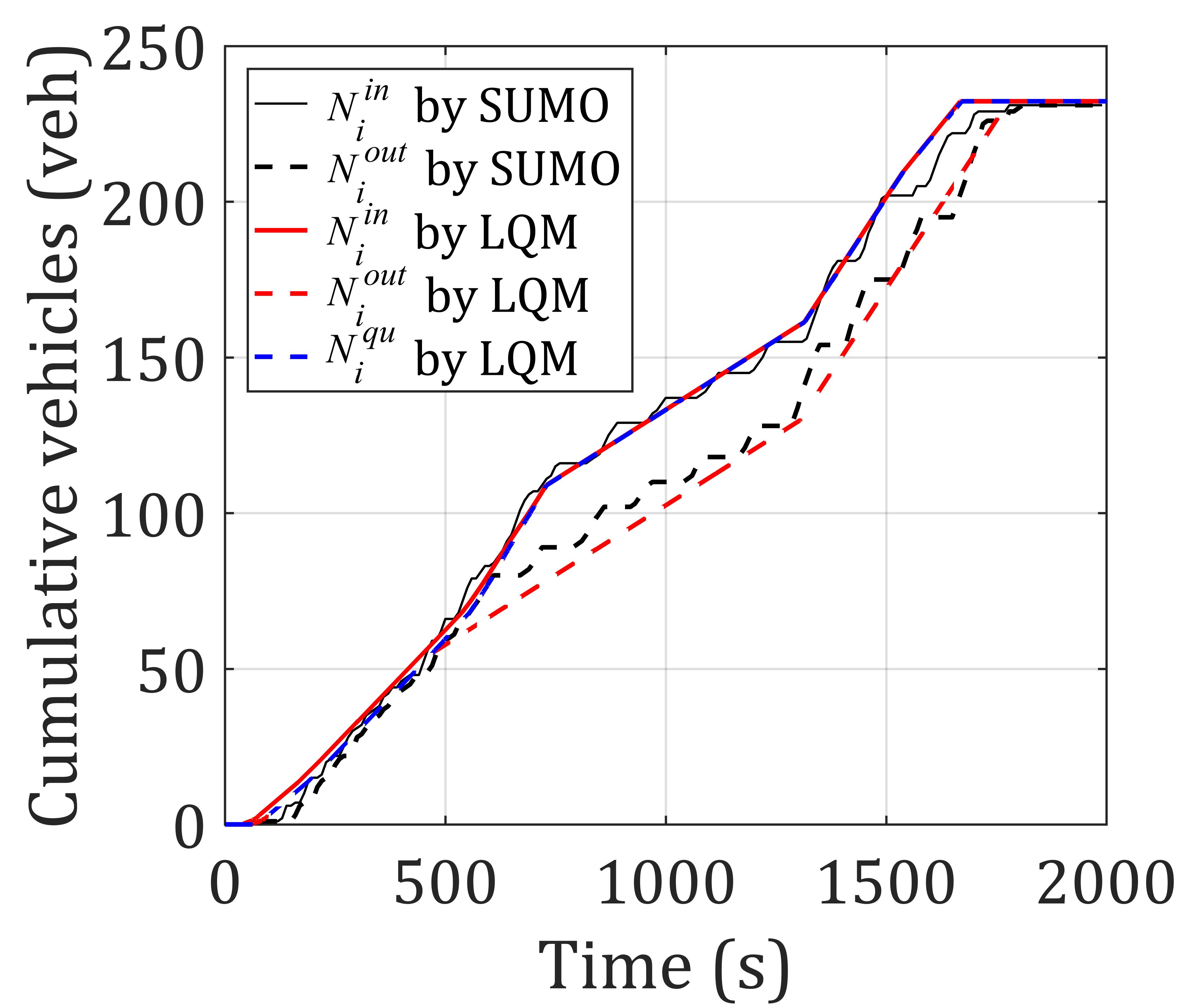}}\subfigure[Flow rate of L66 (veh/s) by LQM]{
\includegraphics[width=4.9cm,height = 4.1cm]{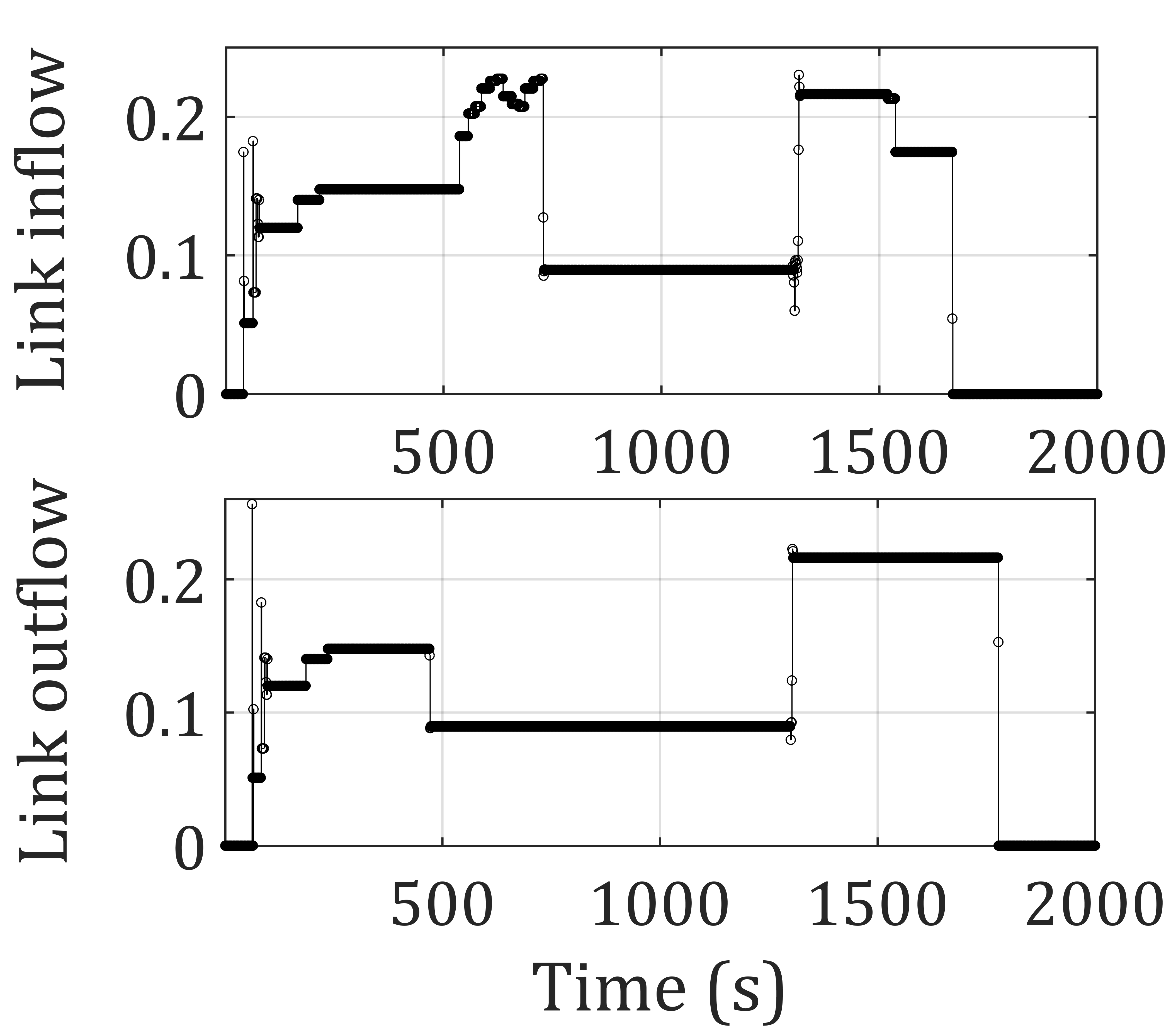} }\subfigure[Queue length of L66]{
\includegraphics[width=4.8cm,height = 4.1cm]{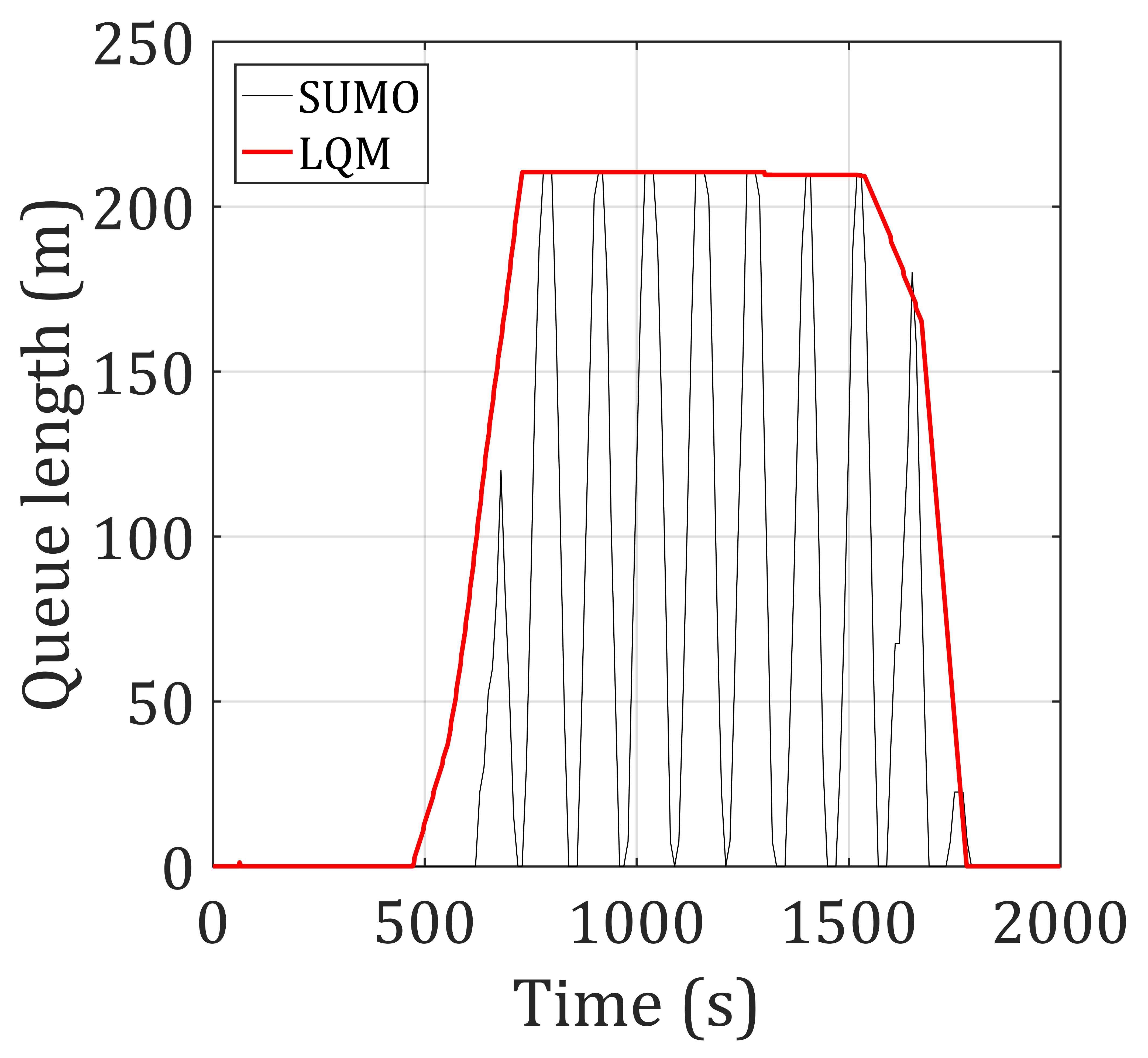}}\\
\subfigure[Cumulative flow of L26]{
\includegraphics[width=4.8cm,height = 4.1cm]{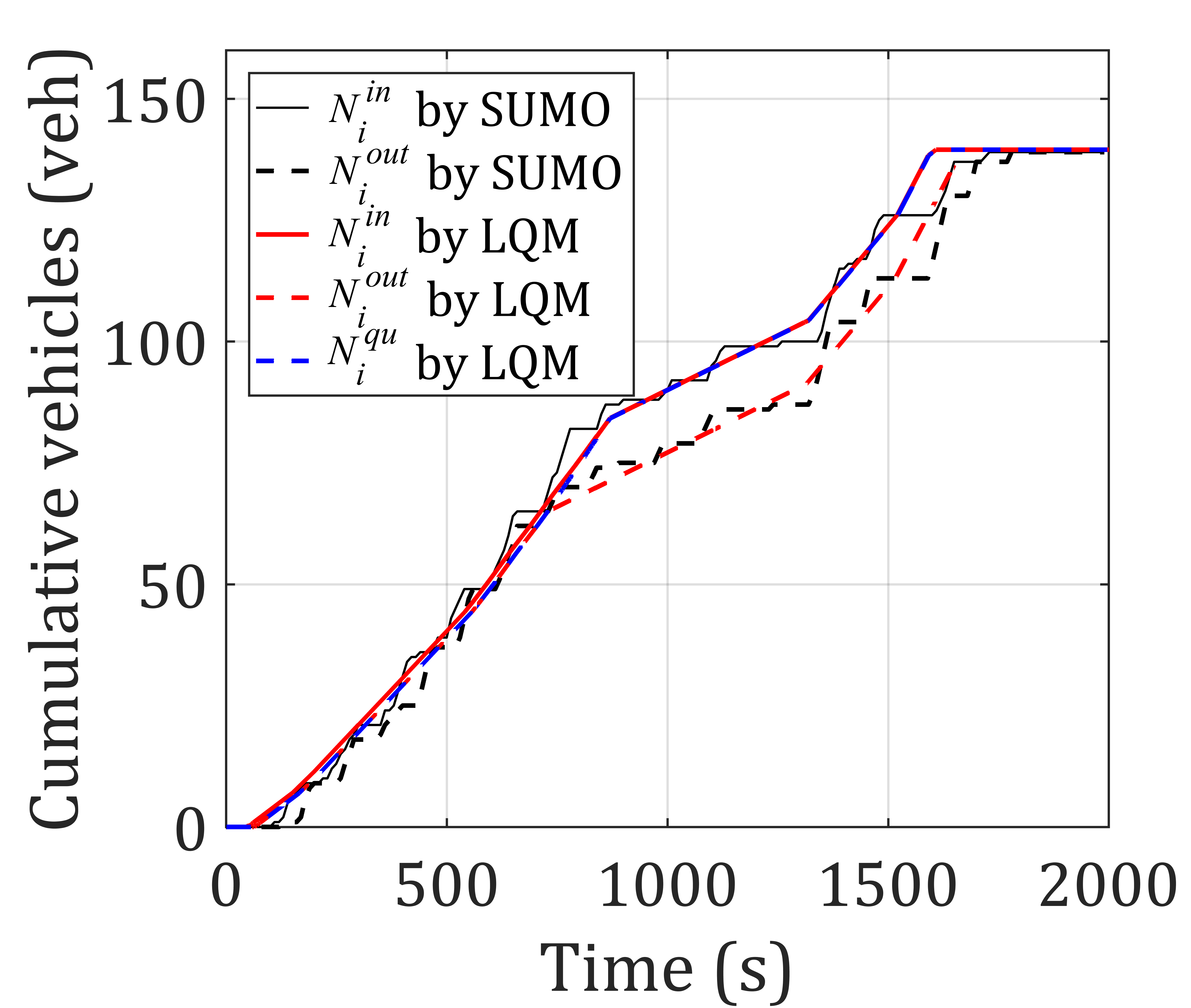}}\subfigure[Flow rate (veh/s) of L26 by LQM]{
\includegraphics[width=4.9cm,height = 4.1cm]{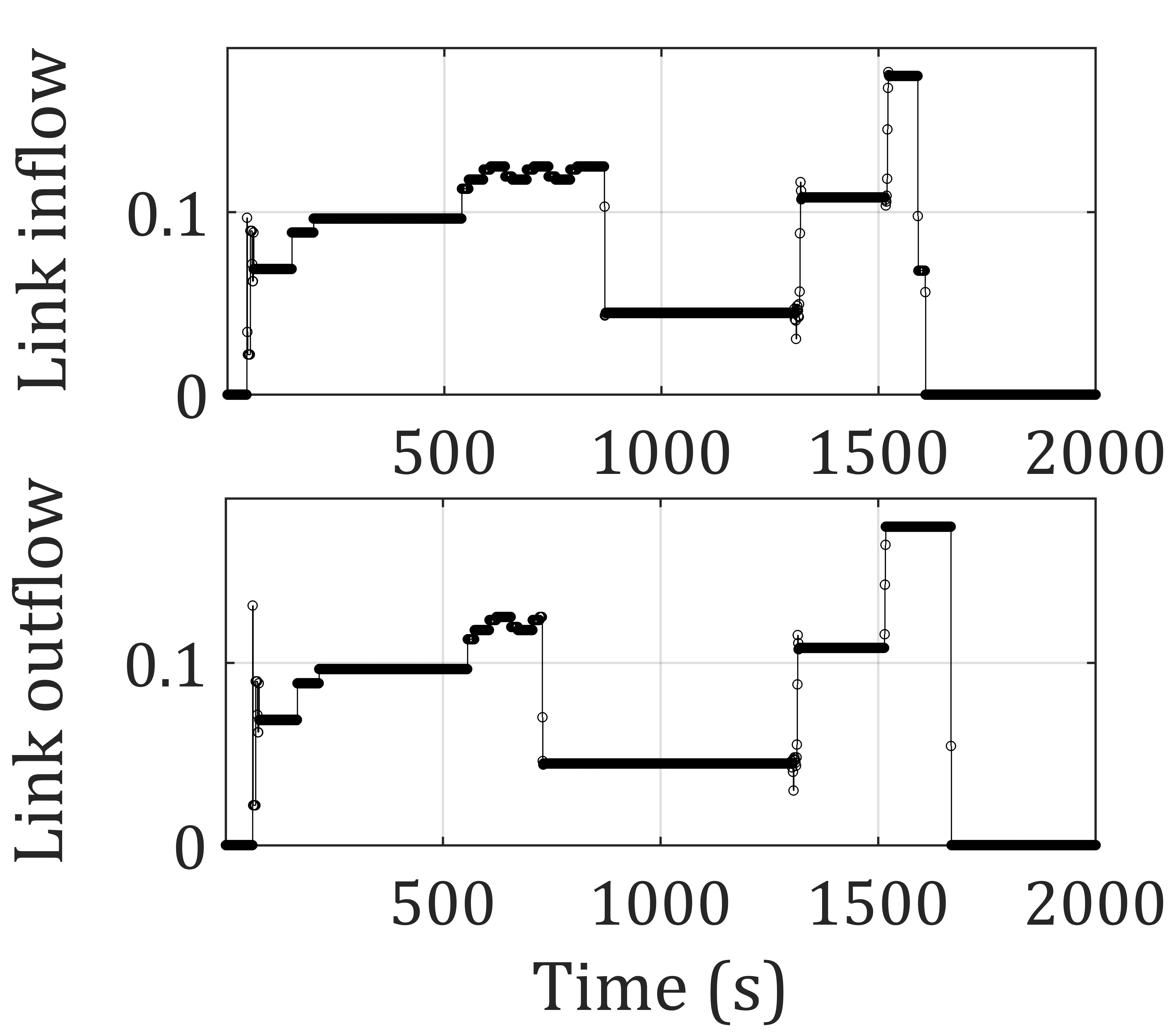} }\subfigure[Queue length of L26]{
\includegraphics[width=4.8cm,height = 4.1cm]{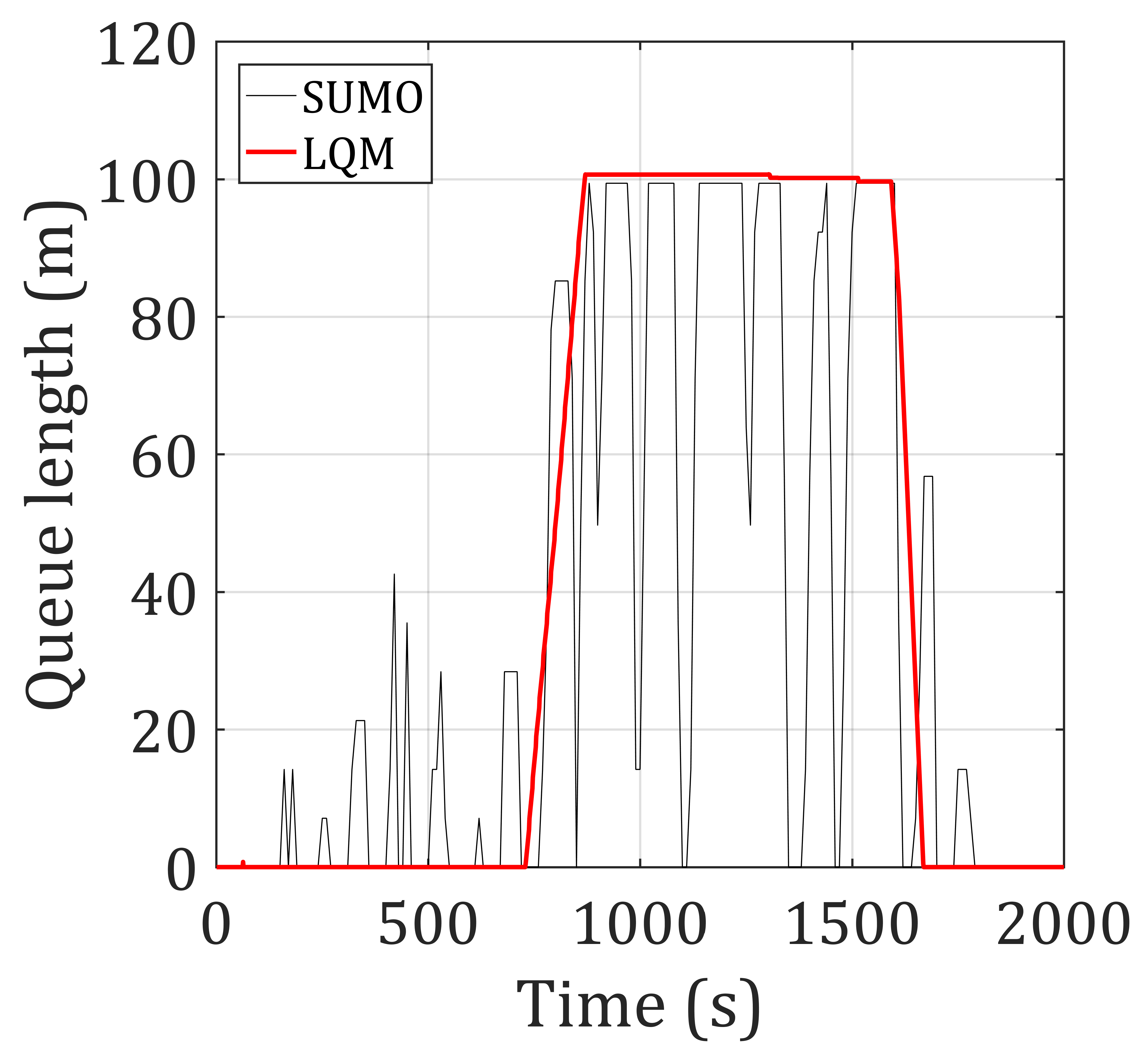}}\\
\subfigure[Cumulative flow of L65]{
\includegraphics[width=4.8cm,height = 4.1cm]{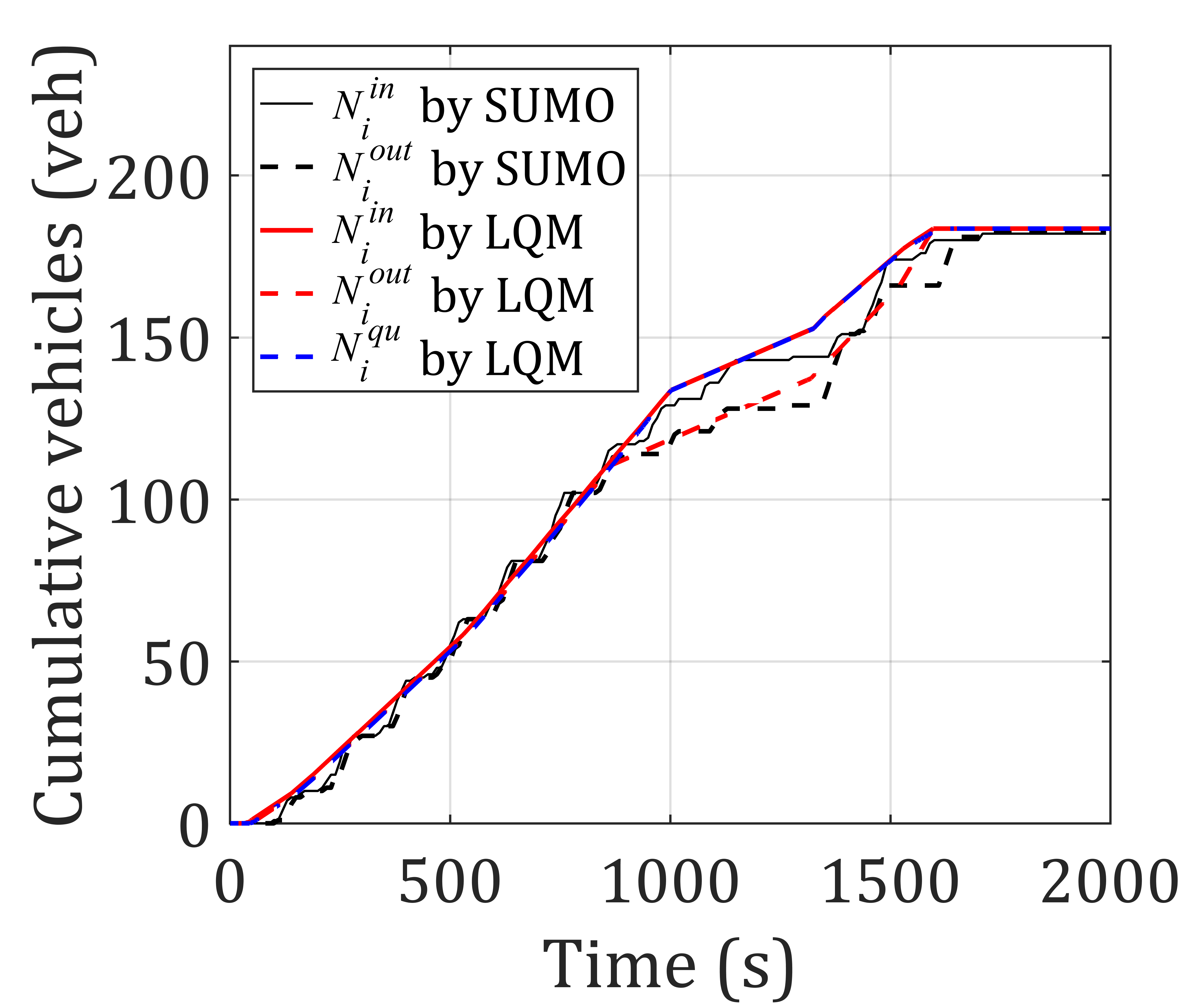}}\subfigure[Flow rate (veh/s) of L65 by LQM]{
\includegraphics[width=4.9cm,height = 4.1cm]{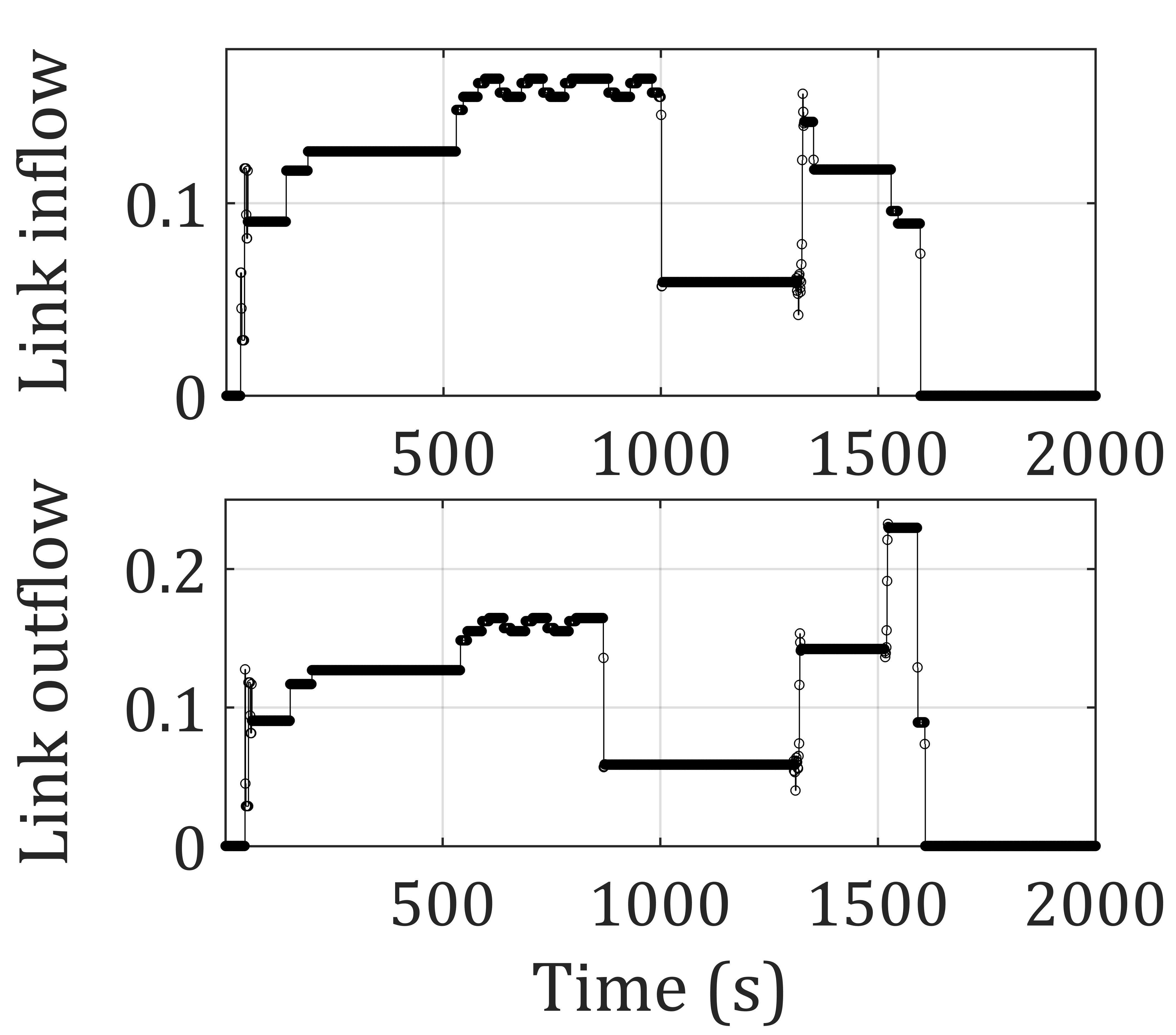} }\subfigure[Queue length of L65]{
\includegraphics[width=4.8cm,height = 4.1cm]{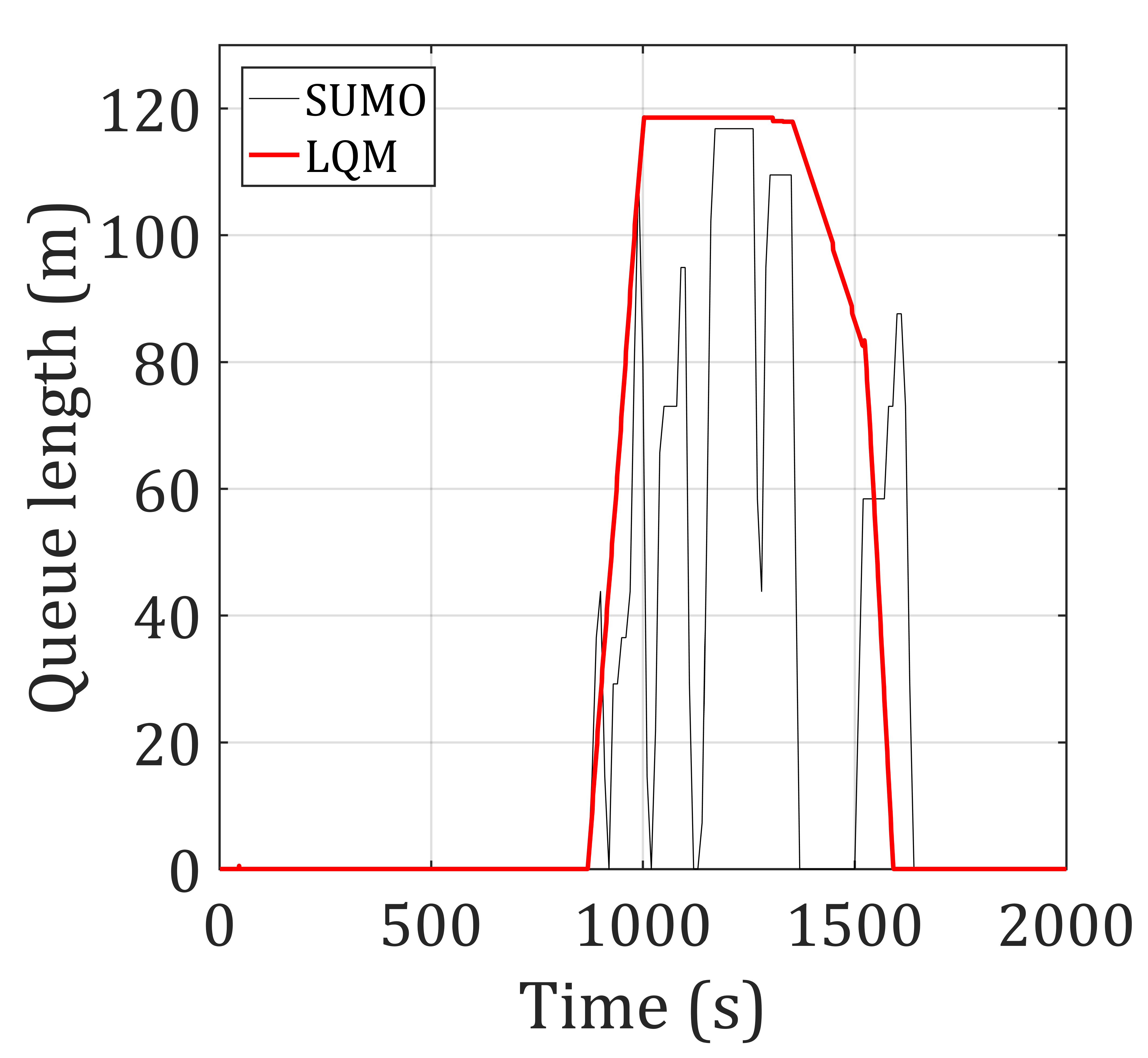}}\\ 
\caption{{Comparison of the bottleneck spillback-related paths simulated by LQM and SUMO}\label{net flows}}
\end{figure}

\begin{figure}[H]
\setcounter{figure}{20} 
\captionsetup{font={small}}
\centering
\setcounter {subfigure} {12}
\subfigure[Cumulative flow of L2]{
\includegraphics[width=4.8cm,height = 4.1cm]{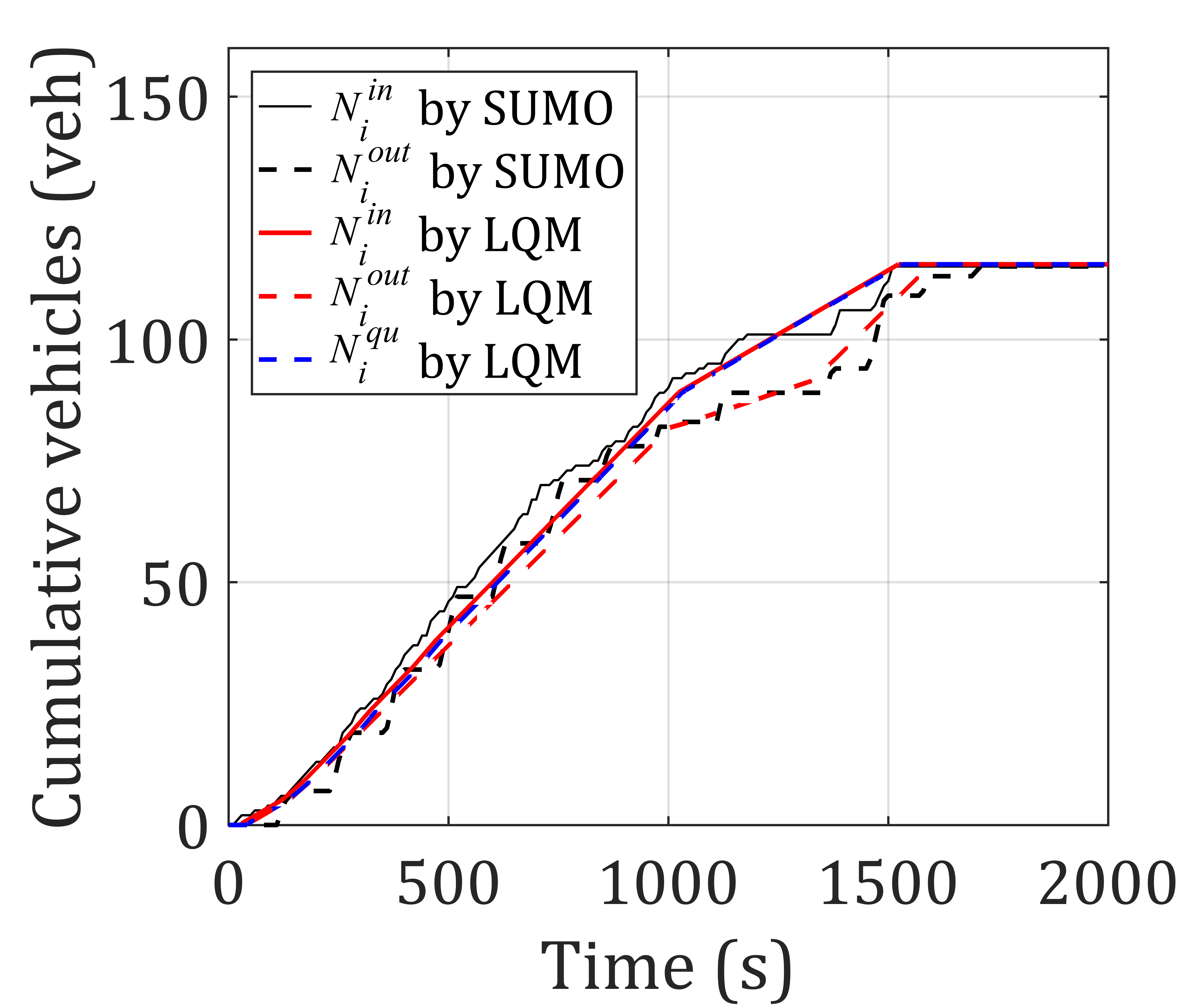}}\subfigure[Flow rate (veh/s) of L2 by LQM]{
\includegraphics[width=4.9cm,height = 4.1cm]{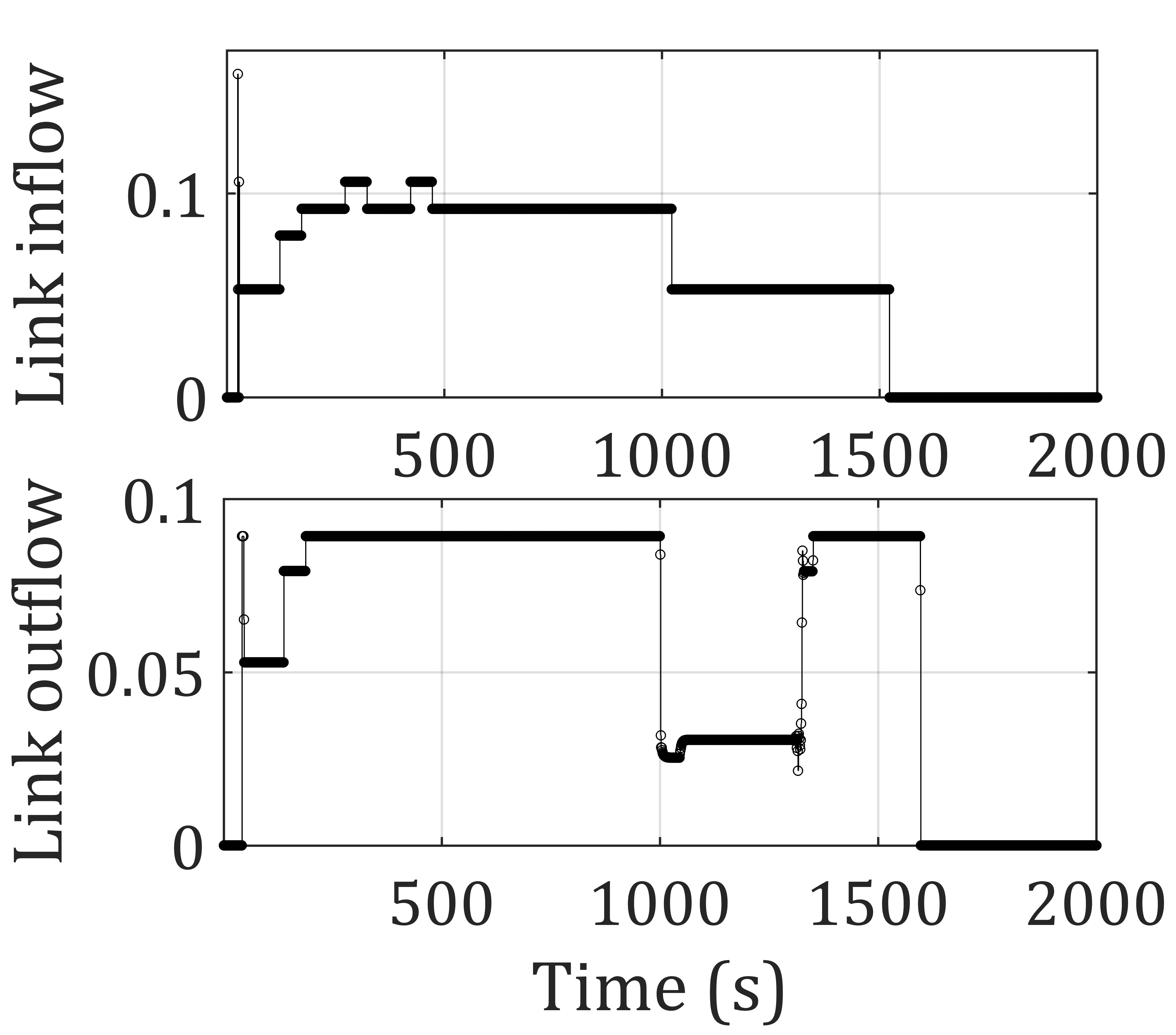} }\subfigure[Queue length of L2]{
\includegraphics[width=4.8cm,height = 4.1cm]{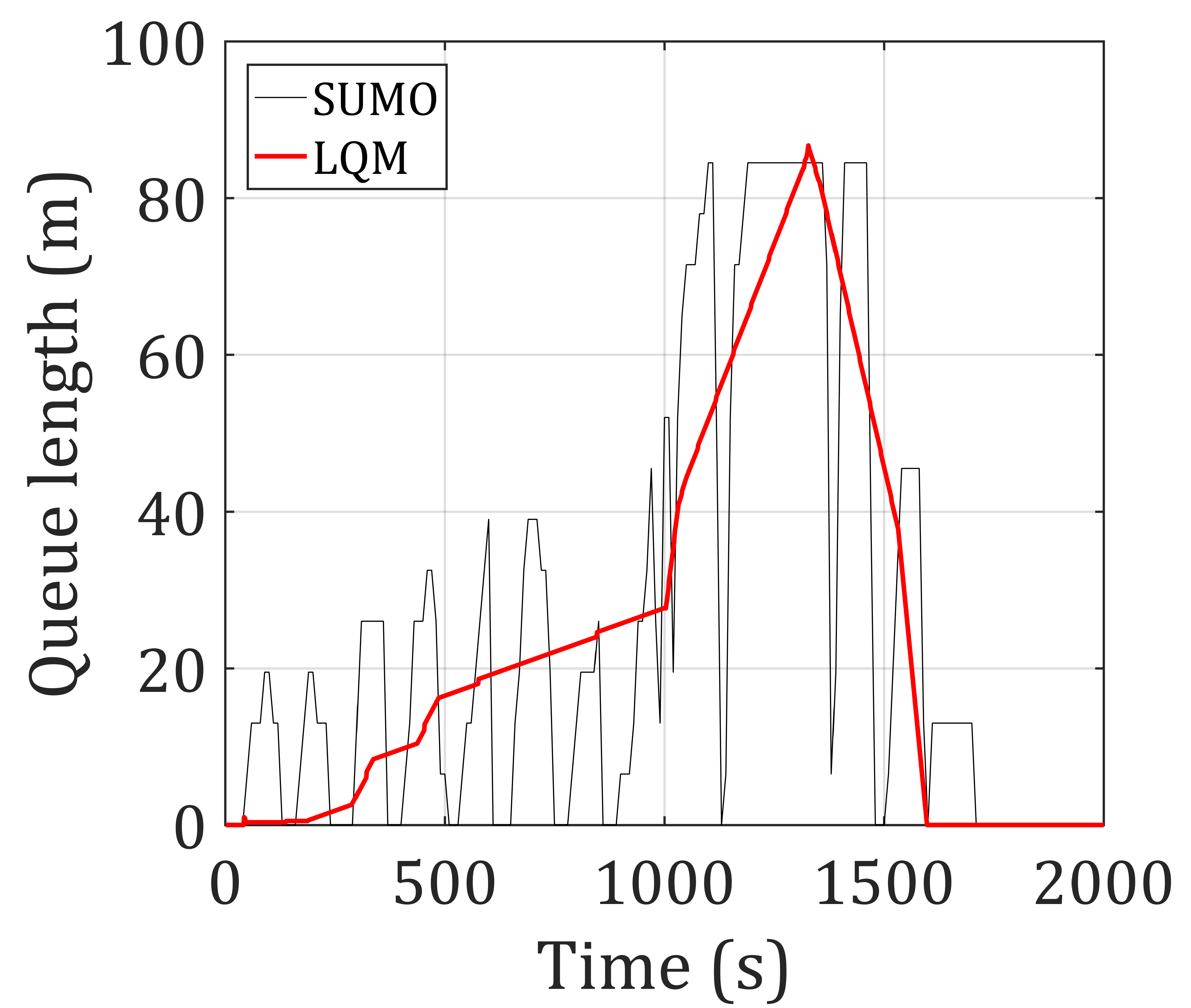}}\\
\subfigure[Cumulative flow of L20]{
\includegraphics[width=4.8cm,height = 4.1cm]{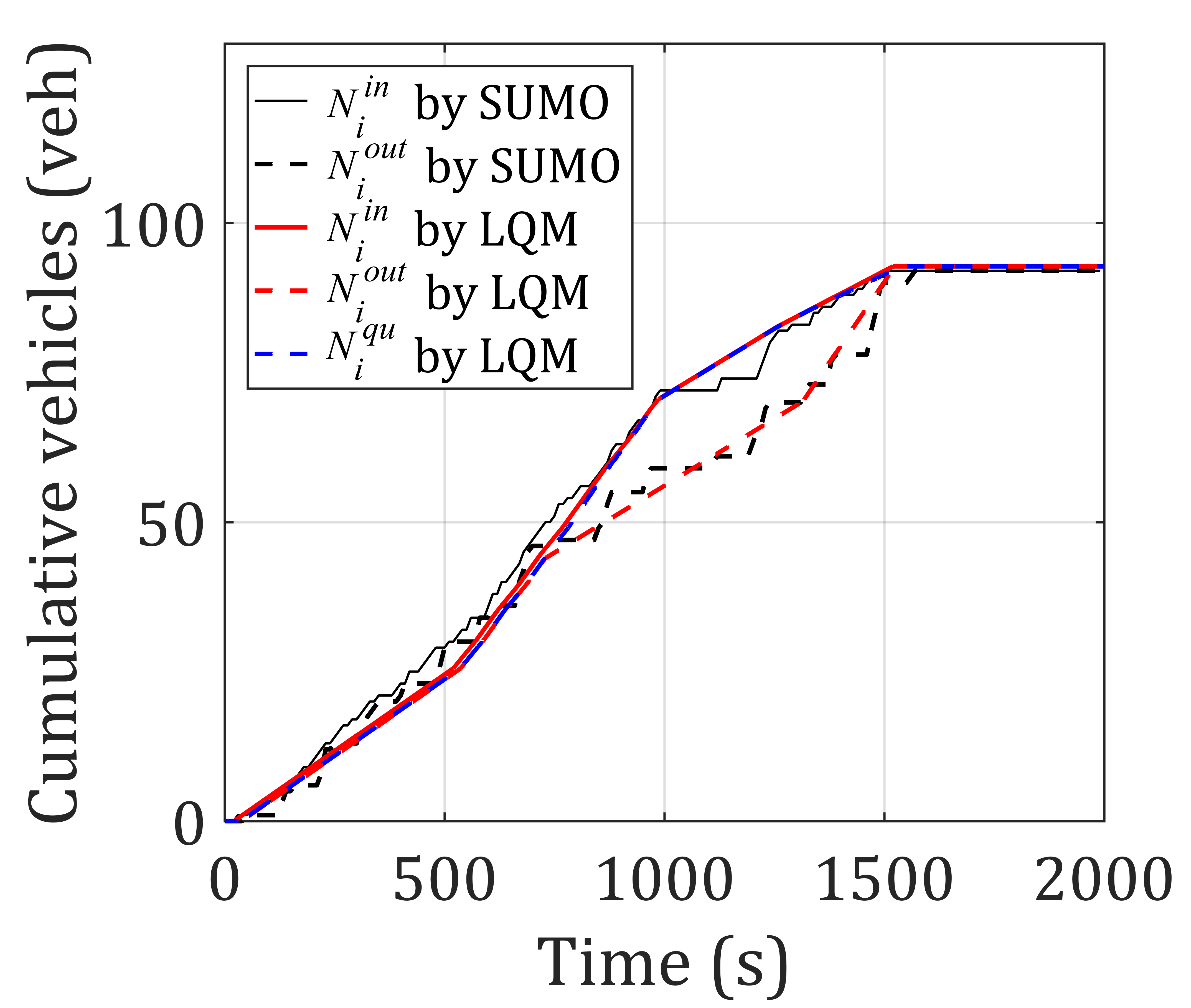}}\subfigure[Flow rate (veh/s) of L20 by LQM]{
\includegraphics[width=4.9cm,height = 4.1cm]{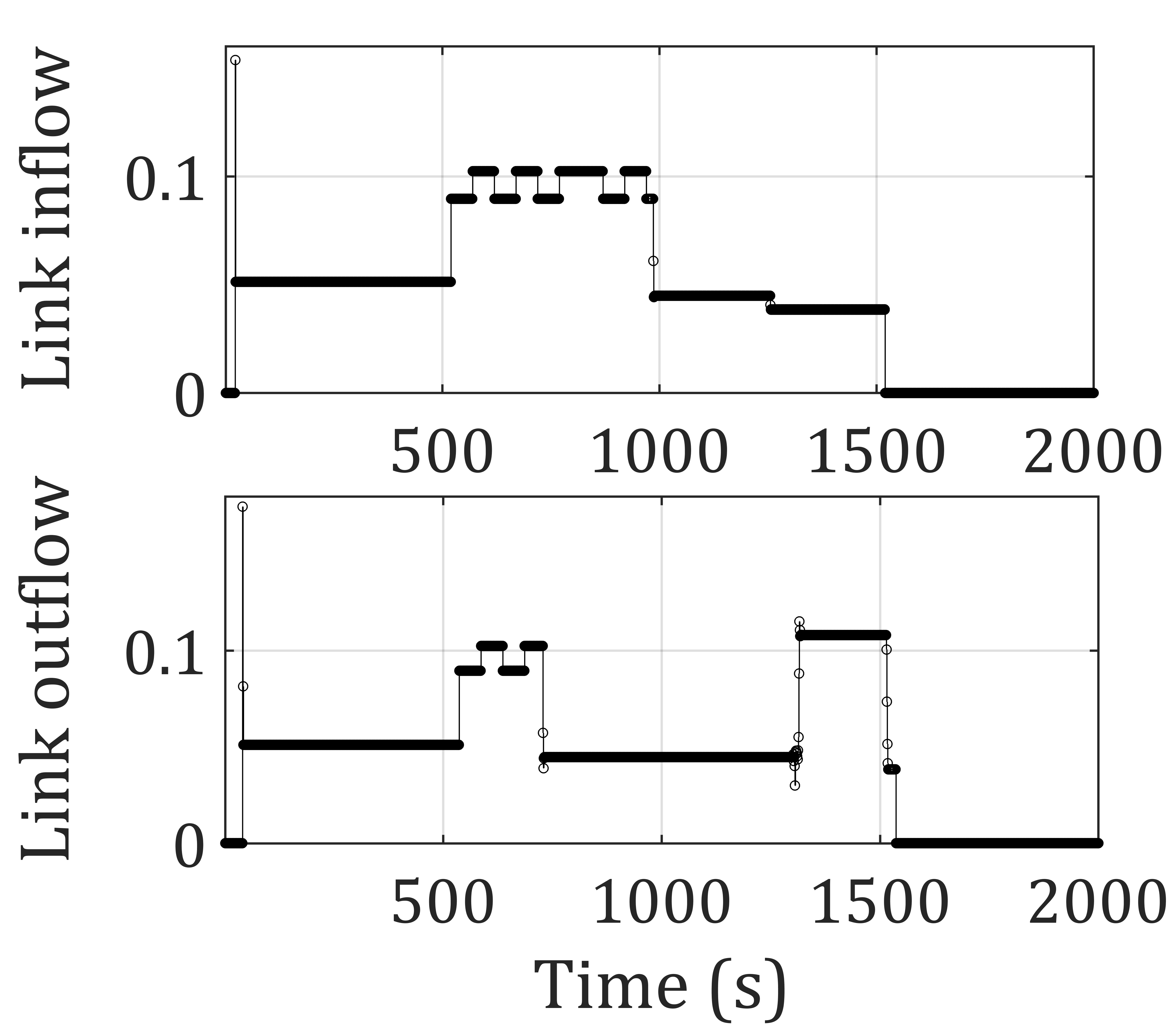} }\subfigure[Queue length of L20]{
\includegraphics[width=4.8cm,height = 4.1cm]{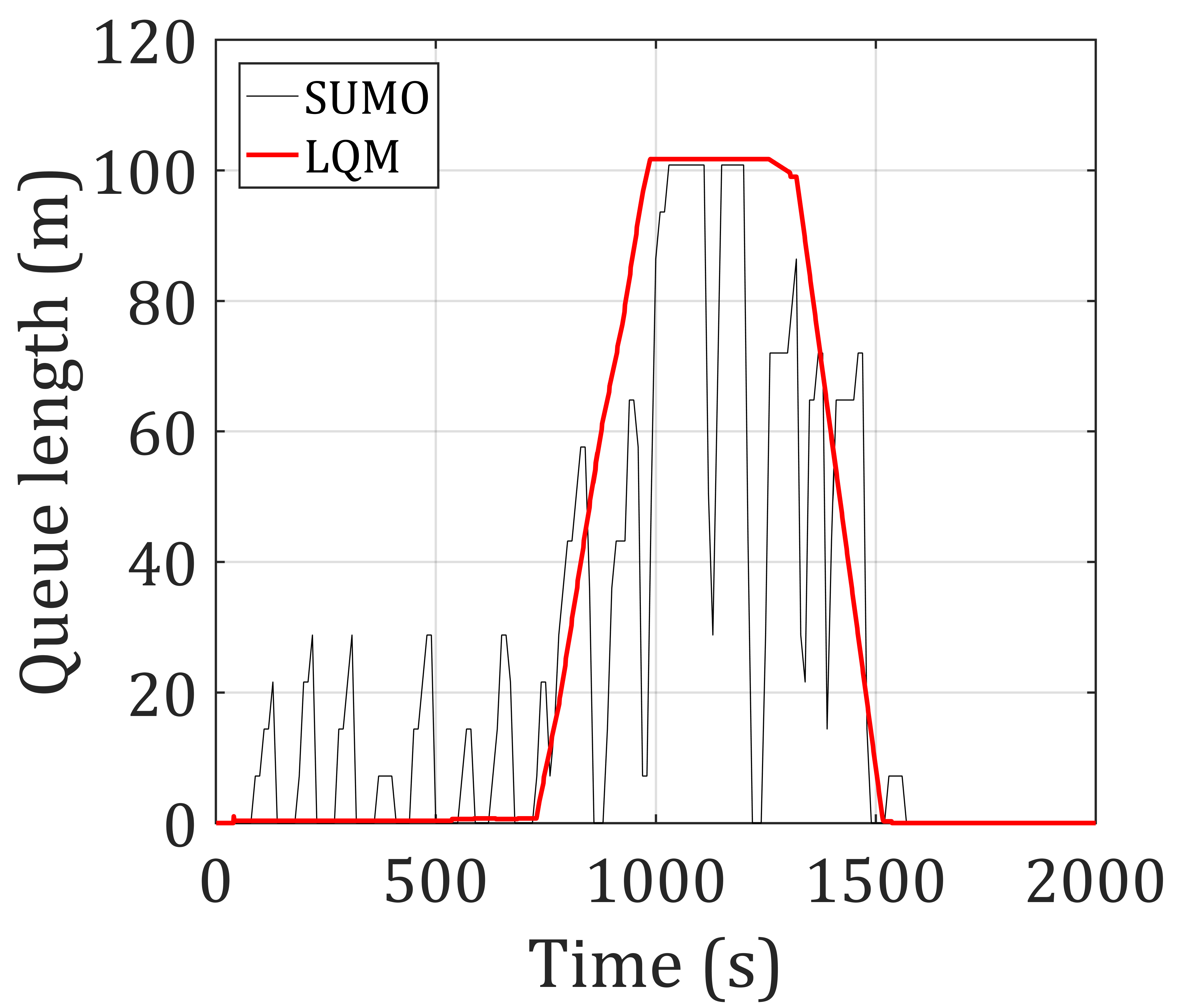}}\\
\caption{{Comparison of the bottleneck spillback-related paths simulated by LQM and SUMO}}
\end{figure}

{As depicted in Figure \ref{net flows}, the cumulative flow curves exhibit remarkable similarity between LQM and SUMO across various links, regardless of their differing lengths. Therefore, although the simulation scale has been expanded, the proposed LQM still reproduces the flow propagation. The flow rates, including both link inflow and outflow rates of each link, occasionally experience fluctuations during some periods, which may be attributed to corresponding variations in input demands, as illustrated in Figure \ref{net demand}, e.g., fluctuations are evident between 0 and 500s of the west and east direction, and between 500 and 1000s of north and south direction.}

{The TFS on Link 28 leads to a decrease in its free-flow speed from 200s onward. Therefore, the queue length on Link 28 experiences a dramatic increase, as depicted in Figure \ref{net flows}(c). Around 490s, spillback emerges on Link 28, causing the queue on its upstream common link (Link 66) to increase from 480s, as shown in Figure \ref{net flows}(f). This spillback propagates along upstream links, including Links 26 and 65, until it reaches Link 2. Consequently, the queue lengths on these links experience significant changes during the simulation process. The consistent changing trends and maximum values of the queue lengths between SUMO and LQM indicate that the proposed model effectively captures flow propagation and queue lengths in the real-sized network. }
 % The changing trends and maximum values of queue length obtained through LQM closely align with those derived from SUMO, indicating that the proposed LQM effectively captures the flow propagation and queue dynamics. 